\documentclass[floatfix,superscriptaddress,a4paper,showpacs,showkeys,nofootinbib,preprint,prc]{revtex4-1}

\usepackage{epsfig}
\usepackage{latexsym}
\usepackage{xspace}
\usepackage[colorlinks=true,linktocpage=true,linkcolor=blue,citecolor=blue,allcolors=blue]{hyperref}
\usepackage[utf8]{inputenc}
\usepackage{indentfirst}
\usepackage{enumerate}
\usepackage{color}
\usepackage{tabularx}

\usepackage{amsmath}
\usepackage{amssymb}
\usepackage[english]{babel}
\usepackage{url}
\newcommand{\eq}[1]{\begin{align} #1 \end{align}}

\begin{document}


\title{
Hagedorn bag-like model with a crossover transition meets lattice~QCD
}

\author{Volodymyr Vovchenko}
\affiliation{
Institut f\"ur Theoretische Physik,
Goethe Universit\"at Frankfurt, Max-von-Laue-Str. 1, D-60438 Frankfurt am Main, Germany}
\affiliation{Frankfurt Institute for Advanced Studies, Giersch Science Center, 
Ruth-Moufang-Str. 1, D-60438 Frankfurt am Main, Germany}

\author{Mark I. Gorenstein}
\affiliation{Frankfurt Institute for Advanced Studies, Giersch Science Center, 
Ruth-Moufang-Str. 1, D-60438 Frankfurt am Main, Germany}
\affiliation{Bogolyubov Institute for Theoretical Physics, 03680 Kiev, Ukraine}

\author{Carsten Greiner}
\affiliation{
Institut f\"ur Theoretische Physik,
Goethe Universit\"at Frankfurt, Max-von-Laue-Str. 1, D-60438 Frankfurt am Main, Germany}

\author{Horst Stoecker}
\affiliation{
Institut f\"ur Theoretische Physik,
Goethe Universit\"at Frankfurt, Max-von-Laue-Str. 1, D-60438 Frankfurt am Main, Germany}
\affiliation{Frankfurt Institute for Advanced Studies, Giersch Science Center, Ruth-Moufang-Str. 1, D-60438 Frankfurt am Main, Germany}
\affiliation{
GSI Helmholtzzentrum f\"ur Schwerionenforschung GmbH, Planckstr. 1, D-64291 Darmstadt, Germany}

\begin{abstract}
Thermodynamic functions, the (higher-order) fluctuations and correlations of conserved charges at $\mu_B = 0$, and the Fourier coefficients of net-baryon density at imaginary $\mu_B$, are considered in the framework of a Hagedorn bag-like model with a crossover transition.
The qualitative behavior of these observables is found to be compatible with lattice QCD results.
Fair quantitative description of the lattice data is obtained when quasiparticle-type quarks and gluons with non-zero masses are introduced into the bag spectrum.
The equation of state of the model exhibits a smooth and wide crossover transition.
\end{abstract}



\maketitle


\section{Introduction}

The empirically known spectrum of hadrons suggests a rapid, possibly exponential, increase of the density of states at large masses~\cite{Patrignani:2016xqp}.
An exponentially rising hadron mass spectrum was first proposed by Hagedorn in the 1960s~\cite{Hagedorn:1965st} in the framework of the statistical bootstrap model, long before the advent of QCD as the fundamental theory of strong interactions.
Evaluations within the MIT bag model~\cite{Chodos:1974je} similarly suggest an exponentially increasing mass spectrum~\cite{Kapusta:1981ay}.
The Hagedorn mass spectrum is characterized by the Hagedorn temperature $T_H \sim 170$~MeV, above which a transition to a new state of matter occurs according to the early ideas~\cite{Cabibbo:1975ig}.
The asymptotic freedom property of QCD suggests a transition to the quark-gluon plasma phase at high temperatures and densities.
First-principle lattice QCD simulations at zero baryochemical potential are consistent with a smooth crossover transition between hadronic and partonic matter~\cite{Aoki:2006we}, characterized by the pseudocritical temperature $T_{pc} \simeq 155$~MeV~\cite{Borsanyi:2010bp,Bazavov:2011nk} obtained from the analysis of chiral observables.

Hagedorn states are possibly created in multi-particle reactions, e.g. during heavy-ion collisions~\cite{Stoecker:1981za}, most abundantly close to the Hagedorn temperature, as was discussed in \cite{NoronhaHostler:2007jf, NoronhaHostler:2009cf}. 
Their appearance can explain the fast chemical equilibration of the hadronic gas, this especially concerns the abundances of (multi-)strange baryons and their anti-particles. 
The presence of Hagedorn states in a hadron resonance gas model provides
also a lowering of the speed of sound and of the shear viscosity over entropy density ratio \cite{NoronhaHostler:2008ju,NoronhaHostler:2012ug}.
Recently also a microcanonical bootstrap description of Hagedorn states with explicit conserved quantum numbers has been
developed: The energy spectra of the resulting hadrons from the decays
of such exotic states follow an exponential law akin with the Hagedorn temperature and thus look thermal by itself~\cite{Beitel:2014kza}. Incorporating those states into a microscopic hadronic transport model,
again fast equilibration of strange and multistrange baryons and mesons has been shown \cite{Beitel:2016ghw}, and the full dynamics of heavy-ion collisions within such an unorthodox picture has also been developed \cite{Gallmeister:2017ths}. 

In the simplest Hagedorn model all hadrons are treated as point particles.
Due to the exponentially increasing hadron mass spectrum, the Hagedorn temperature $T_H$ becomes the limiting temperature above which the partition function diverges -- a behavior which cannot be reconciled with lattice QCD results.
Early on, it has been suggested that hadrons in a statistical system should be treated as spatially extended objects~\cite{Baacke:1976jv,Hagedorn:1980kb,Dixit:1980zj,Gorenstein:1981fa,Kapusta:1982qd}, which essentially corresponds to a van der Waals type excluded volume correction in the partition function of a hadron gas.
The inclusion of the spatial size of hadrons leads to a disappearance of the ``limiting'' temperature under certain conditions~\cite{Kapusta:1982qd}.
Moreover, as shown in Refs.~\cite{Gorenstein:1981fa,Gorenstein:1998am}, there is a possibility of a first-order, second-order, or a crossover transition
in the gas of spatially extended quark-gluon bags, with thermodynamic properties at high temperatures being similar to the MIT bag model equation of state~\cite{Baacke:1976jv}.
Both phases are described within a single partition function.
Different possibilities, such as a crossover transition at zero baryon density and a first-order phase transition at finite baryon density in the gas of quark-gluon bags were explored in various works~\cite{Gorenstein:2005rc,Zakout:2006zj,Zakout:2007nb,Bugaev:2007ww,Ferroni:2008ej,Begun:2009an}.

The temperature dependence of thermodynamic functions at zero chemical potential within the gas of extended quark-gluon bags with a crossover transition was considered in Ref.~\cite{Ferroni:2008ej} in the context of lattice QCD equation of state.
General qualitative features were found to be compatible with lattice QCD, although a quantitative description is lacking.
Recently, a lot of lattice data has appeared on  correlations and fluctuations of conserved charges.
These observables correspond to the derivatives of the partition function with respect to chemical potentials, and they have long been considered sensitive to phase  transitions~\cite{Jeon:2000wg,Asakawa:2000wh}. 
Nowadays, the susceptibilities are actively being used to formulate, test, and constrain various effective QCD models for equation of state at non-zero baryon density~\cite{Albright:2014gva,Albright:2015uua,Vovchenko:2016rkn,Critelli:2017oub,Huovinen:2017ogf,Vovchenko:2017gkg,Motornenko:2018hjw}.
In the present work we explore to what extent the behavior of the conserved charges susceptibilities in the gas of extended hadrons and quark-gluon bags is compatible with lattice QCD.
We also consider an extension of the conventional quark-gluon bag model by introducing the effects of a quasiparticle-like finite, constituent quark and gluon masses, which lead to a substantially improved agreement with the lattice data.

\section{Model description}
\label{sec:model}

\subsection{The partition function}

The model assumes a multi-component system of color neutral objects.
These objects, henceforth referred to as particles, have  finite sizes -- the eigenvolumes.
The particles can carry three abelian charges -- baryon number, electric charge, and strangeness.
These three charges are characterized by the corresponding chemical potentials $\mu_B$, $\mu_Q$, $\mu_S$.
For convenience, we will employ the fugacities, $\lambda_i \equiv e^{\mu_i / T}$.

First, we consider a system with a finite number of different components $f$, a generalization to an infinite number of components will follow later.
The particles under consideration can have arbitrary integer values of baryon charge, electric charge, and strangeness.
It is assumed that particles are non-overlapping -- this constraint is modeled through the excluded-volume correction~\cite{Gorenstein:1981fa,Rischke:1991ke}.
The grand canonical partition functions reads
\eq{\label{eq:ZMultiComp}
Z(T,V,\lambda_B,\lambda_Q,\lambda_S) = 
\sum_{N_1=0}^{\infty} \ldots \sum_{N_f=0}^{\infty} \, 
 \prod_{i=1}^f
\lambda_i^{N_i}~
\frac{[(V~-~\sum_j v_j N_j)\,d_i \, \phi(T;m_i)]^{N_i}}{N_i!}\,\theta(V-\sum_{j=1} v_j  N_j)~.
}
Here $\lambda_i = \lambda_B^{b_i} \, \lambda_Q^{q_i} \, \lambda_S^{s_i}$ where $b_i$, $q_i$, and $s_i$, are the baryon charge, electric charge, and strangeness of particle species $i$, $d_i$ is its degeneracy,
and 
\eq{\label{eq:phi}
\phi(T,m) & = \frac{m^2 T}{2 \pi^2} \, K_2(m/T).
}
The presence of the theta function in~\eqref{eq:ZMultiComp} causes certain technical difficulties.
These technical difficulties can be overcome by considering the isobaric (pressure) ensemble~\cite{Gorenstein:1981fa}.
The isobaric partition function, $\hat{Z}(T,s,\lambda_B,\lambda_Q,\lambda_S)$, is given as the Laplace transform of $Z(T,V,\lambda_B,\lambda_Q,\lambda_S)$~(see details in Ref.~\cite{Gorenstein:1981fa})
\eq{\label{eq:Z}
\hat{Z}(T,s,\lambda_B,\lambda_Q,\lambda_S) & = \int_0^{\infty} Z(T,V,\lambda_B,\lambda_Q,\lambda_S) \, e^{-sV} \, dV = [s - f(T,s,\lambda_B,\lambda_Q,\lambda_S)]^{-1}, \\
\label{eq:f}
f(T,s,\lambda_B,\lambda_Q,\lambda_S) & = \sum_{i=1}^f  \, \lambda_B^{b_i} \, \lambda_Q^{q_i} \, \lambda_S^{s_i} \, d_i\, \phi(T,m_i) \, e^{-v_i \, s}~.
}

In the thermodynamic limit, $V \to \infty$, the grand canonical partition function behaves as $Z(T,V,\lambda_B,\lambda_Q,\lambda_S) \simeq \exp \left[p(T,\lambda_B,\lambda_Q,\lambda_S) \, V/ T \right]$.
The isobaric partition function in the thermodynamic limit has the form
\eq{\label{eq:ZTD}
\hat{Z}(T,s,\lambda_B,\lambda_Q,\lambda_S) & \propto \int_0^{\infty} \exp\left[\frac{V}{T} (p-Ts) \right] \, dV~.
}
It follows from Eq.~\eqref{eq:ZTD} that $\hat{Z}(T,s,\lambda_B,\lambda_Q,\lambda_S)$ has a singularity at $s = s^* = p / T$, and no singularities at $s > s^*$.
The integral representation~\eqref{eq:ZTD} is unconditionally divergent at $s < s^*$, and, therefore, does not directly provide information about the behavior of the isobaric partition function in that region.
Thus, there is a possibility that $\hat{Z}$ has singularities at $s < s^*$.
These considerations lead to the conclusion that the system pressure $p(T,\lambda_B,\lambda_Q,\lambda_S)$ in the thermodynamic limit is defined by the \emph{farthest-right} singularity $s^*$ of the isobaric partition function $\hat{Z}$, i.e. 
\eq{\label{eq:ps*}
p(T,\lambda_B,\lambda_Q,\lambda_S) = T s^*.
}
The exact nature of the farthest-right singularity $s^*$ depends on the input particle spectrum. 

\subsection{The mass-volume density of states}

In the isobaric ensemble, the input particle spectrum enters through Eq.~\eqref{eq:f} only.
This permits a generalization to a system with an infinite number of different components, characterized by some density function.
First, let us rewrite Eq.~\eqref{eq:f} in the following form
\eq{\label{eq:fbqs}
f(T,s,\lambda_B,\lambda_Q,\lambda_S) & = \sum_{b = -\infty}^{\infty} \, \sum_{q = -\infty}^{\infty} \, \sum_{s = -\infty}^{\infty} \sum_{i \in \{b,q,s\} }     \, \lambda_B^{b} \, \lambda_Q^{q} \, \lambda_S^{s} \, d_i \, \phi(T,m_i) \, e^{-v_i \, s}~.
}
Here the sum $i$ goes through all particles which carry the specific baryon number $b$, the electric charge $q$, and strangeness $s$.
One can now introduce a mass-volume density of states $\rho(m,v;b,q,s)$, which determines the number of particle states carrying fixed quantum numbers $b$, $q$, and $s$, in the mass-volume interval $[m,v; m + dm, v + dv]$.
Equation~\eqref{eq:fbqs} is generalized to
\eq{\label{eq:fbqs2}
f(T,s,\lambda_B,\lambda_Q,\lambda_S) & = \sum_{b = -\infty}^{\infty} \, \sum_{q = -\infty}^{\infty} \, \sum_{s = -\infty}^{\infty} \int dv \int dm \, \lambda_B^{b} \, \lambda_Q^{q} \, \lambda_S^{s} \, \rho(m,v;b,q,s) \, \phi(T,m)  \, e^{-v \, s}~.
}
Equation~\eqref{eq:fbqs2} is reduced to~\eqref{eq:fbqs} for $\rho(m,v;b,q,s) = \sum_{i \in \{b,q,s\}} \, d_i \, \delta(m - m_i) \, \delta(v - v_i)$.

Finally, let us rewrite Eq.~\eqref{eq:fbqs2} as follows
\eq{
\label{eq:full}
f(T,s,\lambda_B,\lambda_Q,\lambda_S) & = \int dv \, \int dm \, \rho(m,v; \lambda_B,\lambda_Q,\lambda_S)  \, \phi(T,m) \, e^{-v \, s},
}
where we have introduced the generalized fugacity dependent mass-volume density of states:
\eq{\label{eq:full}
\rho(m,v; \lambda_B,\lambda_Q,\lambda_S) = \sum_{b = -\infty}^{\infty} \, \sum_{q = -\infty}^{\infty} \, \sum_{s = -\infty}^{\infty} \, \lambda_B^{b} \, \lambda_Q^{q} \, \lambda_S^{s} \, \rho(m,v;b,q,s).
}
Note that $\rho(m,v) \equiv \rho(m,v; \lambda_B = 1,\lambda_Q = 1,\lambda_S = 1)$ is the mass-volume density of all states irrespective of their quantum numbers.

We follow the picture presented in Refs.~\cite{Gorenstein:1998am,Gorenstein:2005rc}.
The particle spectrum
is assumed to
consist of two contributions: 
\begin{enumerate}
    \item the established, low mass hadrons and resonances listed in Particle Data Tables~\cite{Patrignani:2016xqp}; 
    \item an exponential Hagedorn spectrum of the heavy quark-gluon bags.
\end{enumerate}
Therefore,
\eq{\label{eq:rho}
\rho(m,v; \lambda_B,\lambda_Q,\lambda_S) & = \rho_H(m,v; \lambda_B,\lambda_Q,\lambda_S) + \rho_Q(m,v; \lambda_B,\lambda_Q,\lambda_S), 
}
where $\rho_H$ corresponds to the established hadrons listed in PDG, and $\rho_Q$ corresponds to the quark-gluon bags.
The PDG hadrons form a discrete spectrum, therefore $\rho_H$ is given as a finite sum of $\delta$ functions:\footnote{We neglect here the effects of finite resonance widths.  These can have an important effect in precision thermal model applications, such as thermal fits~\cite{Vovchenko:2018fmh}, but are not very relevant for the mostly qualitative aspects of the equation of state studied here.  }
\eq{\label{eq:rhoH} 
\rho_H(m,v; \lambda_B,\lambda_Q,\lambda_S) & = \sum_{i \in \textrm{HRG}} \lambda_B^{b_i} \, \lambda_Q^{q_i} \, \lambda_S^{s_i} \,  d_i \, \delta(m - m_i) \, \delta(v - v_i). 
}

Each of the PDG hadrons is assumed to have a finite eigenvolume parameter $v_i$.
In the spirit of the bag model, we assume here that the eigenvolumes of the PDG hadrons are proportional to their mass: $v_i = m_i / \varepsilon_0$, where $\varepsilon_0 = 4B$ unless stated otherwise.
Here $B$ is the bag constant.
In principle, one can consider a different parametrization of $v_i$, e.g. based on  phenomenological knowledge of some hadron-hadron interactions, the only requirement here is that all eigenvolumes are non-vanishing, i.e. $v_i > 0$. 

\subsection{Quark-gluon bags}

The mass-volume spectrum of the quark-gluon bags, $\rho_Q$, is the crucial ingredient of the model, which determines some of its most qualitative features.
The form of $\rho_Q$ depends strongly on the assumptions regarding the internal color-flavor structure of the bags~(see, e.g., Refs.~\cite{Zakout:2006zj,Zakout:2007nb}).
In the region where both $m$ and $v$ are large, the spectrum can be described in the framework of the bag model~\cite{Chodos:1974je}.
The mass-volume density of states was computed assuming bags filled with non-interacting massless quarks and gluons, at zero chemical potentials~\cite{Kapusta:1981ay,Gorenstein:1982if,Gorenstein:1983rm}, and also for finite baryon chemical potential~\cite{Gorenstein:1982ua}.
One obtains
\eq{
\rho_Q(m,v; \lambda_B,\lambda_Q,\lambda_S) & = C \, v^{\gamma} \, (m - Bv)^{\delta} \, \exp \left\{ \frac{4}{3} [\sigma_{Q}(\lambda_B,\lambda_Q,\lambda_S)]^{1/4} \, v^{1/4} \, (m-Bv)^{3/4} \right\} \nonumber \\ 
& \quad \times \theta(v-V_0) \, \theta(m - Bv - M_0).
\label{eq:rhoQ} 
}
Here $V_0$ is a model parameter which is given a sufficiently large value, $B$ is the bag constant, and $M_0 > 0$ is a parameter which regularizes the mass-volume density close to the lower threshold~\cite{Gorenstein:2005rc}. As will be shown, the exact value of $M_0$ has no significance for applications considered in this paper. 
$\sigma_Q$ corresponds to the energy density (or three times the pressure) of the non-interacting gas of massless quarks and gluons.
Here this quantity is taken as a function of all three chemical potentials in (2+1)-flavor QCD:
\eq{\label{eq:SB}
\sigma_{Q}(\lambda_B,\lambda_Q,\lambda_S) & = \frac{19 \pi^2}{12} +
\sum_{f = u,d,s} \, \left[ \frac{3}{2} \, (\log \lambda_f)^2 + \frac{3}{4 \pi^2} \, (\log \lambda_f)^4 \right]
}
with $\lambda_u = \lambda_B^{1/3} \, \lambda_Q^{2/3}$, $\lambda_d = \lambda_B^{1/3} \, \lambda_Q^{-1/3}$, and $\lambda_s = \lambda_B^{1/3} \, \lambda_Q^{-1/3} \, \lambda_S^{-1}$.

Equation~\eqref{eq:rhoQ} implies that the eigenvolume of a QGP bag with a given mass is fluctuating.
These fluctuations are given by the distribution $\rho_Q(m,v; \lambda_B)$.
The presence of volume fluctuations is crucial for obtaining a transition to quark-gluon plasma: assuming the fixed mass-volume relation, e.g. $m = 4Bv$ from the MIT bag model~\cite{Chodos:1974je}, leads to a constant energy density at high temperature, which is incompatible with lattice QCD~(see Refs.~\cite{Zakout:2006zj,Zakout:2007nb,Ferroni:2008ej} for more details).

The values of parameters $C$, $\gamma$, and $\delta$ in the
pre-exponential factor in Eq.~\eqref{eq:rhoQ} depend strongly
on the details of the bag model calculation. 
For example, they depend on whether the colourlessness constraint for the bags is implemented~\cite{Gorenstein:1982ib,Gorenstein:1982ua,Gorenstein:1983rm}, as well as on other internal symmetry constraints considered~\cite{Zakout:2006zj,Zakout:2007nb}.
Therefore, these parameters are usually treated as free model parameters.
Such a philosophy is considered in the present work as well.
Similar arguments apply to the possible dependence of $C$, $\gamma$, and $\delta$ on fugacities $\lambda_B$, $\lambda_Q$, $\lambda_S$. 
In the absence of a detailed knowledge, we omit the possible dependence of these parameters on fugacities in the present study. 

\paragraph*{The Hagedorn temperature.}

The exponential hadron mass spectrum was first introduced by Hagedorn~\cite{Hagedorn:1965st}. It has the general form 
\eq{\label{eq:rhoHagedorn}
\rho_{\rm hag}(m) = A \, m^{-\alpha} \, \exp(m/T_H)~,
}
with the parameter $T_H$ called the Hagedorn temperature.
The mass-volume relation~\eqref{eq:rhoQ} for quark-gluon bags employed in the present work also implies this same exponential asymptotic mass spectrum.
The mass spectrum for the quark-gluon bags reads
\eq{\label{eq:rhoQm}
\rho_Q(m) = \int dv \, \rho_Q(m,v) = C \int_{V_0}^{(m-M_0)/B} dv \, v^\gamma \, (m-Bv)^{\delta} \,  \exp  [w(v;m)]~,
}
with $w(v;m) =  \frac{4}{3} [\sigma_{Q}]^{1/4} \, v^{1/4} \, (m-Bv)^{3/4}$.
The integral converges as long as $M_0 > 0$.
The function $w(v;m)$ has a peak for large values of $m$.
Therefore, the integral in \eqref{eq:rhoQ} can be approximated for large $m$ with the Laplace's method.
One has $w(v;m) \approx w(v_0;m) + \frac{1}{2} w''_{vv}(v_0;m) \, (v-v_0)^2$ where $v_0$ satisfies the condition $w'_v (v_0;m) = 0$. This yields:
\eq{\label{eq:v0}
v_0 & = \frac{m}{4B}~, \\
w(v_0;m) & = \left( \frac{\sigma_Q}{3B} \right)^{1/4} \, m~, \\
w''_{vv}(v_0;m) & = - \frac{16 \, (\sigma_Q)^{1/4} \, B^{7/4}}{3^{5/4} \, m}~.
}
Equation~\eqref{eq:v0} is the familiar relation between the mass of a quark-gluon bag and its average volume from the MIT bag model with massless quarks~\cite{Chodos:1974je}.
Applying Laplace's method to Eq.~\eqref{eq:rhoQm} one obtains the asymptotic mass spectrum
\eq{\label{eq:rhoQm2}
\rho_Q(m) \simeq \frac{\sqrt{2\pi} \, 3^{\delta+5/8} \, C}{4^{\gamma+\delta+1}\,B^{\gamma+7/8} \, \sigma_Q^{1/8}} \, m^{\gamma + \delta + 1/2} \, \exp\left[ \frac{m \, \sigma_Q^{1/4}}{(3B)^{1/4}} \right].
}
This form coincides with the Hagedorn mass spectrum~\eqref{eq:rhoHagedorn} with
\eq{\label{eq:TH}
A = \frac{\sqrt{2\pi} \, 3^{\delta+5/8} \, C}{4^{\gamma+\delta+1}\,B^{\gamma+7/8} \, \sigma_Q^{1/8}}, \qquad \qquad \alpha = -\left(\gamma + \delta + \frac{1}{2}\right), \qquad \qquad T_H = \left(\frac{3B}{\sigma_Q}\right)^{1/4}~.
}

The accuracy of Eq.~\eqref{eq:rhoQm2} for given $m$ depends on the values of the parameters of the mass-volume density.
For the parameter sets used in the present work, Eq.~\eqref{eq:rhoQm2} is accurate to within 10\% relative error for $m \gtrsim 8-10$~GeV$/c^2$, this is illustrated in Appendix~\ref{app:A}.
Note that Eq.~\eqref{eq:rhoQm2} is derived here solely for the purpose of illustrating the appearance of the familiar Hagedorn mass spectrum form. Equation~\eqref{eq:rhoQm2} is not used in further applications presented in this paper.

\subsection{The pressure function}
\label{sec:press}

The pressure function~[Eq.~\eqref{eq:ps*}] is defined by the farthest-right singularity $s^*$ of the isobaric partition function~[Eq.~\eqref{eq:Z}],
\eq{
\hat{Z} = \frac{1}{s - f(T,s,\lambda_B,\lambda_Q,\lambda_S)},
} 
where $f$ is given by Eq.~\eqref{eq:f}.
Function $\hat{Z}$ has a pole singularity at $s_H = f(T,s_H,\lambda_B,\lambda_Q,\lambda_S)$.
Another possibility is a singularity $s_Q$ in the function $f(T,s,\lambda_B,\lambda_Q,\lambda_S)$ itself. 

Let us compute the function $f$ for the particular mass-volume density given by~\eqref{eq:rho}-\eqref{eq:rhoQ}.
First, we split it into two parts $f = f_H + f_Q$ with
\eq{\label{eq:fH}
f_H(T,s,\lambda_B,\lambda_Q,\lambda_S) & = \sum_{i \in \rm HRG} d_i \, \phi(T,m) \, \lambda_B^{b_i} \, \lambda_Q^{q_i} \, \lambda_S^{s_i} \, \exp\left(- \frac{m_i s}{4 B} \right), \\
\label{eq:fQ}
f_Q(T,s,\lambda_B,\lambda_Q,\lambda_S) & = C \, \int_{V_0} dv \, v^\gamma \, \exp\left(-v s \right) \int_{Bv + m_0} dm \, (m - Bv)^{\delta} \nonumber \\ 
& \qquad \times \exp \left\{ \frac{4}{3} [\sigma_{Q}(\lambda_B,\lambda_Q,\lambda_S)]^{1/4} \, v^{1/4} \, (m-Bv)^{3/4} \right\} \, \phi(T,m) \, .
}

The quark-gluon bags in Eq.~\eqref{eq:fQ} are heavy, $m \gtrsim 2$~GeV, therefore one can use the non-relativistic approximation:
\eq{\label{eq:nonrel}
\phi(T,m) \stackrel{m \gg T}{\simeq} \left(\frac{mT}{2\pi}\right)^{3/2} \, \exp\left(-\frac{m}{T}\right). 
}
This approximation has a relative accuracy of 10\% or better for $m/T > 20$, a condition which is realized in the applications considered in this work.
The error due to the non-relativistic approximation for quark-gluon bags in the resulting pressure is even smaller, see Appendix~\ref{app:B}.

The expression for $f_Q$ then simplifies to
\eq{\label{eq:fQ2}
f_Q & \simeq C \int_{V_0} dv \, v^\gamma \, \exp\left(-vs \right) \int_{Bv + M_0} dm \, (m - Bv)^{\delta} \, \left(\frac{mT}{2\pi}\right)^{3/2} \, \exp[g(m)],
}
where 
\eq{\label{eq:g}
g(m) = -\frac{m}{T} + \frac{4}{3} [\sigma_{Q}(\lambda_B,\lambda_Q,\lambda_S)]^{1/4} \, v^{1/4} \, (m-Bv)^{3/4}.
}

The integration over $m$ in Eq.~\eqref{eq:fQ2} can be carried out approximately using Laplace's method.
One has
\eq{
g(m) \approx g(m_0) + \frac{1}{2} g''(m_0) \, (m-m_0)^2,
}
where $m_0$ satisfies the equation $g'(m_0) = 0$.
The dominant part of the contribution of the QGP bags with volume $v$ to the thermodynamics of the system is given by those bags with mass $m \simeq m_0$.
Using Eq.~\eqref{eq:g} $m_0$ is obtained explicitly
\eq{\label{eq:m0}
m_0 = B \,v + \sigma_{Q}(\lambda_B,\lambda_Q,\lambda_S) \, v \, T^4.
}
One can invert the above relation to obtain
\eq{\label{eq:m0}
v(m_0) = \frac{m_0}{B + \sigma_{Q}(\lambda_B,\lambda_Q,\lambda_S) \, T^4}.
}
One can see that $v(m_0)$ is a decreasing function of temperature. This elucidates the so-called effect of thermal compression of bags.

Let us note that
\eq{\label{eq:gm0}
g(m_0) & = \frac{1}{3} \, \sigma_{Q}(\lambda_B,\lambda_Q,\lambda_S) \, v \, T^3 - \frac{Bv}{T}, \\
g''(m_0) & = -\frac{1}{4 \,  \sigma_{Q}(\lambda_B,\lambda_Q,\lambda_S) \, v \, T^5}.
}

Applying Laplace's method to Eq.~\eqref{eq:fQ2} one obtains
\eq{\label{eq:fQ3}
f_Q & \simeq \frac{C}{\pi} \, T^{4+4\delta} \, [\sigma_Q]^{\delta+1/2} \, [B+\sigma_Q T^4]^{3/2} \int_{V_0} dv \, v^{2+\gamma+\delta} \, \exp[-v (s - s_B)].
}
Here $s_B$ corresponds to $s_B = p_B /T$, where $p_B(T,\lambda_B,\lambda_Q,\lambda_S)$ coincides with the pressure in the MIT bag model equation of state~\cite{Baacke:1976jv}:
\eq{\label{eq:pBag}
p_B(T,\lambda_B,\lambda_Q,\lambda_S) = \frac{\sigma_Q(T,\lambda_B,\lambda_Q,\lambda_S)}{3} T^4 - B.
}

Recalling the definition of the ``upper'' incomplete gamma function
\eq{\label{eq:gamminc}
\Gamma(\alpha, x) = \int_{x}^{\infty} \, t^{\alpha-1} \, e^{-t} \, d t,
}
one can perform the integration over $v$ in Eq.~\eqref{eq:fQ3} explicitly
\eq{\label{eq:fQ4}
f_Q & \simeq \frac{C}{\pi} \, T^{4+4\delta} \, [\sigma_Q]^{\delta+1/2} \, [B+\sigma_Q T^4]^{3/2} \,
\left(s - s_B\right)^{-(\gamma+\delta+3)} \, \Gamma\left[\gamma+\delta+3, (s - s_B) V_0 \right].
}
Note that dependence on $\lambda_B$, $\lambda_Q$, $\lambda_S$ in the above relation enters through $\sigma_Q \equiv \sigma_Q(T,\lambda_B,\lambda_Q,\lambda_S)$
and $s_B \equiv s_B(T,\lambda_B,\lambda_Q,\lambda_S)$.
Also note that the application of Laplace's method eliminates the dependence of the final result on the parameter $M_0$ from Eq.~\eqref{eq:rhoQ}.

The function $f$ has a singularity at $s_Q = s_B$, as follows from \eqref{eq:fQ4}.
Thus, the system pressure is defined at given temperature and chemical potentials as
\eq{\label{eq:pmax}
p(T,\lambda_B,\lambda_Q,\lambda_S) = T \, \max\{s_H(T,\lambda_B,\lambda_Q,\lambda_S), s_Q(T,\lambda_B,\lambda_Q,\lambda_S)\}.
}
The model may contain a phase transition, defined as a ``collision'' of singularities $s_H$ and $s_Q$ at particular values of the thermodynamic parameters, i.e. $s_H(T_c) = s_Q(T_c)$ at the critical temperature $T_c$ of the phase transition.
This mechanism was first described in Ref.~\cite{Gorenstein:1981fa}.
In this case $s_H(T) > s_Q(T)$ for $T < T_c$ and $s_H(T) < s_Q(T)$ for $T > T_c$. 
A detailed analysis performed in Ref.~\cite{Begun:2009an} reveals that a phase transition as described above is only realized when $\gamma + \delta < -3$ and $\delta < -7/4$. 
For other values of these parameters a crossover-type transition is realized, i.e. $s_H(T) > s_Q(T)$ for all $T$.
If the crossover transition takes place, then $p/T^4 \stackrel{T \to \infty} \to p_B / T^4$~\cite{Begun:2009an}\footnote{Note that $p/T^4 \stackrel{T \to \infty} \to p_B / T^4$ does \emph{not} necessarily imply $p \stackrel{T \to \infty} \to p_B$. Whether this is the case depends on the values of $\gamma$ and $\delta$~\cite{Begun:2009an}.}, i.e. the system proceeds to the phase which has thermodynamic properties similar to that of the quark-gluon plasma described by the MIT bag model equation of state.

Lattice QCD simulations at physical quark masses reveal that the transition from hadronic to partonic degrees of freedom at $\mu_B = 0$ is of a crossover type~\cite{Aoki:2006we,Borsanyi:2010bp,Bazavov:2011nk}.
Therefore, in this work we focus only on the case where the crossover transition is realized.
The possibility of a real phase transition at finite $\mu_B$ will be considered in a separate publication.

The farthest-right singularity of the isobaric partition function $\hat{Z}$ is equal to $s_H$ for all possible values of thermodynamic parameters when the crossover scenario is realized.
The pressure, $p = T s_H$, satisfies the following transcendental equation:
\eq{\label{eq:pfull}
& p(T,\lambda_B,\lambda_Q,\lambda_S) =
T \sum_{i \in \rm HRG} d_i \, \phi(T,m) \, \lambda_B^{b_i} \, \lambda_Q^{q_i} \, \lambda_S^{s_i} \, \exp\left(- \frac{m_i p}{4 B T} \right) \nonumber \\
& \quad + \frac{C}{\pi} \, T^{5+4\delta} \, [\sigma_Q]^{\delta+1/2} \, [B+\sigma_Q T^4]^{3/2} \,
\left(\frac{T}{p-p_B}\right)^{\gamma+\delta+3} \, \Gamma\left[\gamma+\delta+3, \frac{(p-p_B)V_0}{T} \right].
}

The first term in Eq.~\eqref{eq:pfull} corresponds to the discrete part of the particle spectrum. It equals the sum of the partial pressures evaluated self-consistently for an ideal Boltzmann gas with shifted chemical potentials $\mu_i^* = \mu_i - \frac{m_i p}{4BT}$.
The second term corresponds to the contribution of the quark-gluon bags.
The two terms are not independent -- they both depend self-consistently on the total system pressure to which they both contribute.
Equation~\eqref{eq:pfull} is solved numerically in the present work.
The energy density, the entropy density, the speed of sound and the various susceptibilities are obtained from the pressure function as derivatives with respect to $T$ or $\lambda_{B,Q,S}$, through the standard thermodynamic relations.

One should note that the application of the Laplace's method used to perform the mass integration in the derivation of Eq.~\eqref{eq:pfull} is, strictly speaking, most accurate in the limit of large masses, i.e. when $m_0$~[Eq.~\eqref{eq:m0}] is large.
The method is expected to be less accurate if contributions of ``small'' quark-gluon bags close to the mass-volume density threshold $V_0$, are significant. 
This situation can take place in the vicinity of the crossover transition, where the bags start to appear in addition to the PDG hadrons. 
In Appendix~\ref{app:B} we show numerically that, for the parameter sets used in the present study, Laplace's method allows to evaluate the pressure with a precision of better than 2\%, for all temperature values considered. 
We therefore adopt Laplace's method in all our subsequent calculations, as that method remedies some technical difficulties and presents a clear physical picture.
In a more elaborate study one may omit the Laplace's method approximation altogether.

It should be pointed out that the mass-volume density~\eqref{eq:rhoQ} of quark-gluon bags is obtained for asymptotically large masses and volumes, whereas the lower end of the mass-volume spectrum is regulated by the cut-off parameters $V_0$ and $M_0$ only.
One may ask how this lower end of the mass-volume spectrum matches with the spectrum of the PDG hadrons at even lower masses.
In Appendix~\ref{app:A} we show that the spectra of PDG hadrons and quark-gluon bags can indeed be merged rather smoothly for the model parameter sets under consideration here.

\subsection{Other quantities}
\label{sec:aux}

\subsubsection{Particle number density}

The particle number density is the average number of particles per unit volume.
It is the sum (integral) of individual densities corresponding to different particle species:
\eq{\label{eq:ndef}
n = \sum_i n_i,
}
where the sum goes over all species.
As the considered particle spectrum  includes the continuous spectrum of the quark-gluon bags, the sum in Eq.~\eqref{eq:ndef} in general corresponds to the integral over the mass-volume density of states, i.e. $\sum_i \lambda_B^{b_i} \, \lambda_Q^{q_i} \, \lambda_S^{S_i} \, d_i \equiv \int dv \, \int dm \, \rho(m,v;\lambda_B,\lambda_Q,\lambda_S)$.

The individual densities can be computed by introducing the fictitious fugacities $\lambda_i$ for all species into the partition function.
$n_i$ are given by the standard expression for the multi-component excluded-volume model~\cite{Yen:1997rv}:
\eq{\label{eq:ni}
n_i = \frac{n^{\rm id}_i \, e^{-v_i \, p / T} }{1 + \sum_j \, v_j \, n^{\rm id}_j \, e^{-v_j \, p / T} }.
}
Here
\eq{
n^{\rm id}_i = d_i \, \phi(T,m_i)
\, \lambda_B^{b_i} \, \lambda_Q^{q_i} \, \lambda_S^{S_i}~,
}
and $p$ is the system pressure.

The total hadron density reads
\eq{\label{eq:ntot}
n = \frac{\sum_i n^{\rm id}_i \, e^{-v_i \, p / T} }{1 + \sum_j \, v_j \, n^{\rm id}_j \, e^{-v_j \, p / T} }
= \frac{n^{\rm id}}{1 + \kappa},
}
with
\eq{\label{eq:nidHag}
n^{\rm id} & = \sum_i n^{\rm id}_i \, e^{-v_i \, p / T} =  \int dv \, \int dm \, \rho(m,v;\lambda_B,\lambda_Q,\lambda_S) \, \phi(T,m) \, e^{-v p / T}, \\
\label{eq:kappaHag}
\kappa & = \sum_i v_i \, n^{\rm id}_i \, e^{-v_i \, p / T} =  \int dv \, \int dm \, v \, \rho(m,v;\lambda_B,\lambda_Q,\lambda_S) \, \phi(T,m) \, e^{-v p / T}.
}

For the particle spectrum~\eqref{eq:rho} consisting of the PDG hadrons and quark-gluon bags the above quantities can be computed explicitly.
The calculation proceeds in the same fashion as done for the pressure function in Sec.~\ref{sec:press}:
the PDG part of the mass-volume density is computed explicitly, whereas for the
quark-gluon part
one first applies Laplace's method
to perform the integration over the mass in Eqs.~\eqref{eq:nidHag} and \eqref{eq:kappaHag}, the remaining integrals over the volume can be expressed in terms of the incomplete Gamma function. The result is:
\eq{
n^{\rm id} & = \sum_{i \in \rm HRG} d_i \, \phi(T,m_i) \, \lambda_B^{b_i} \, \lambda_Q^{q_i} \, \lambda_S^{s_i} \, \exp\left(- \frac{m_i p}{4 B T} \right) \nonumber \\
& \quad + \frac{C}{\pi} \, T^{4+4\delta} \, [\sigma_Q]^{\delta+1/2} \, [B+\sigma_Q T^4]^{3/2} \,
\left(\frac{T}{p-p_B}\right)^{\gamma+\delta+3} \, \Gamma\left[\gamma+\delta+3, \frac{(p-p_B)V_0}{T} \right], \\
\kappa & = \sum_{i \in \rm HRG} d_i \, \frac{\phi(T,m_i) \, m_i}{4 B} \, \lambda_B^{b_i} \, \lambda_Q^{q_i} \, \lambda_S^{s_i} \, \exp\left(- \frac{m_i p}{4 B T} \right) \nonumber \\
& \quad + \frac{C}{\pi} \, T^{4+4\delta} \, [\sigma_Q]^{\delta+1/2} \, [B+\sigma_Q T^4]^{3/2} \,
\left(\frac{T}{p-p_B}\right)^{\gamma+\delta+4} \, \Gamma\left[\gamma+\delta+4, \frac{(p-p_B)V_0}{T} \right].
}

\subsubsection{Filling fraction}

Another interesting quantity is the filling fraction (f.f.) -- the ratio between the average total volume occupied by hadrons over the system volume.
The definition of this quantity reads
\eq{\label{eq:ff}
f.f. = \frac{ \sum_i v_i \, \langle N_i \rangle }{V}
= \sum_i v_i \, n_i = \frac{\sum_i v_i \, n^{\rm id}_i \, e^{-v_i \, p / T}}{1 + \sum_j v_j \, n^{\rm id}_j \, e^{-v_j \, p / T} } = \frac{\kappa}{1 + \kappa}~.
}

\subsubsection{Average particle eigenvolume}
The average eigenvolume of a particle in the thermal system can be computed as follows:
\eq{\label{eq:vav}
\langle v \rangle \equiv \frac{\sum_i v_i n_i}{\sum_i n_i} = \frac{\kappa}{n^{\rm id} }~ = \frac{f.f.}{n}.
}

\subsubsection{Average particle mass}
The average mass of a particle in the thermal system is given by
\eq{\label{eq:mav}
\langle m \rangle \equiv \frac{\sum_i m_i n_i}{\sum_i n_i} = \frac{ \int dv \, \int dm \, m \, \rho(m,v;\lambda_B,\lambda_Q,\lambda_S) \, \phi(T,m) \, e^{-v p / T}, }{n}~.
}

The explicit calculation, employing the Laplace integration over $m$ and the incomplete Gamma function to express the integral over $v$, yields
\eq{
n \langle m \rangle & = \sum_{i \in \rm HRG} d_i \, \phi(T,m_i) \, m_i \, \lambda_B^{b_i} \, \lambda_Q^{q_i} \, \lambda_S^{s_i} \, \exp\left(- \frac{m_i p}{4 B T} \right) \nonumber \\
& \quad + \frac{C}{\pi} \, T^{4+4\delta} \, [\sigma_Q]^{\delta+1/2} \, [B+\sigma_Q T^4]^{5/2} \,
\left(\frac{T}{p-p_B}\right)^{\gamma+\delta+4} \, \Gamma\left[\gamma+\delta+4, \frac{(p-p_B)V_0}{T} \right].
}

\section{Calculation results for bags with massless quarks and gluons}

Calculations here are performed for the following set of parameters:
\eq{\label{eq:ParamSetScan}
\gamma = 0, \quad  -3 \leq \delta \leq -\frac{1}{2},  \quad B^{1/4} = 250~\textrm{MeV}, \quad C = 0.03~\text{GeV}^{-\delta + 2}, \quad V_0 = 4~\textrm{fm}^3.
}

All model parameters are fixed, the only exception is the $\delta$ exponent.
In the present study we fix $\gamma = 0$, for simplicity.
Non-zero $\gamma$ values can be considered equally well.
For $\gamma = 0$, the crossover-type transition is realized if $\delta \geq -3$~\cite{Begun:2009an}.
Therefore, the $\delta$ exponent is varied here in the range $-3 \leq \delta \leq -\frac{1}{2}$, in the steps of $\frac{1}{2}$.
This variation of $\delta$ corresponds to the range $0 \leq \alpha \leq \frac{5}{2}$ for the exponent $\alpha$ in the Hagedorn mass spectrum~[see Eq.~\eqref{eq:TH}].
The scan in $\delta$ performed here is, therefore, similar to the study presented in Ref.~\cite{Ferroni:2008ej}.

The value of the bag constant $B$ determines the Hagedorn temperature through Eq.~\eqref{eq:TH}.
The $B^{1/4} = 250$~MeV value corresponds to $T_H \simeq 165$~MeV -- a sensible value for the Hagedorn mass spectrum.

The constant $C$ determines the overall normalization of the exponential spectrum of quark-gluons bags.
In the spirit of the quark-gluon bag model, it is usually considered as a free parameter. 
It can, in principle, be determined microscopically, e.g. as the solution of the bootstrap equation, see Ref.~\cite{Beitel:2014kza} for an illustration.
The qualitative features of the resulting equation of state are found to be rather insensitive to variations in $C$.

The parameter $V_0$ determines the lower mass-volume cut-off for the quark-gluon bag spectrum.
This value should be sufficiently large to avoid an overlap between the quark-gluon bag spectrum and the ground state hadrons.
This ensures that the ground state hadrons determine the equation of state at low temperatures and/or densities.
On the other hand, a too large $V_0$ value would create a large gap between the spectrum of established hadrons and that of the quark-gluon bags.
The $V_0 = 4$~fm$^3$ value was found to be sufficient with regard to the above considerations.

\begin{figure*}[t]
  \centering
  \includegraphics[width=.49\textwidth]{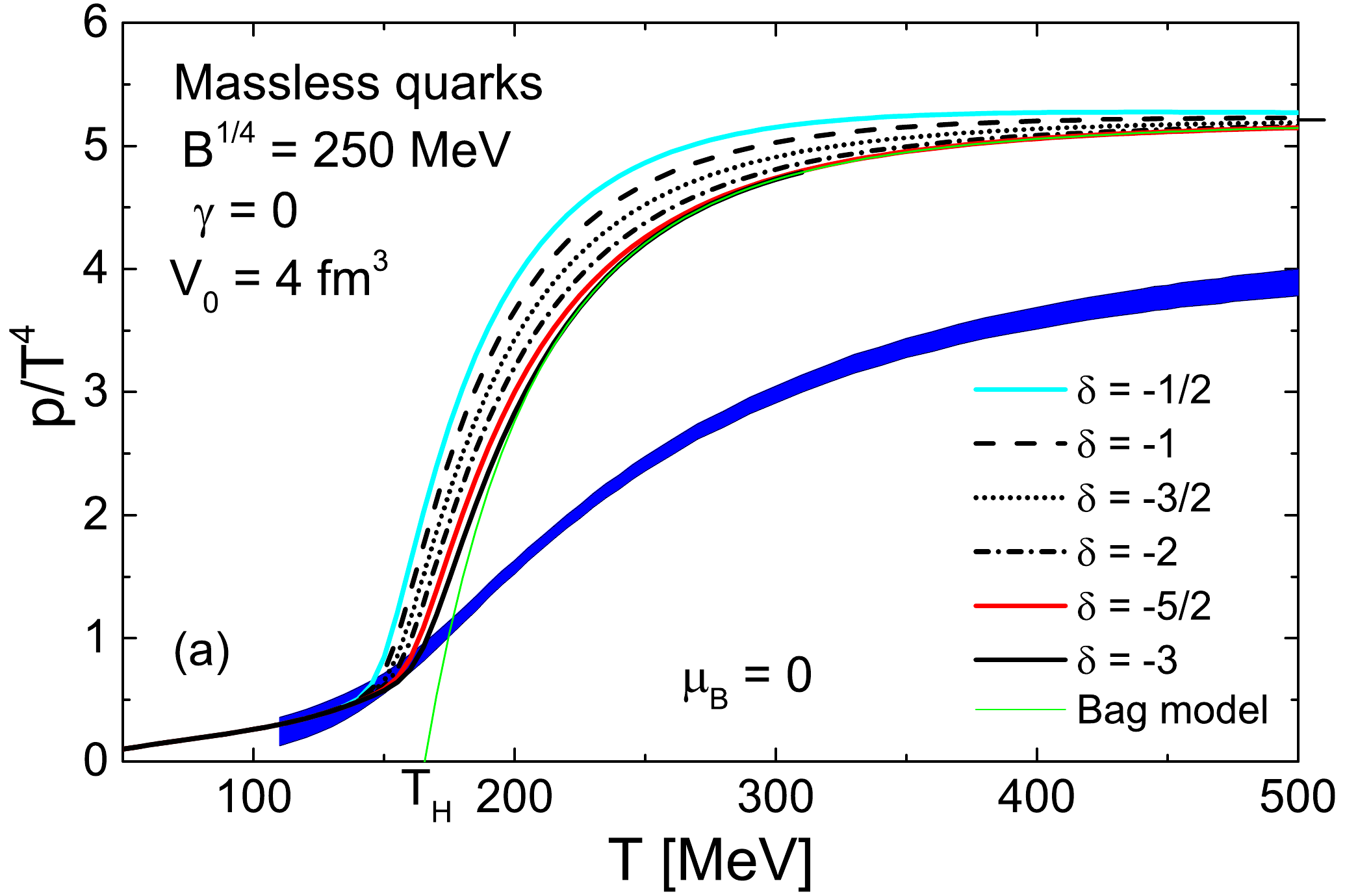}
  \includegraphics[width=.49\textwidth]{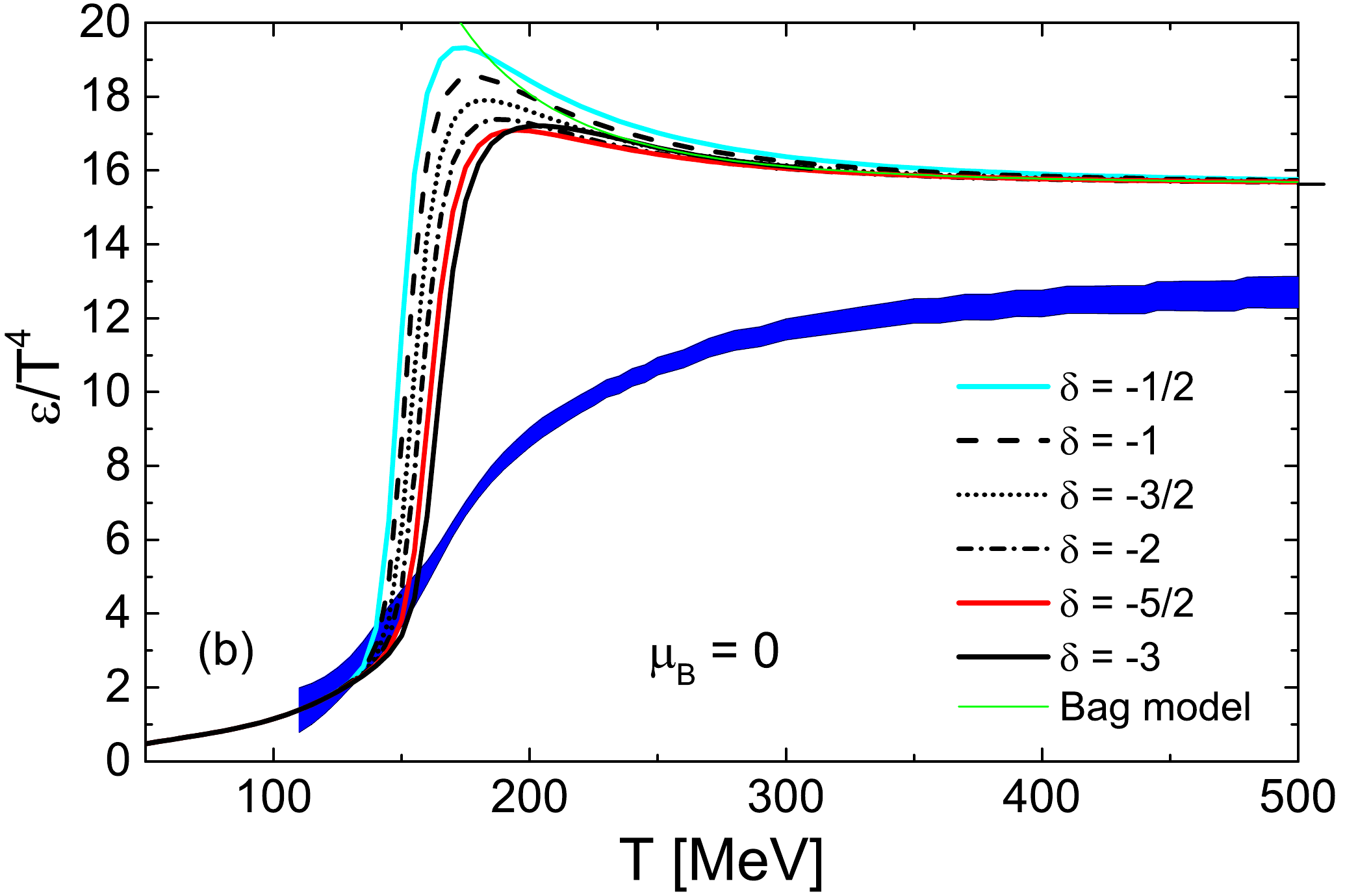}
  \caption{Temperature dependence of (a) scaled pressure $p/T^4$ and (b) scaled energy density $\varepsilon/T^4$ calculated in the Hagedorn model with quark-gluon bags filled with massless quarks and gluons. Lattice QCD data of the Wuppertal-Budapest collaboration~\cite{Borsanyi:2013bia} are shown by the blue bands.
  The short horizontal lines depict the Stefan-Boltzmann limiting values.
  }
  \label{fig:MITBag}
\end{figure*}

The temperature dependence of the scaled pressure $p/T^4$ and the scaled energy density $\varepsilon / T^4$ is depicted in Fig.~\ref{fig:MITBag}.
The model shows a crossover transition, the functions plotted approach the Stefan-Boltzmann limit of massless quarks at high temperatures.
The results are compared with the lattice QCD data of the Wuppertal-Budapest collaboration\footnote{Similar lattice results were also obtained by the HotQCD collaboration~\cite{Bazavov:2014pvz}.}~(blue bands)~\cite{Borsanyi:2013bia}. 
On a quantitative level, the agreement of the model with the lattice data is not very good.
This especially true for the energy density: the model predicts a peak in the temperature dependence of $\varepsilon / T^4$ -- a qualitative feature not seen in lattice simulations.
The main reason for this disagreement is that the QGP phase is described by the MIT bag model with massless quarks, which is known to provide only a rough description of QCD thermodynamics at large temperatures.

\begin{figure*}[t]
  \centering
  \includegraphics[width=.49\textwidth]{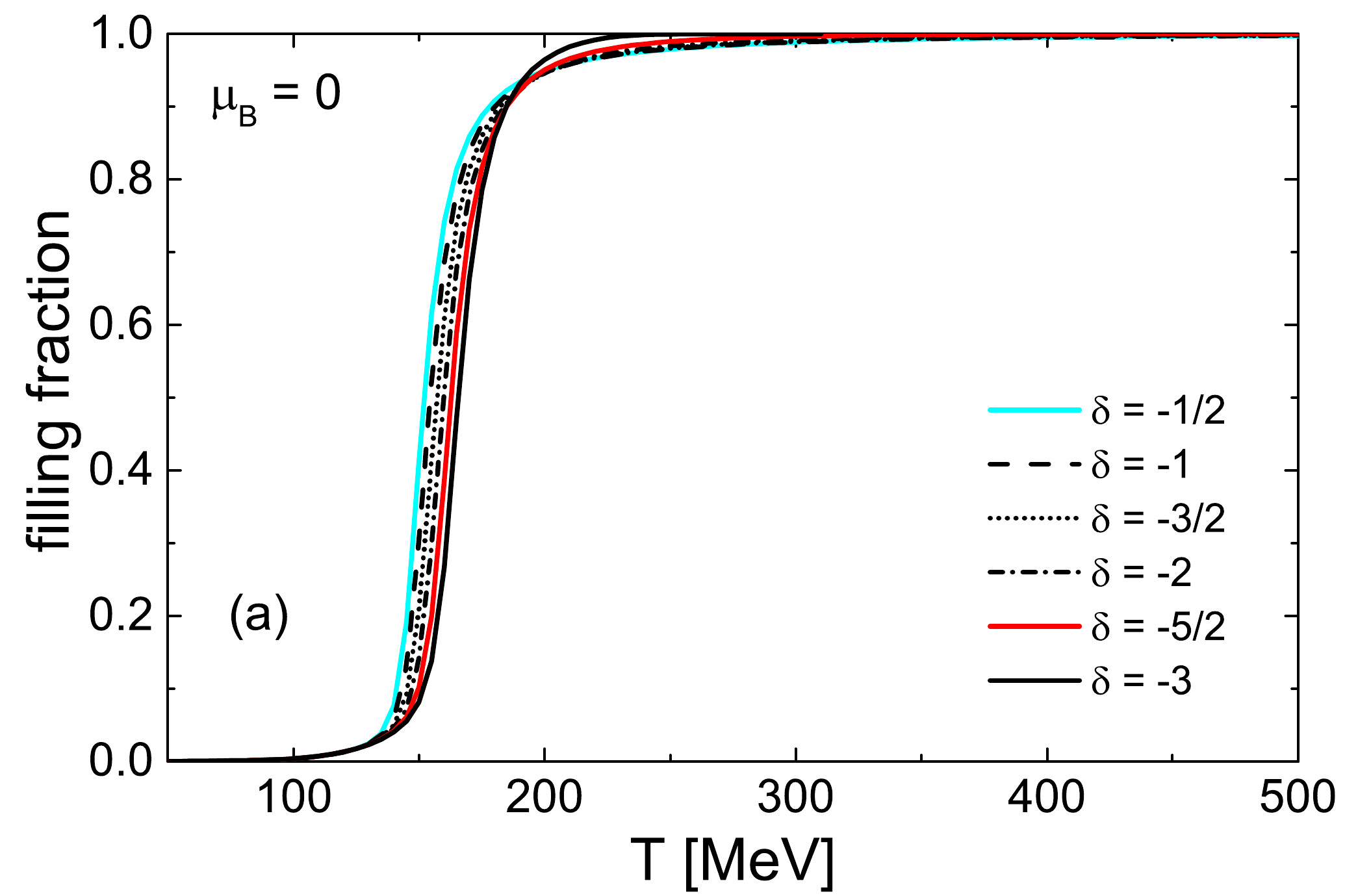}
  \includegraphics[width=.49\textwidth]{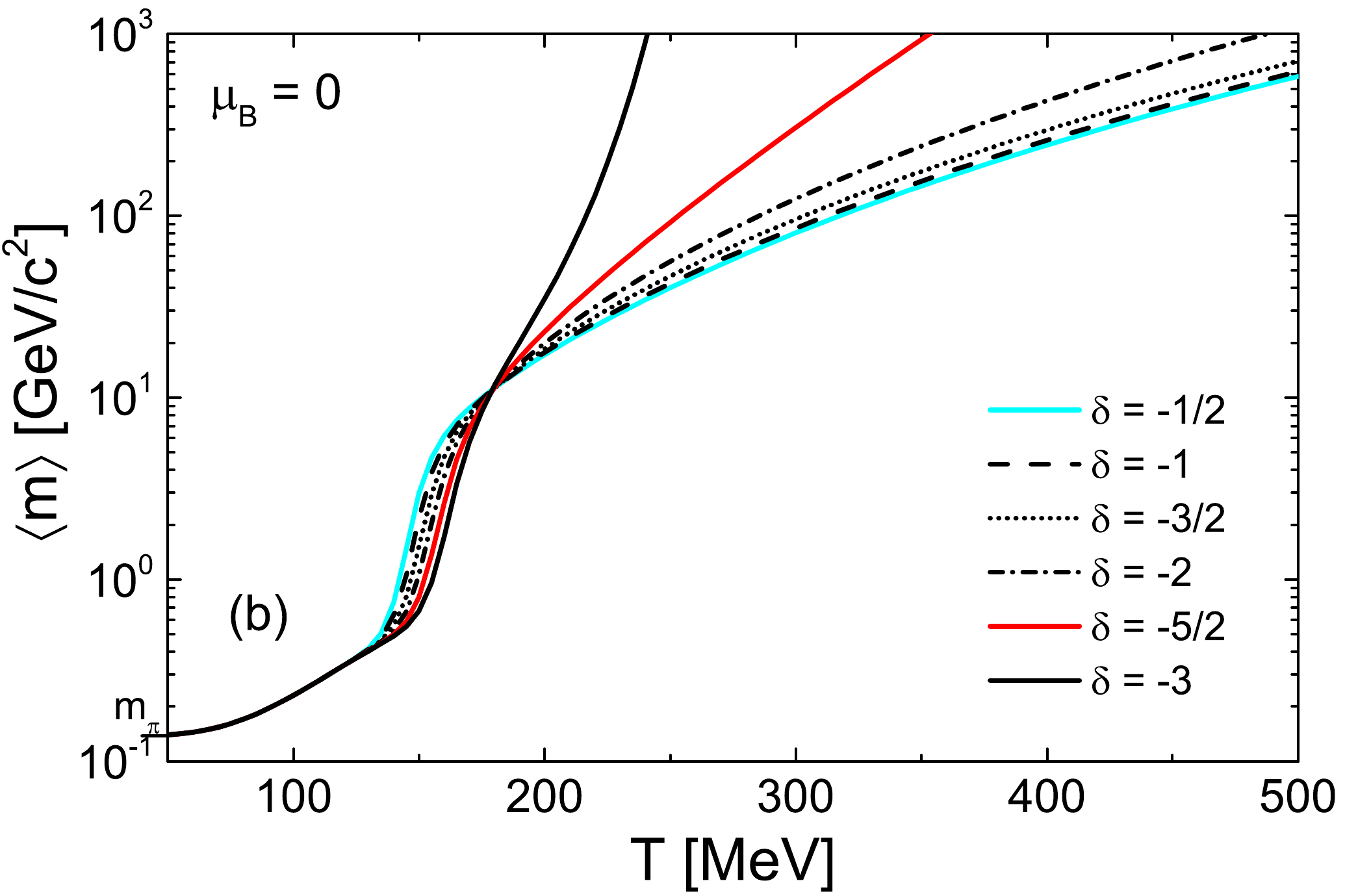}
  \includegraphics[width=.49\textwidth]{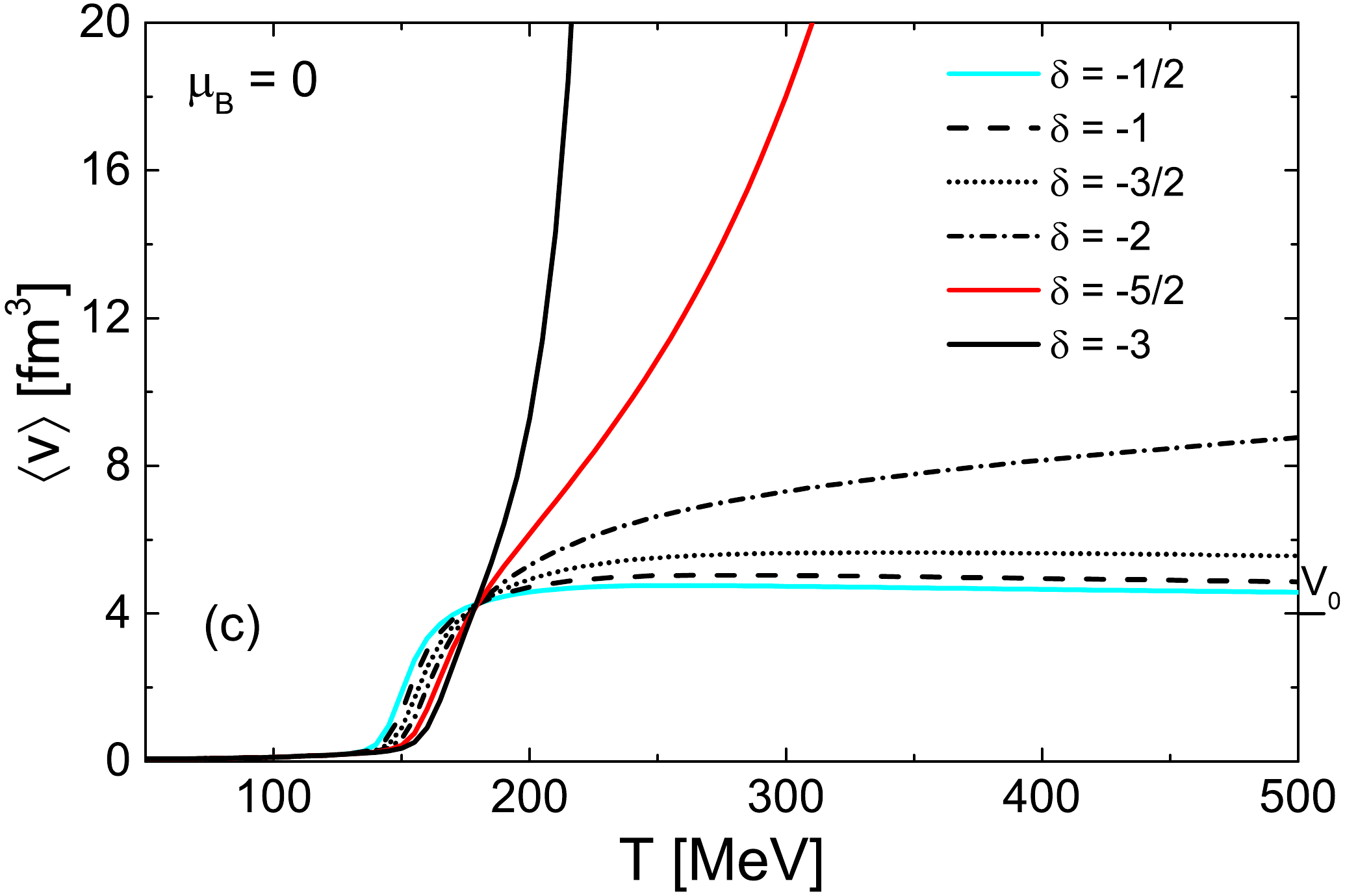}
  \includegraphics[width=.49\textwidth]{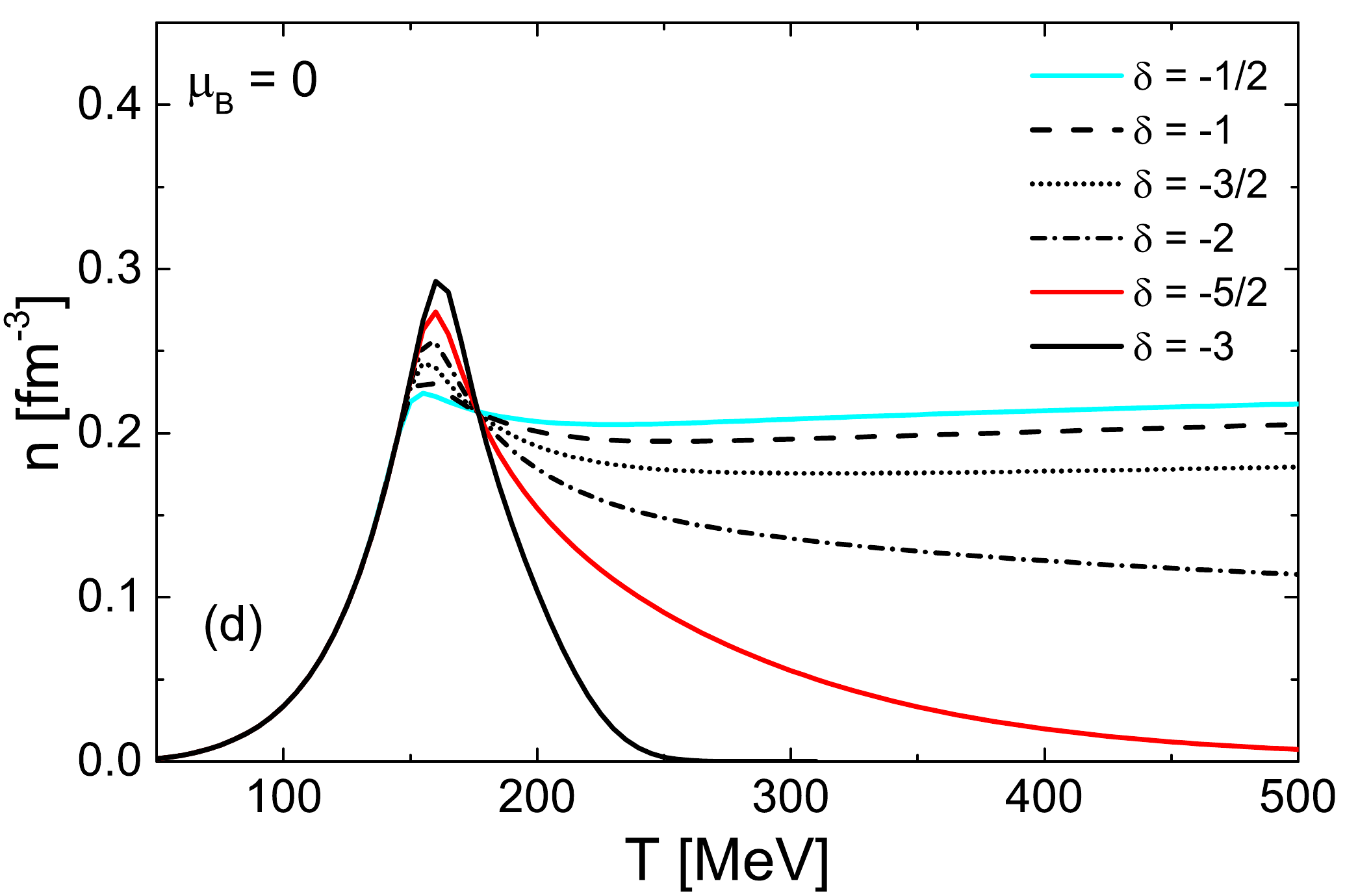}
  \caption{The temperature dependence of
  (a) the filling fraction (f.f.),
  (b) the average particle mass $\langle m \rangle$,
  (c) the average particle volume $\langle v \rangle$, and
  (d) the particle number density $n$, calculated in the Hagedorn model with quark-gluon bags filled with massless quarks and gluons.
  }
  \label{fig:MITBagAux}
\end{figure*}

The auxiliary quantities introduced in Sec.~\ref{sec:aux} are depicted in Fig.~\ref{fig:MITBagAux}.
The filling fraction (Fig.~\ref{fig:MITBagAux}a) shows a monotonic increase with temperature, from small values ($f.f. \simeq 0$) at small temperatures towards $f.f. \simeq 1$ at high temperatures.
This implies that almost the whole volume is occupied by the finite-sized 
particles at high temperatures.

The particle chemistry at different temperatures can be clarified by studying the temperature dependence of the mean hadron mass $\langle m \rangle$ (Fig.~\ref{fig:MITBagAux}b).
$\langle m \rangle$ is a monotonically increasing function of temperature, for all values of $\delta$ considered. At small temperatures, $T \lesssim 70$~MeV, one has $\langle m \rangle \simeq m_\pi \simeq 138$~MeV/$c^2$.
This means that the system consists mainly of pions at low temperatures and $\mu_B = 0$, as expected.
$\langle m \rangle$ increases rapidly in the vicinity of the Hagedorn temperature $T_H \simeq 165$~MeV, signalling that the particle chemistry becomes dominated by quark-gluon bags.
$\langle m \rangle$ continues to increase at high temperatures, the rate of increase depends on the value of $\delta$: the smaller $\delta$ is, the stronger is the increase of $\langle m \rangle$.

The temperature dependence of the mean hadron volume $\langle v \rangle$ depends non-trivially on the value of $\delta$~(Fig.~\ref{fig:MITBagAux}c).
For $-3 \leq \delta \leq -2$, $\langle v \rangle$ shows a fairly fast monotonic increase at high temperatures.
For $-3/2 \leq \delta \leq -1/2$, on the other hand, $\langle v \rangle$ exhibits slow monotonic \emph{decrease} at high temperatures.
The $\gamma$ and $\delta$ dependence of the behavior of $\langle v \rangle$ at asymptotically high temperatures was studied in Ref.~\cite{Begun:2009an}: taking $\gamma = 0$ one has $\langle v \rangle \to \infty$ for $\delta < -7/4$ and $\langle v \rangle \to V_0$ for $\delta > -7/4$.
The present numerical results are consistent with this analytic expectation.

The non-trivial behavior of $\langle v \rangle$ with respect to $\delta$ similarly implies a non-trivial behavior of the particle number density $n$~(Fig.~\ref{fig:MITBagAux}d).
Indeed, as follows from Eqs.~\eqref{eq:nidHag},~\eqref{eq:ff}, and \eqref{eq:vav}, the mean hadron density can be expressed as $n = f.f. / \langle v \rangle$.
As $f.f. \simeq 1$ at high temperatures irrespective of the value of $\delta$, the asymptotic behavior of $n$ is determined by the corresponding behavior of $\langle v \rangle$. For $\delta > -7/4$ one has $\langle v \rangle \to V_0$ and therefore $n \to 1 / V_0$.
On the other hand, at $\delta < -7/4$ one has $\langle v \rangle \to \infty$ which implies $n \to 0$.
The numerical results shown in Fig.~\ref{fig:MITBagAux}d are consistent with these considerations.
In fact, the vanishing hadron number density for $\delta < -7/4$ implies that an arbitrary large but finite subvolume of the system is occupied at high temperatures by a single bag filled with quark-gluon plasma.

\begin{figure*}[t]
  \centering
  \includegraphics[width=.69\textwidth]{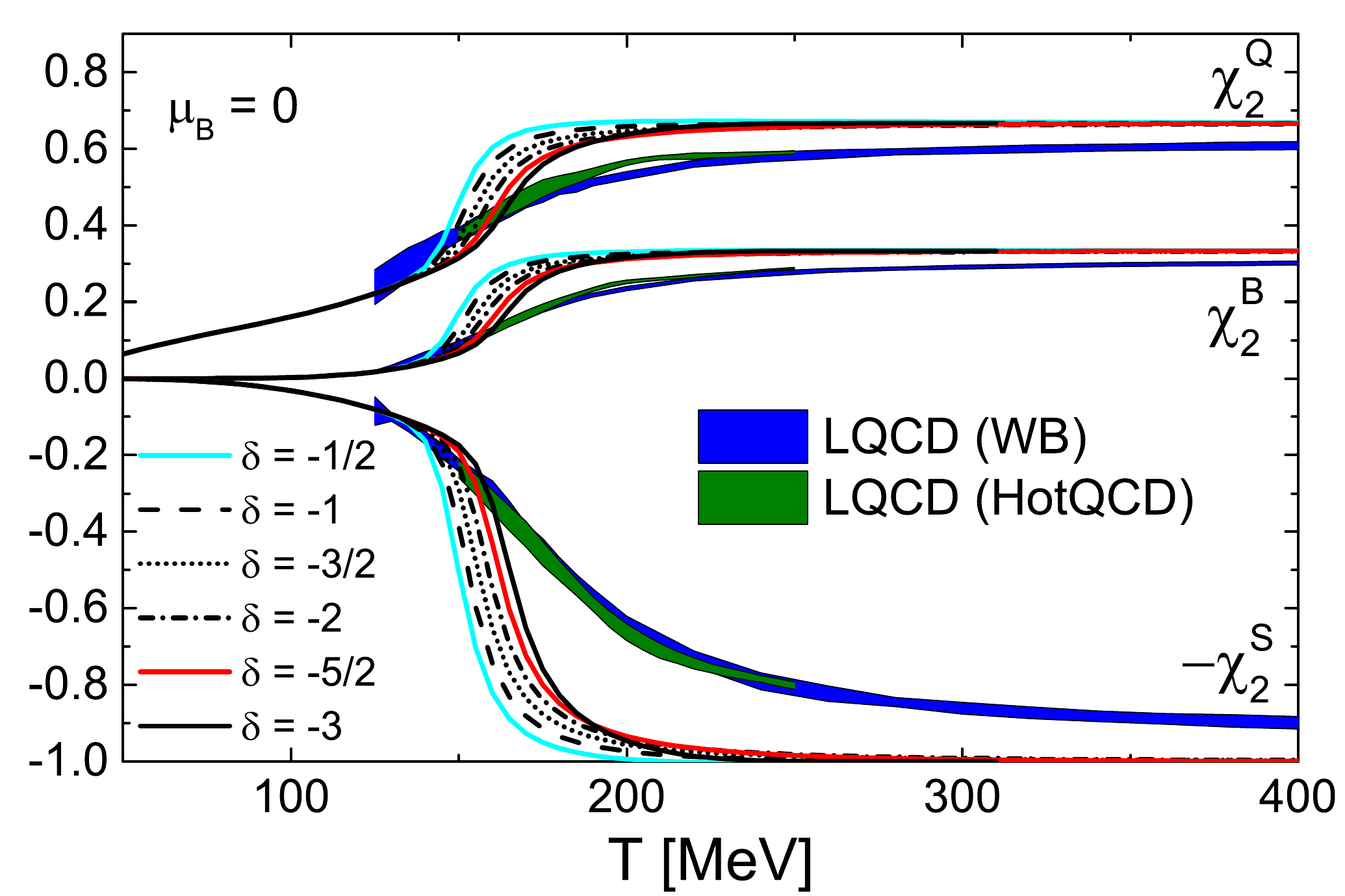}
  \caption{The temperature dependencies of the second order diagonal susceptibilities, $\chi_2^B$, $\chi_2^Q$, and $-\chi_2^S$, calculated in the Hagedorn model with quark-gluon bags filled with massless quarks and gluons at $\mu_B = 0$, and compared with the lattice QCD data from Refs.~\cite{Borsanyi:2011sw,Bazavov:2012jq}.
  }
  \label{fig:Masslesschis}
\end{figure*}

Fluctuations and correlations of conserved charges are another observables, accessible with lattice QCD, suggested long ago to be sensitive to the parton-hadron transition~\cite{Jeon:2000wg,Asakawa:2000wh}.
These observables, henceforth referred to as susceptibilities of conserved charges, are defined by the derivatives of the pressure function with respect to the corresponding chemical potentials:
\begin{equation}\label{BSQ}
\chi_{lmn}^{BSQ}~=~\frac{\partial^{l+m+n}p/T^4}{\partial(\mu_B/T)^l \,\partial(\mu_S/T)^m \,\partial(\mu_Q/T)^n}~\,.
\end{equation}

The matrix of the second order conserved charges susceptibilities has been studied in lattice QCD simulations at the physical point in Refs.~\cite{Borsanyi:2011sw,Bazavov:2012jq}.
Lattice simulations agree well with the predictions of the ideal HRG model at temperatures below the pseudocritical one, and show a behavior consistent with an approach towards the Stefan-Boltzmann limit at high temperatures.
The temperature dependencies of the second order diagonal susceptibilities, $\chi_2^B$, $\chi_2^Q$, and $\chi_2^S$, are shown in Fig.~\ref{fig:Masslesschis} in comparison with the lattice QCD data.
The behavior of the susceptibilities in the model is qualitatively compatible with lattice QCD results. From a quantitative point of view, one sees that the approach to the Stefan-Boltzmann limiting values is too fast in the model compared to the lattice data.

\section{Quark-gluon bags with massive quarks and gluons}
\label{chap:massive}

\subsection{Modification of the model}

While the simple bag model picture above appears to describe many qualitative features seen in lattice data, the quantitative description of the main thermodynamical functions, such as pressure, energy density, interaction measure, and the speed of sound, is obviously not very good.
This description cannot be notably improved solely by a variation of the parameters in Eq.~\eqref{eq:ParamSetScan}.
The main reason for the discrepancy is the inaccuracy of the standard MIT bag model equation of state~\eqref{eq:pBag} for describing the strongly coupled  quark-gluon plasma.
Therefore, an improvement of the model can be achieved by an appropriate generalization of Eq.~\eqref{eq:pBag} to describe the thermodynamics of high-temperature QCD more accurately.
At the same time, it is desirable to preserve the overall bag model picture when generalizing~\eqref{eq:pBag}.

Equation~\eqref{eq:pBag} assumes massless quark and gluon degrees of freedom.
Meanwhile, it is well known that quarks and gluons attain sizable ``thermal'' masses in the quasiparticle model of the equation of state of the quark-gluon plasma~\cite{Peshier:1994zf,Gorenstein:1995vm,Peshier:1995ty,Levai:1997yx,Peshier:1999ww,Bluhm:2007cp,Plumari:2011mk}, quark or gluon thermal masses up to GeV are possible in the temperature range of interest.
A notably improved description of the high-temperature lattice data was reported for a bag model with finite constant masses of quarks and gluons~\cite{Ivanov:2005be}.

Here we adopt a similar strategy and consider constant, finite values of quark and gluon masses:
\eq{\label{eq:QuarkMasses}
m_u = m_d = 300~\text{MeV}, \qquad m_s = 350~\text{MeV}, \qquad m_g = 800~\text{MeV}.
}
These values are taken here as a representative case, other combinations of the thermal quark and gluon masses are certainly possible.

Equation~\eqref{eq:SB} for $\sigma_Q$ should be modified to reflect \emph{massive} quarks and gluons in the bag model equation of state.
The modified $\sigma_Q$ is temperature-dependent and reads
\eq{\label{eq:MassiveQuarks}
\sigma_{Q}(T,\lambda_B,\lambda_Q,\lambda_S) & = 
\frac{8}{\pi^2 \, T^4} \, \int_{0}^{\infty} \, dk \frac{k^4}{\sqrt{k^2 + m_g^2}} \, \left[ \exp\left(\frac{\sqrt{k^2 + m_g^2}}{T}\right) - 1 \right]^{-1}
\nonumber
\\
& \qquad + \sum_{f=u,d,s} \frac{3}{\pi^2 \, T^4} \, \int_{0}^{\infty} \, dk \frac{k^4}{\sqrt{k^2 + m_f^2}} \, \left[ \lambda_f^{-1} \, \exp\left(\frac{\sqrt{k^2 + m_f^2}}{T}\right) + 1 \right]^{-1}
\nonumber
\\
& \qquad + \sum_{f=u,d,s} \frac{3}{\pi^2 \, T^4} \, \int_{0}^{\infty} \, dk \frac{k^4}{\sqrt{k^2 + m_f^2}} \, \left[ \lambda_f \, \exp\left(\frac{\sqrt{k^2 + m_f^2}}{T}\right) + 1 \right]^{-1}~.
}
With this modification of $\sigma_{Q}$, Eq.~\eqref{eq:pBag} reproduces the \emph{heavy-bag} model studied in Ref.~\cite{Ivanov:2005be}.
Here, for simplicity, we only consider temperature-independent quark and gluon masses.
A more involved model may take into account their temperature dependence, as typically done in quasiparticle models. 
This can be achieved by employing, e.g., a hard-thermal-loop description~\cite{Braaten:1991gm,Andersen:1999sf} for the intrinsic thermal pressure of the bags.
An explicit temperature dependence of the quark and gluon masses, however, will require careful considerations regarding the thermodynamic consistency of the model~\cite{Gorenstein:1995vm}, and will be considered elsewhere.

\subsection{Modification of the parameters}

The Equation~\eqref{eq:TH} which relates the Hagedorn temperature $T_H$ to the bag constant $B$ should be modified for quarks and gluons with finite thermal masses. This is because the quantity $\sigma_Q$ is modified and now depends on the temperature, i.e. $\sigma_Q = \sigma_Q(T)$.
The Hagedorn temperature can therefore be obtained as the solution of the following transcendental equation:
\eq{\label{eq:THMassive}
T_H = \left[ \frac{3B}{\sigma_Q(T_H)} \right]^{1/4}~.
}

\begin{figure*}[t]
  \centering
  \includegraphics[width=.69\textwidth]{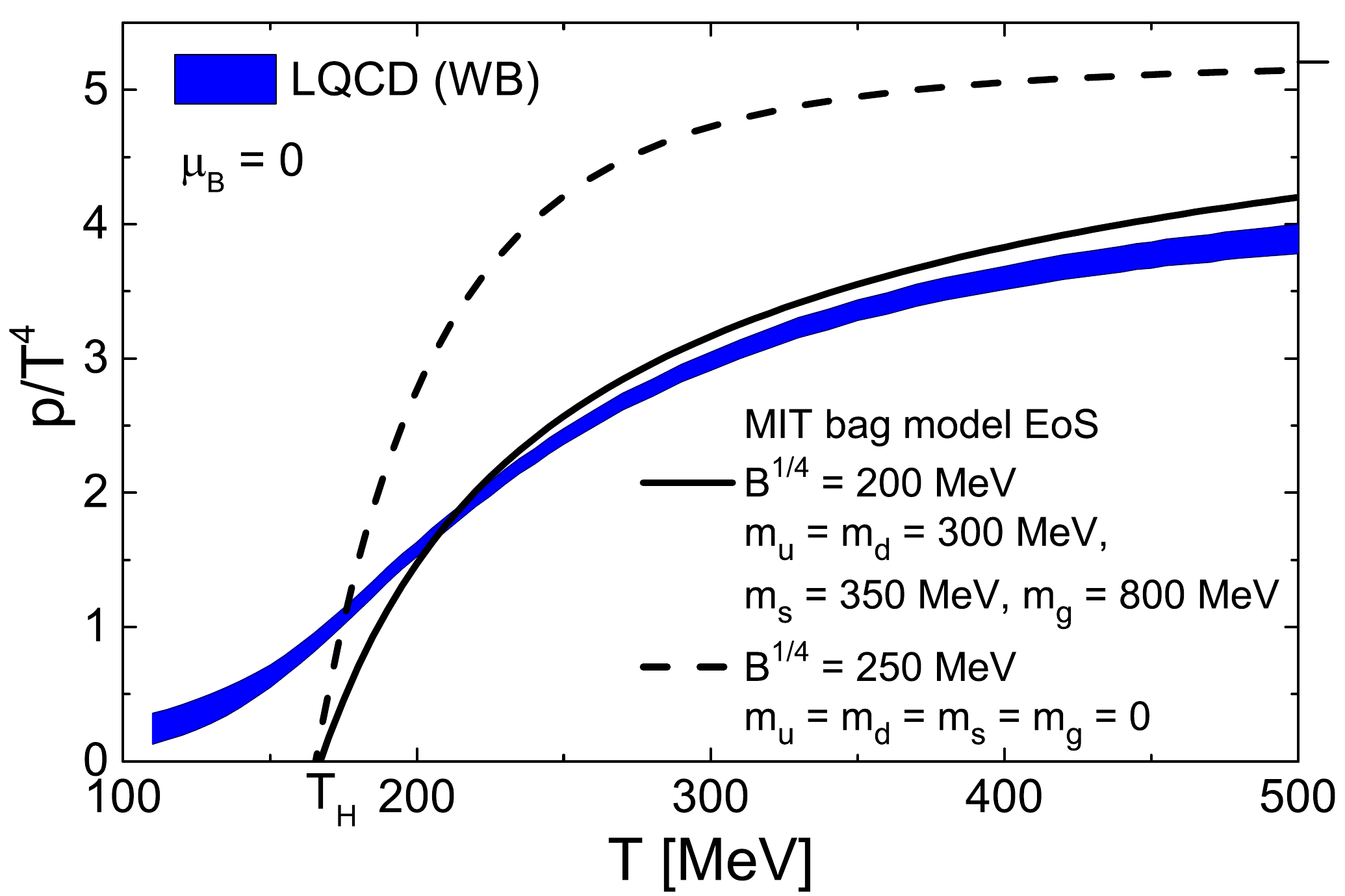}
  \caption{The temperature dependence of the scaled pressure $p/T^4$, calculated within the bag model equation of state~[Eq.~\eqref{eq:pBag}] for massive quarks~[Eqs.~\eqref{eq:QuarkMasses} and~\eqref{eq:MassiveQuarks}] with $B^{1/4} = 200$~MeV~(solid line), and for massless quarks~[Eq.~\eqref{eq:SB}] with $B^{1/4} = 250$~MeV~(dashed line).
  The lattice QCD data of the Wuppertal-Budapest collaboration~\cite{Borsanyi:2013bia} are shown by the blue band.
  }
  \label{fig:MassiveBagModelEoS}
\end{figure*}

For $B^{1/4} = 250$~MeV and for thermal masses of quarks and gluons given by Eq.~\eqref{eq:QuarkMasses} one obtains $T_H \simeq 199.5$~MeV, a rather high value compared to the massless quarks case before.
A smaller value of the bag constant $B^{1/4} = 200$~MeV is thus used here to yield $T_H \simeq 167.1$~MeV.
The resulting bag model equation of state with massive quarks~[Eq.~\eqref{eq:pBag}] is shown in Fig.~\ref{fig:MassiveBagModelEoS} by the solid line for this choice of parameters.
One sees a clear improvement in the description of the lattice data at high temperatures compared to the previously employed bag model with massless quarks~(dashed line in Fig.~\ref{fig:MassiveBagModelEoS}).

The decreased value of the bag constant necessitates an increase of the value of the parameter $V_0$, to avoid a large overlap of the spectrum of stable hadrons and quark-gluon bags.
We, therefore, adopt the value $V_0 = 8$~fm$^3$ in the following.

As before, we set $\gamma = 0$ and $C = 0.03$~GeV$^{-\delta + 2}$.
We do not vary the value of $\delta$ but settle here for the value $\delta = -2$. In this case one expects a crossover transition to a gas of infinitely large quark-gluon bags in the limit of high temperatures, as elaborated in the previous section.

\subsection{Thermodynamic functions}

\begin{figure*}[t]
  \centering
  \includegraphics[width=.49\textwidth]{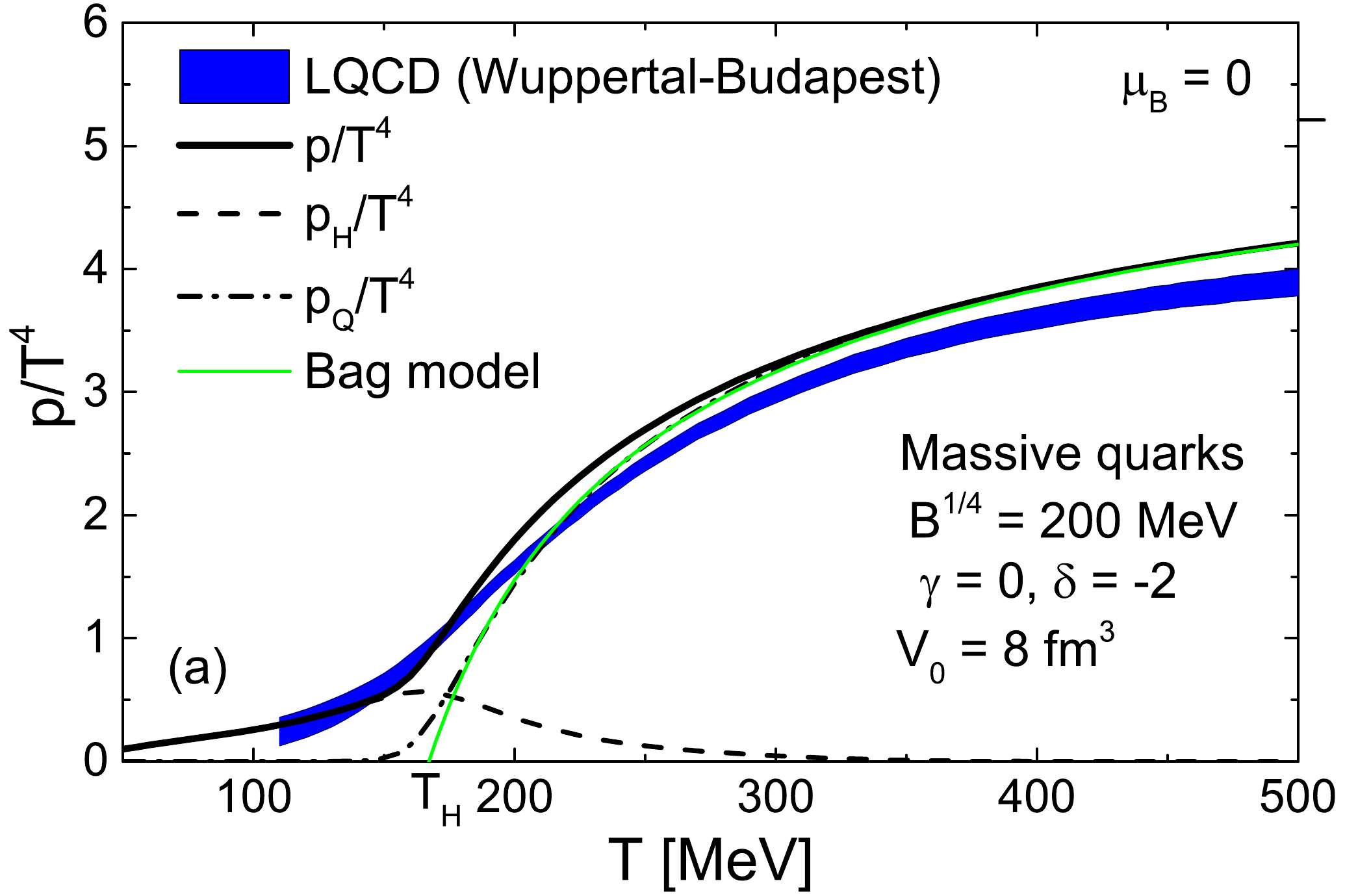}
  \includegraphics[width=.49\textwidth]{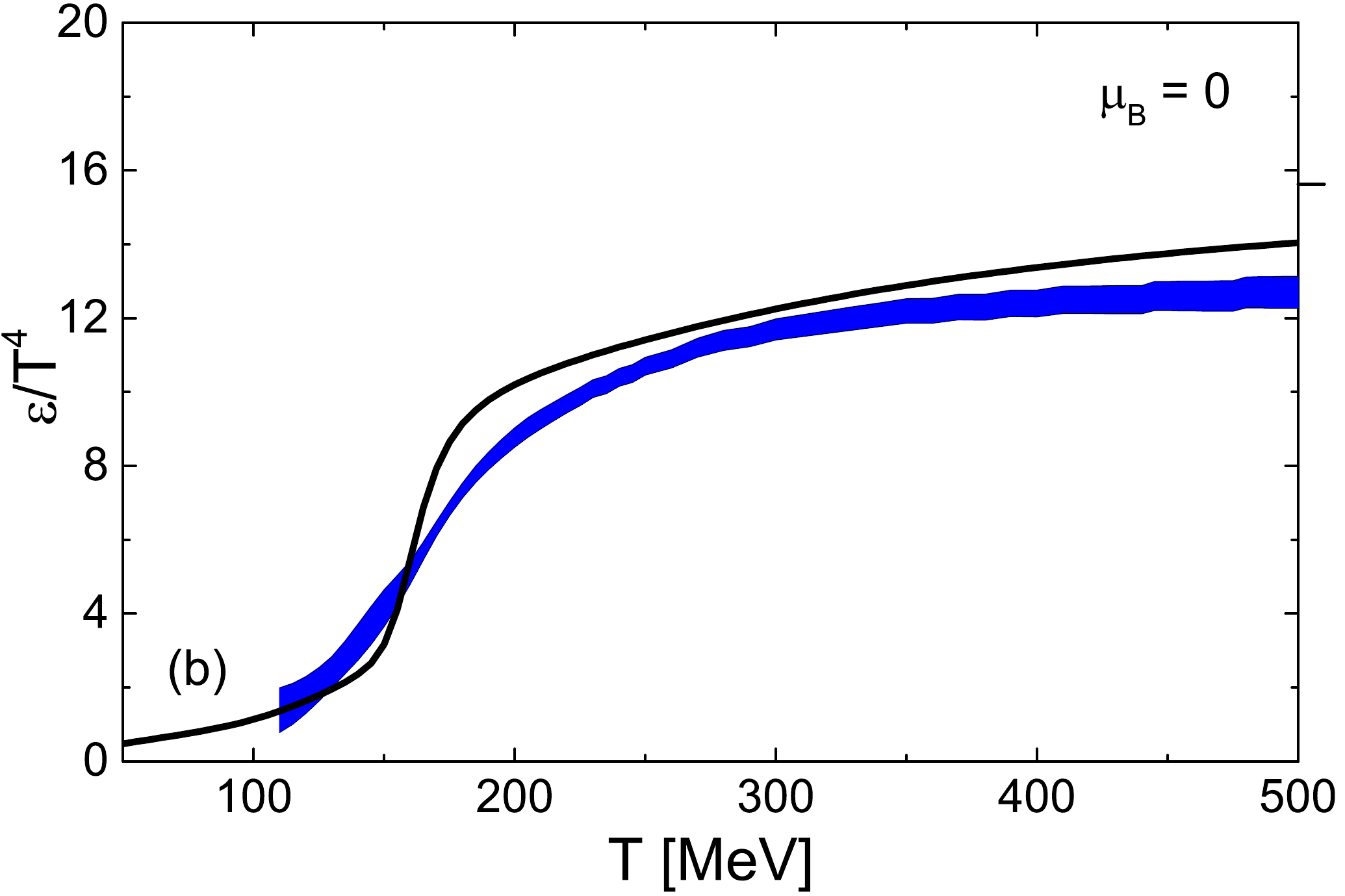}
  \includegraphics[width=.49\textwidth]{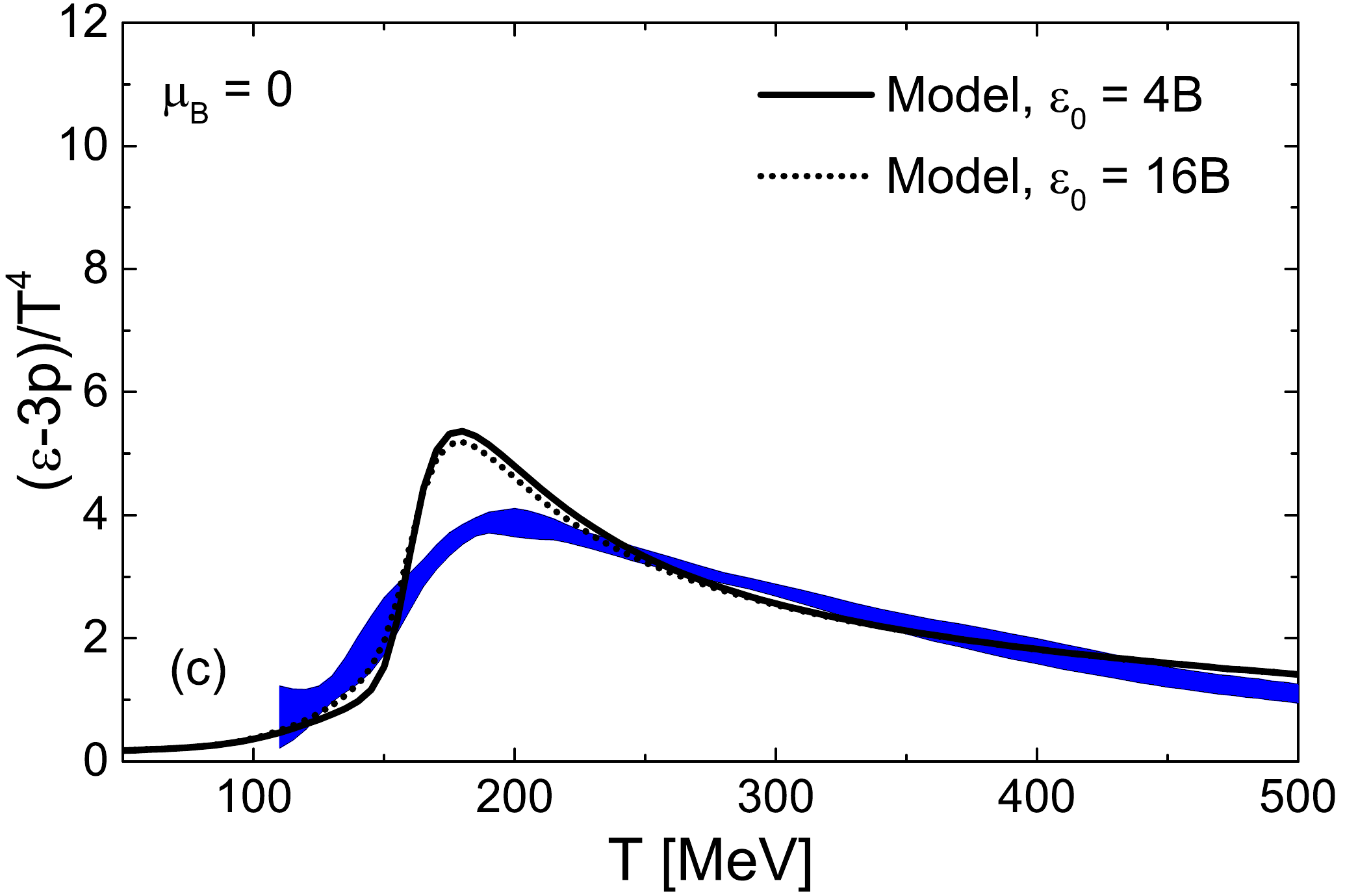}
  \includegraphics[width=.49\textwidth]{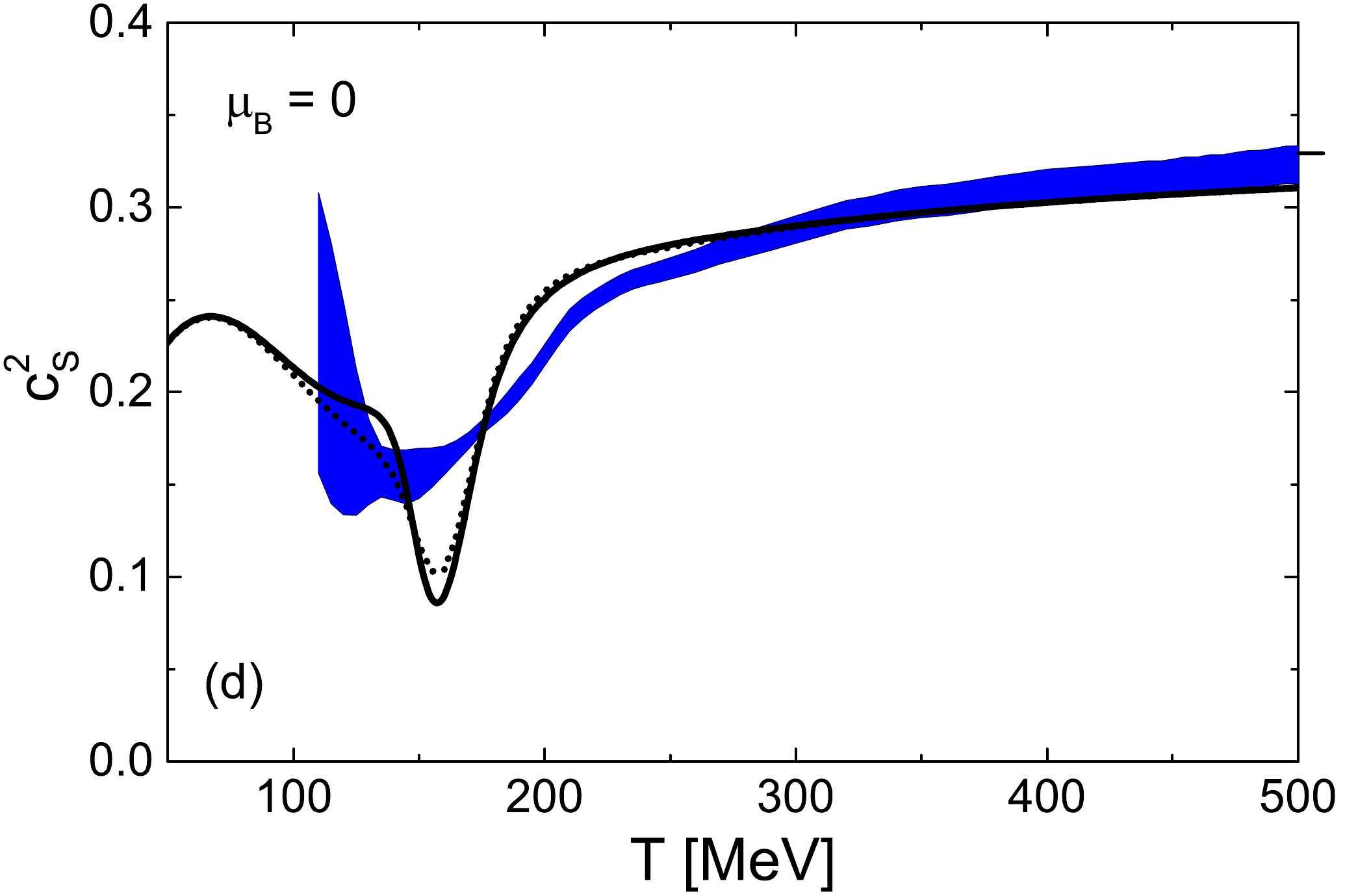}
  \caption{The temperature dependence of
  (a) the scaled pressure $p/T^4$,
  (b) the scaled energy density  $\varepsilon/T^4$,
  (c) the trace anomaly $(\varepsilon - 3p)/T^4$,
  and
  (d) the speed of sound squared $c_s^2$,
  computed in the Hagedorn model with
  quark-gluon bags filled with massive quarks and gluons.
  The dashed and dash-dotted lines in (a) depict, respectively, the first and second terms of Eq.~\eqref{eq:pfull} for the total pressure whereas the green line depicts the pressure from the bag model equation of state.
  The dotted curves in (c) and (d) depict  calculation results where the eigenvolumes of the PDG hadrons were taken to be four times smaller, i.e. $\varepsilon_0 = 16B$. The lattice QCD data of the Wuppertal-Budapest collaboration~\cite{Borsanyi:2013bia} are shown by the blue bands.
  }
  \label{fig:MassiveThermod}
\end{figure*}

The temperature dependence of the scaled pressure $p/T^4$,  the scaled energy density $\varepsilon / T^4$, the interaction measure $(\varepsilon -3p) / T^4$, and the speed of sound squared $c_s^2 = dp/dT$ at $\mu_B = 0$ is depicted in Fig.~\ref{fig:MassiveThermod}.
All quantities are in a rather good
agreement with the lattice QCD data of the Wuppertal-Budapest collaboration~\cite{Borsanyi:2013bia}.
It is particularly notable that the scaled energy density shows a monotonic behavior, consistent with lattice QCD, in contrast to the previous simple model where the bags are filled with massless quarks and gluons.
The behavior of the auxiliary quantities from Sec.~\ref{sec:aux} is found here to be quite similar to the one shown in Fig.~\ref{fig:MITBagAux} for the massless quarks and gluons case.

The model describes the lattice data on a semi-quantitative level.
Sizeable deviations are seen close to $T_H$ only for the trace anomaly and the speed of sound squared, which is sensitive to the second temperature derivative of the pressure function.
An improved description of the data can be achieved by variations of the free parameters of the model. 
However, in light of the general limitations of the present model this appears to be rather unnecessary.
We proceed, instead, by studying, in this model, the behavior of those lattice observables, which are considered to be sensitive probes of the nature of the quark-hadron transition.

\subsection{Second order correlations and fluctuations of conserved charges}

\begin{figure*}[t]
  \centering
  \includegraphics[width=.32\textwidth]{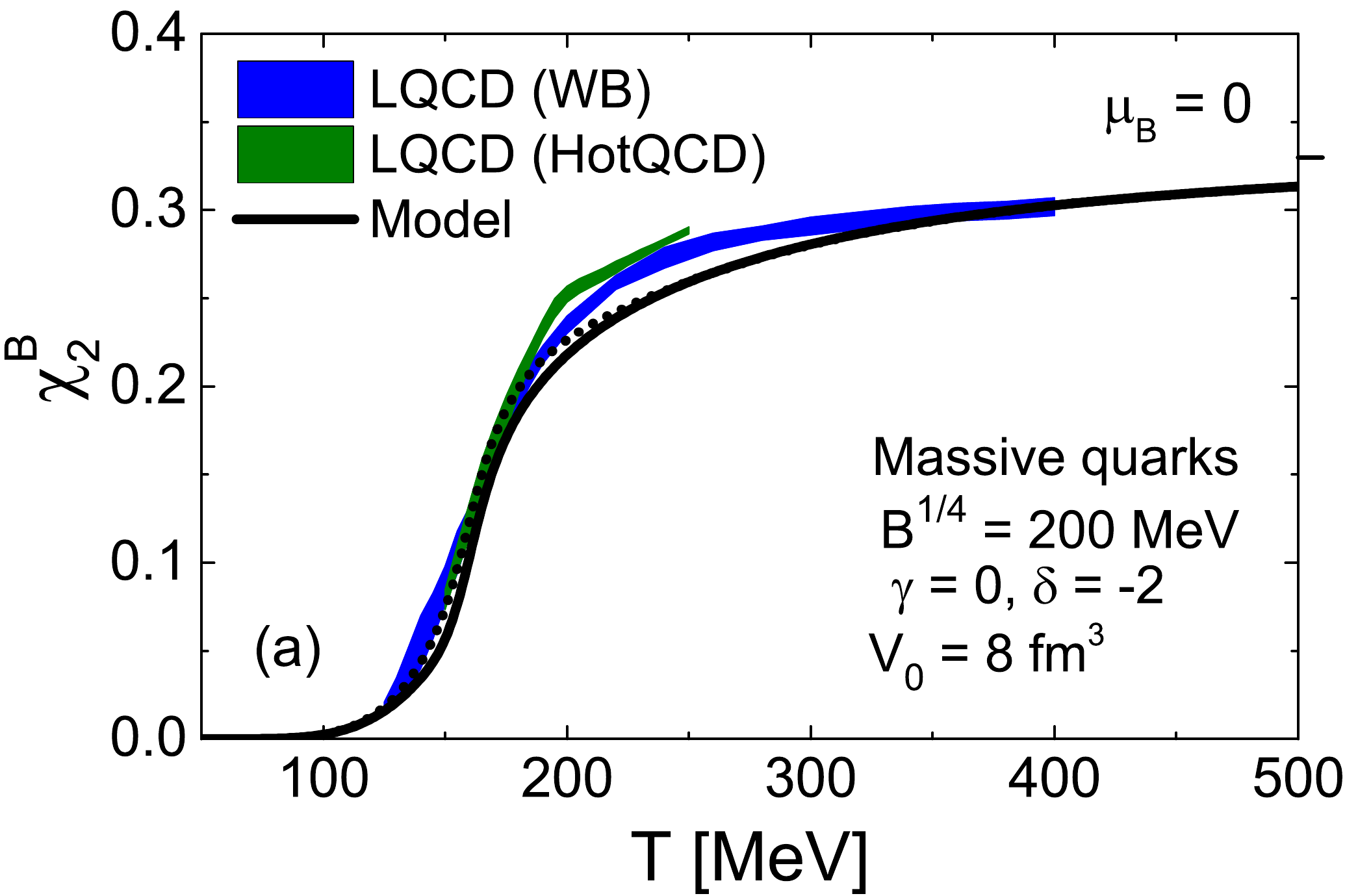}
  \includegraphics[width=.32\textwidth]{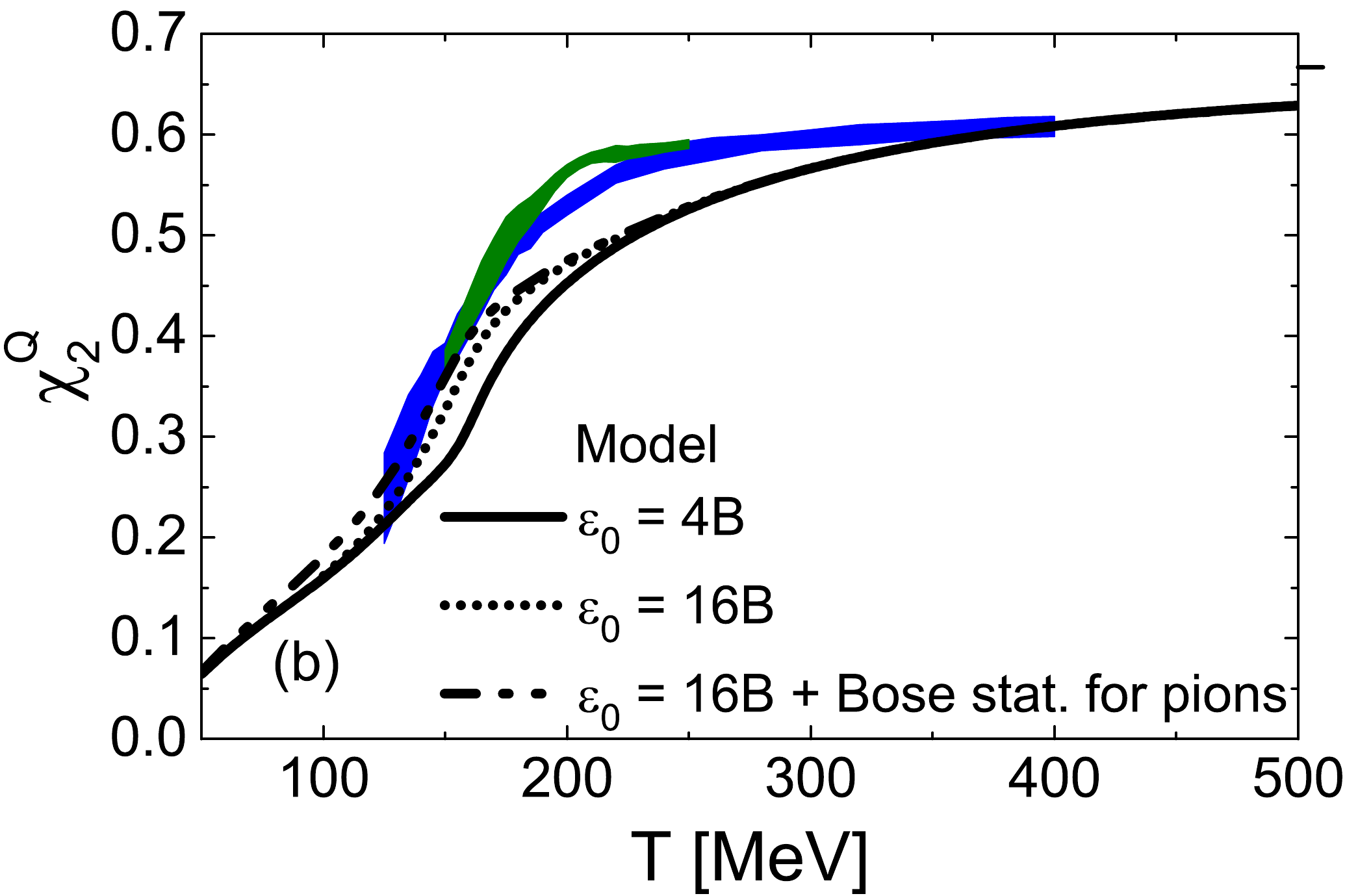}
  \includegraphics[width=.32\textwidth]{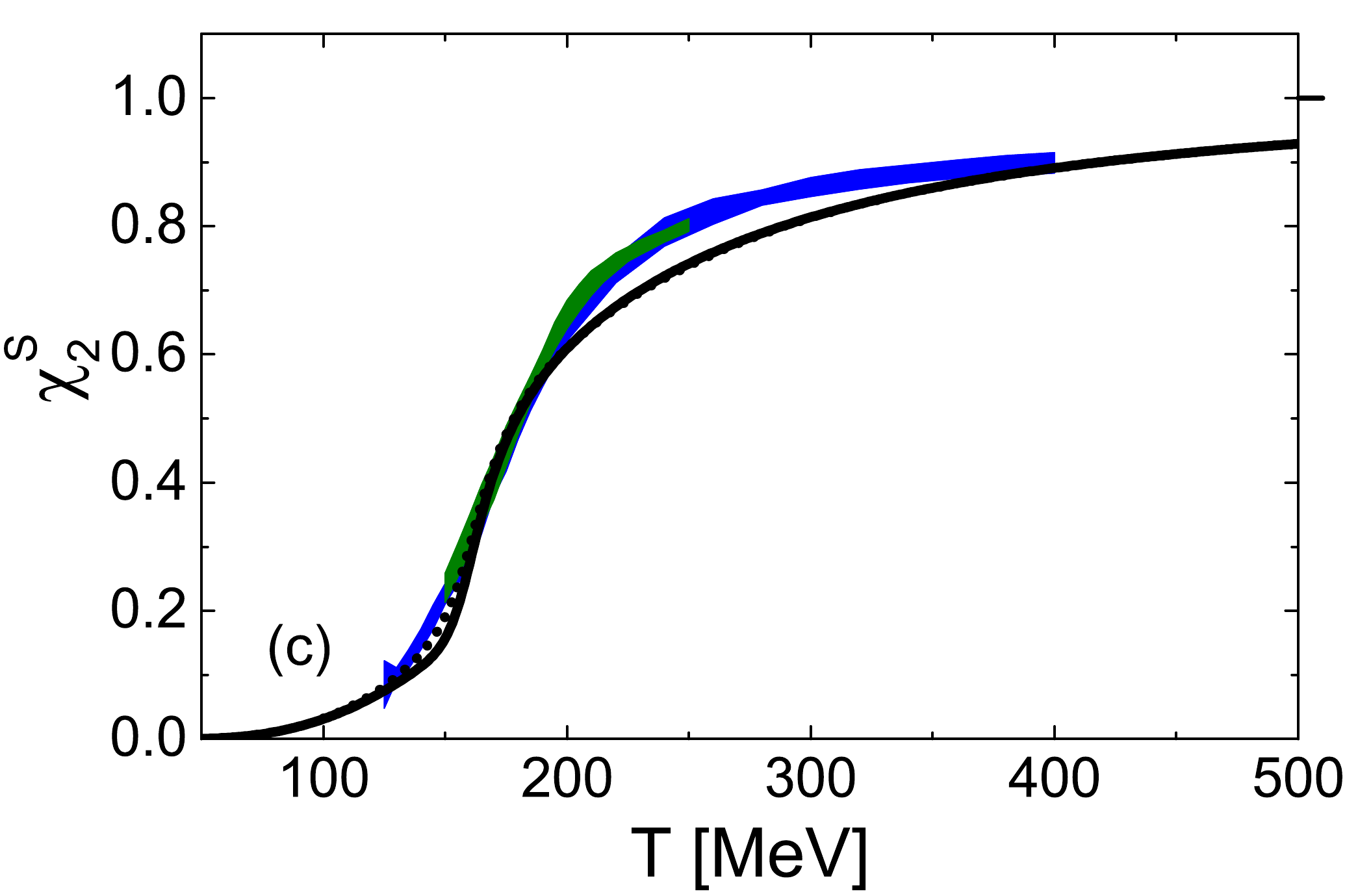}
  \includegraphics[width=.32\textwidth]{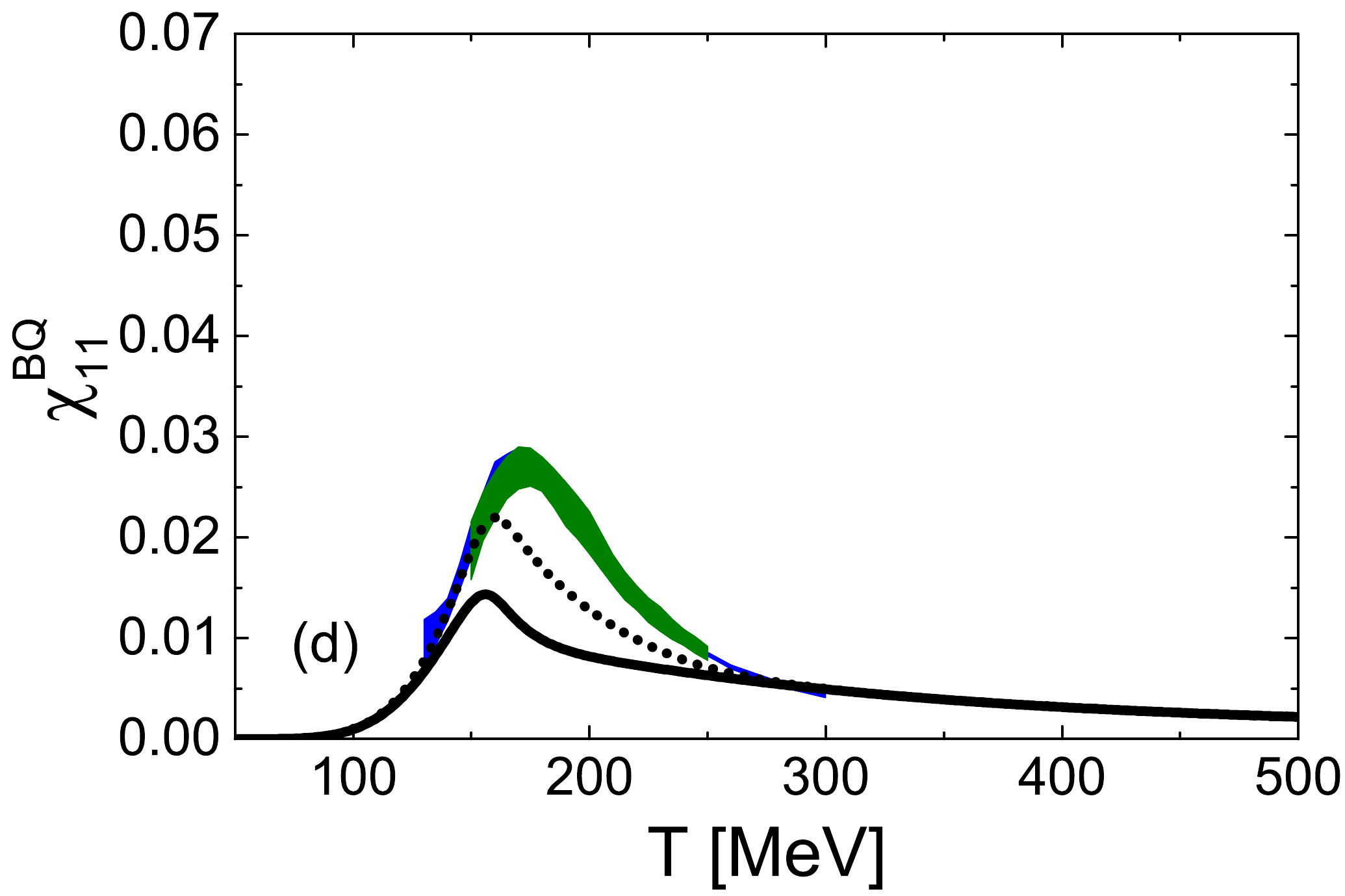}
  \includegraphics[width=.32\textwidth]{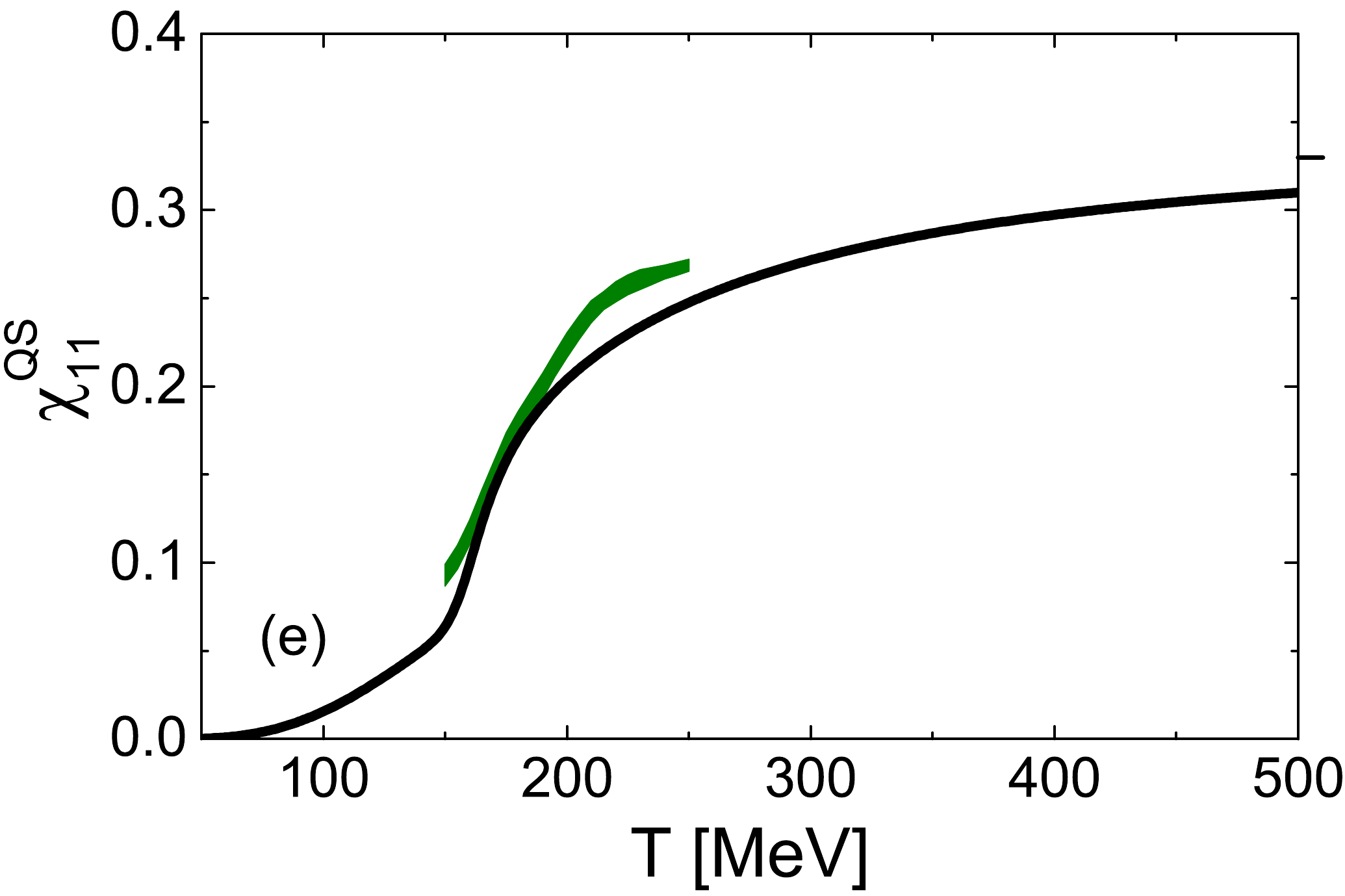}
  \includegraphics[width=.32\textwidth]{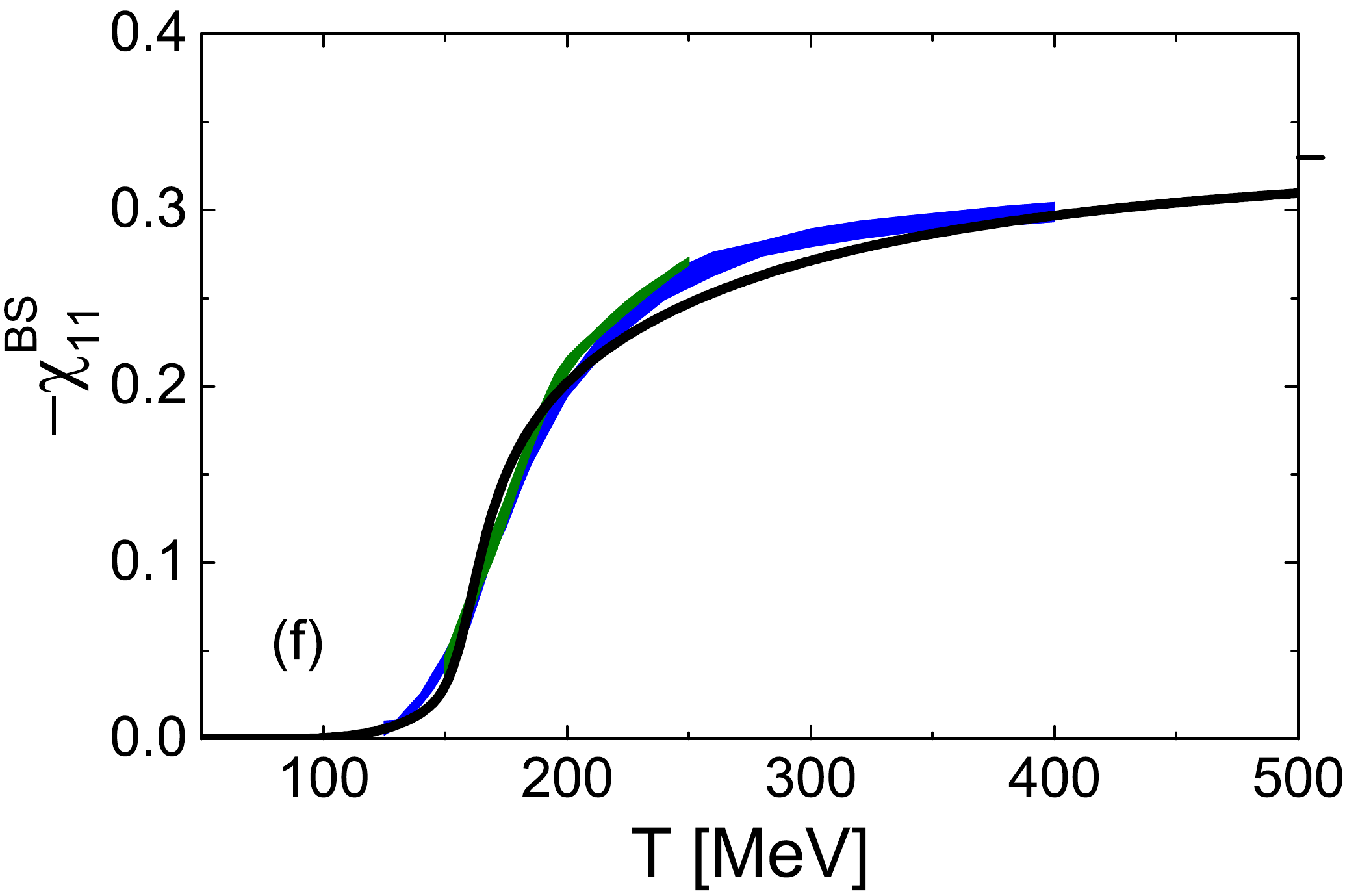}
  \caption{The temperature dependence of the second order conserved charges susceptibilities:
  (a) $\chi_2^B$,
  (b) $\chi_2^Q$,
  (c) $\chi_2^S$,
  (d) $\chi_{11}^{BQ}$,
  (e) $\chi_{11}^{QS}$,
  (f) $-\chi_{11}^{BS}$,
  computed in the Hagedorn model with
  quark-gluon bags filled with massive quarks and gluons.
  Lattice QCD data of the Wuppertal-Budapest~\cite{Borsanyi:2011sw} and HotQCD collaborations~\cite{Bazavov:2012jq} are shown by the
  blue and green bands, respectively.
  }
  \label{fig:MassiveSusc}
\end{figure*}

We turn now to the behavior of the susceptibilities of conserved charges.
The temperature dependence of the matrix of the second order conserved charges susceptibilities in the present model is depicted in Fig.~\ref{fig:MassiveSusc}.
These are compared to the lattice QCD data of the Wuppertal-Budapest~\cite{Borsanyi:2011sw} and HotQCD collaborations~\cite{Bazavov:2012jq}.
The model predictions agree qualitatively with the lattice data for all observables considered.
A notable underestimation of the lattice data is observed for $\chi_2^Q$ and $\chi_{11}^{BQ}$ in the vicinity and also above the crossover temperature region.
We argue that these observables are sensitive to the eigenvolume values assumed for the discrete, PDG part of the hadronic spectrum. In order to illustrate this sensitivity, we additionally depict in Fig.~\ref{fig:MassiveSusc} the calculation results in the case where the eigenvolumes of the PDG hadrons were taken to be four times smaller, i.e. $\varepsilon_0 = 16B$ instead of the standard choice of $\varepsilon_0 = 4B$.
This modification results in an notably improved description of $\chi_2^Q$ and $\chi_{11}^{BQ}$, as well in a slightly better agreement for some other observables such $(\varepsilon-3p)/T^4$, $c_s^2$, and $\chi_2^B$, see the dotted curves in Figs.~\ref{fig:MassiveThermod} and \ref{fig:MassiveSusc}. It also does not break the existing agreement for other observables.

The net charge susceptibility $\chi_2^Q$ is furthermore sensitive to the quantum statistical effects for charged pions, owing to their small masses and to the fact they carry electric charge~\cite{Karsch:2015zna}.
The Bose-Einstein statistics of pions enhances the fluctuation observables.
To illustrate this, we have performed the calculation where we have substituted the Boltzmann pressures of the three pions in Eq.~\eqref{eq:pfull} by the corresponding pressures of the ideal Bose-Einstein gas.
The resulting effect on $\chi_2^Q$ is depicted in Fig.~\ref{fig:MassiveSusc}(b) by the dash-dotted line. The inclusion of the Bose statistics for pions improves the agreement with the lattice data for $\chi_2^Q$ in the crossover temperature region, other observables are almost unaffected.

\begin{figure*}[t]
  \centering
  \includegraphics[width=.69\textwidth]{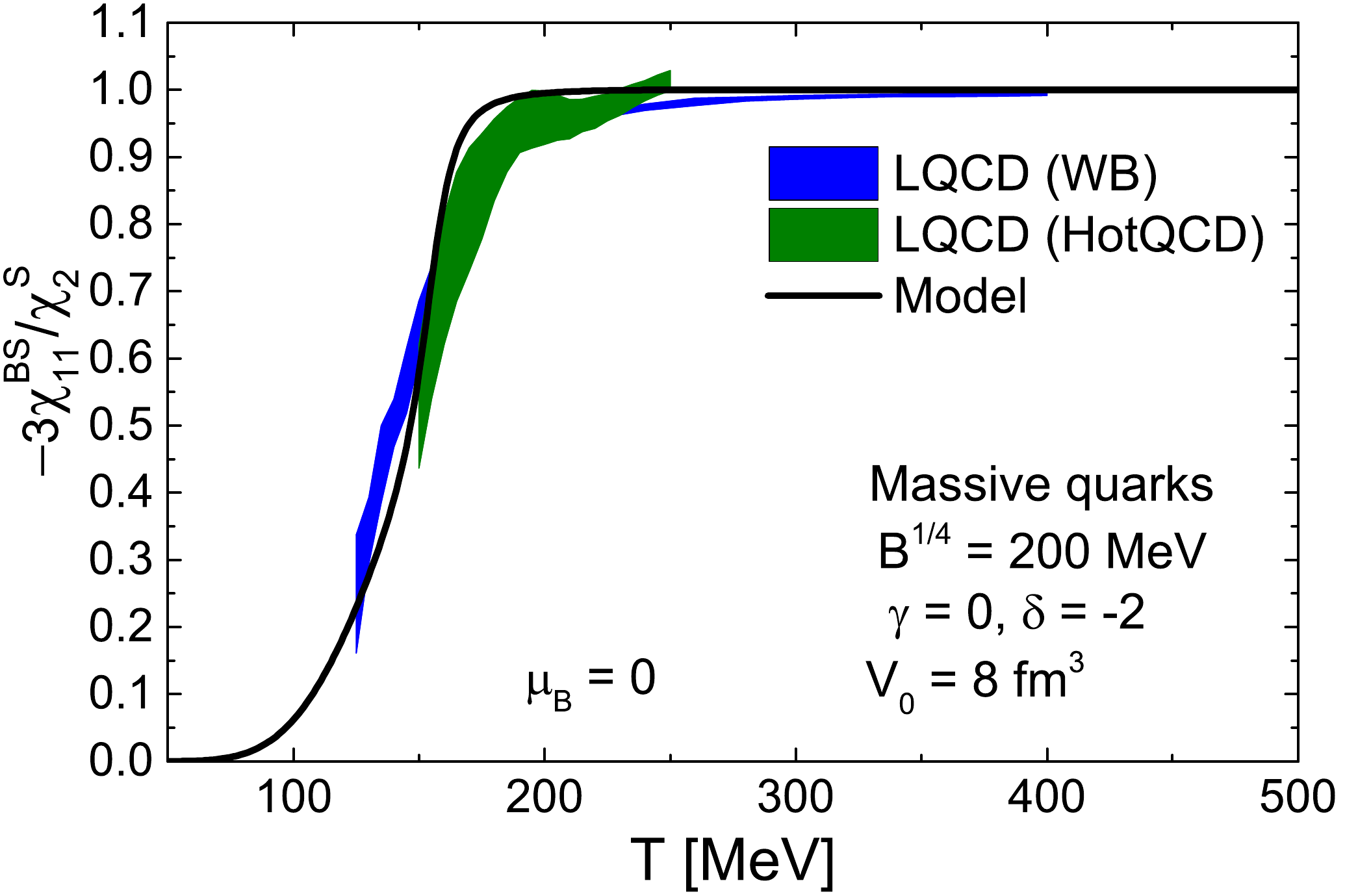}
  \caption{The temperature dependence of the baryon-strangeness correlator ratio $C_{BS} = -3 \frac{\chi_{11}^{BS}}{\chi_2^S}$,
  computed in the Hagedorn bag-like model with
  quark-gluon bags filled with massive quarks and gluons.
  Lattice QCD data of the Wuppertal-Budapest~\cite{Borsanyi:2011sw} and HotQCD collaborations~\cite{Bazavov:2012jq} are shown by the
  blue and green bands, respectively.
  }
  \label{fig:MassiveCBS}
\end{figure*}

We also consider the baryon-strangeness correlator ratio, 
\eq{
C_{BS} = -3\frac{\chi_{11}^{BS}}{\chi_2^S},
}
introduced in Ref.~\cite{Koch:2005vg} as a useful diagnostic of QCD matter.
An uncorrelated gas of hadrons and resonances yields a strong dependence of $C_{BS}$ on both the temperature and the baryochemical potential.
The QGP phase, on the other hand, is characterized by $C_{BS} \simeq 1$ at both zero and finite $\mu_B$.
The temperature dependence of $C_{BS}$ at $\mu_B = 0$ is depicted in Fig.~\ref{fig:MassiveCBS}, together with the lattice QCD data.
$C_{BS}$ shows a quick increase at small temperatures followed by a quick saturation just above the Hagedorn temperature $T_H$ in the model.
This behavior is consistent with the lattice QCD data.

\subsection{Higher-order susceptibilities}

\begin{figure*}[t]
  \centering
  \includegraphics[width=.49\textwidth]{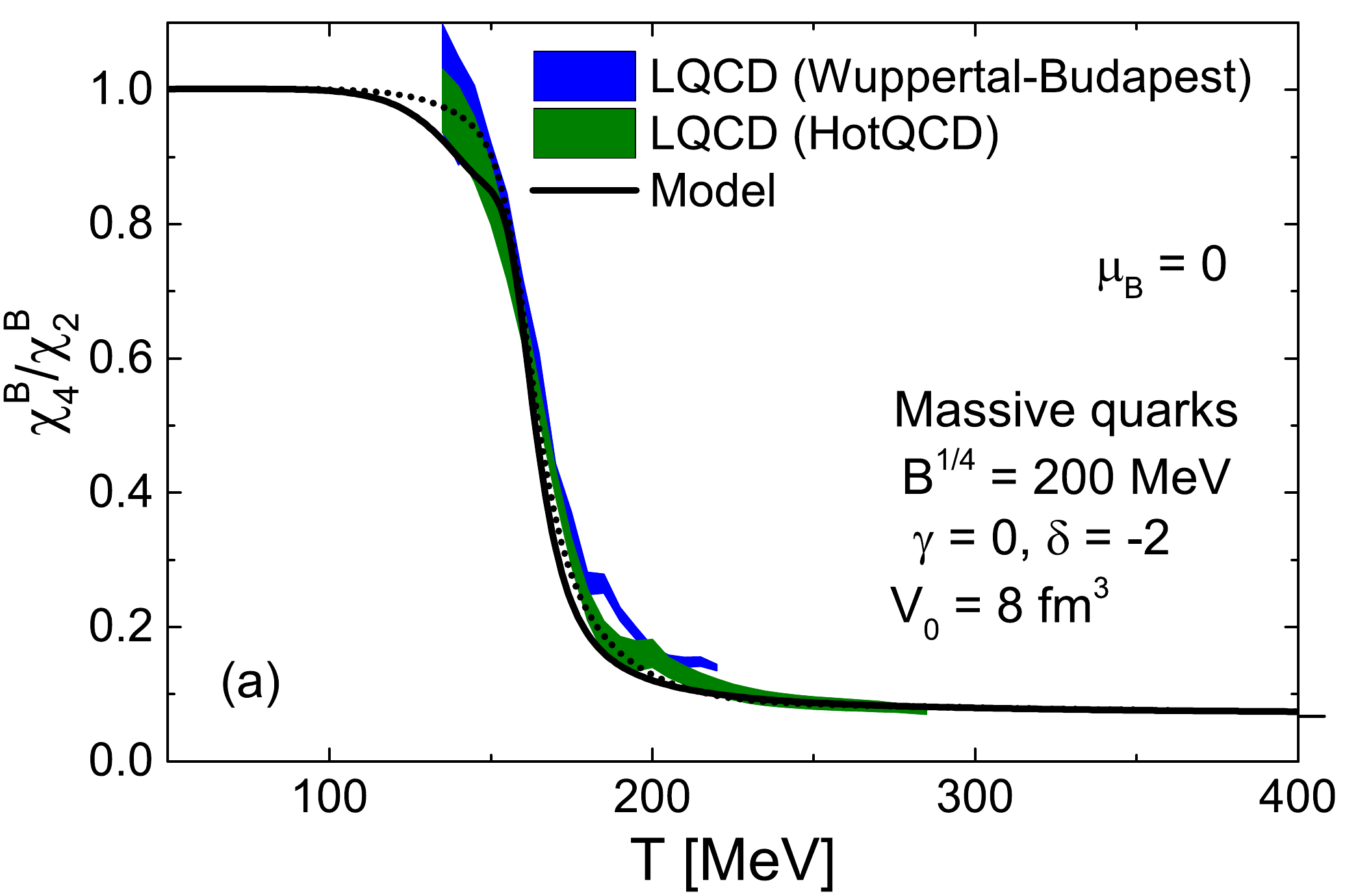}
  \includegraphics[width=.49\textwidth]{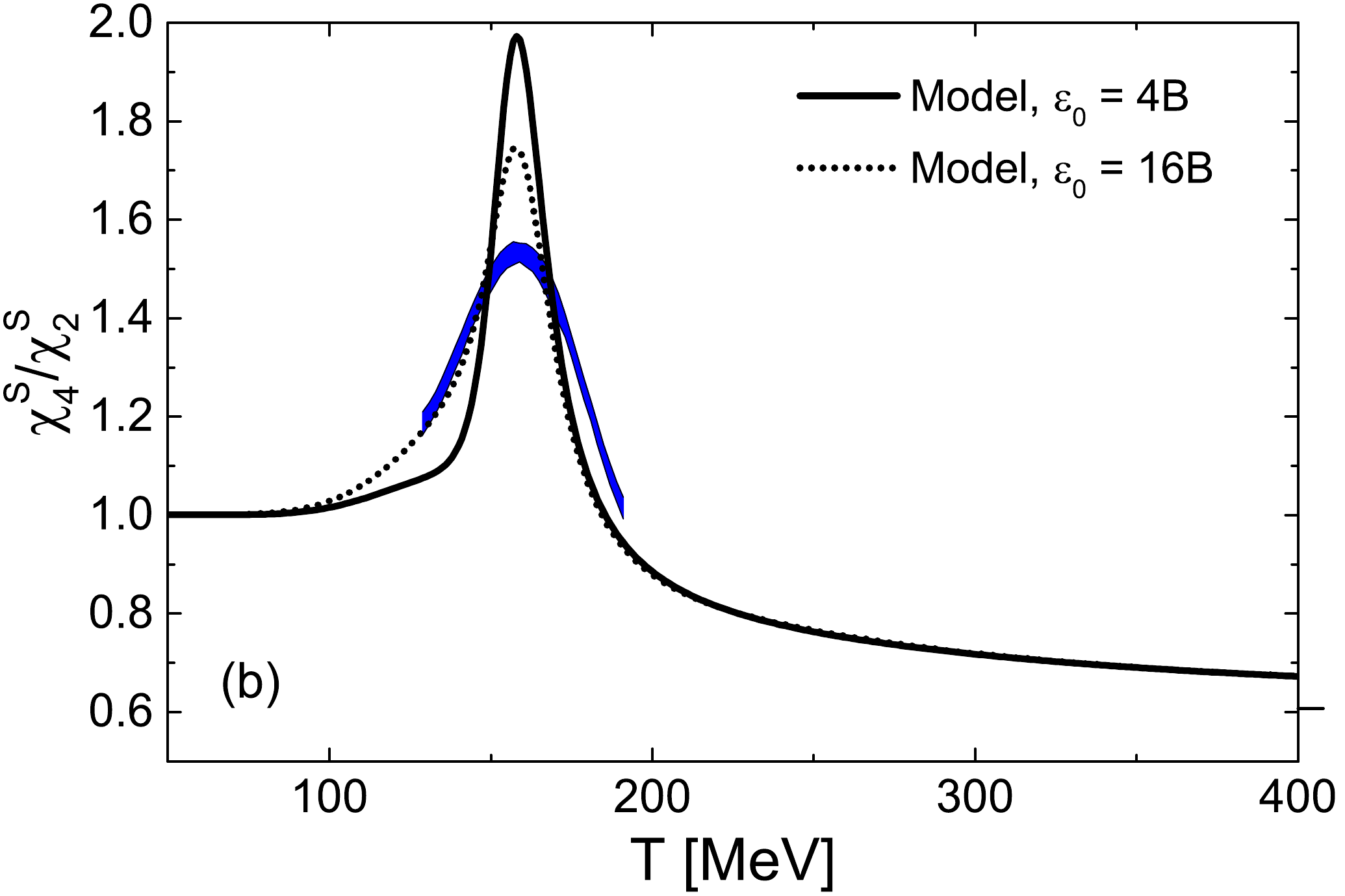}
  \includegraphics[width=.49\textwidth]{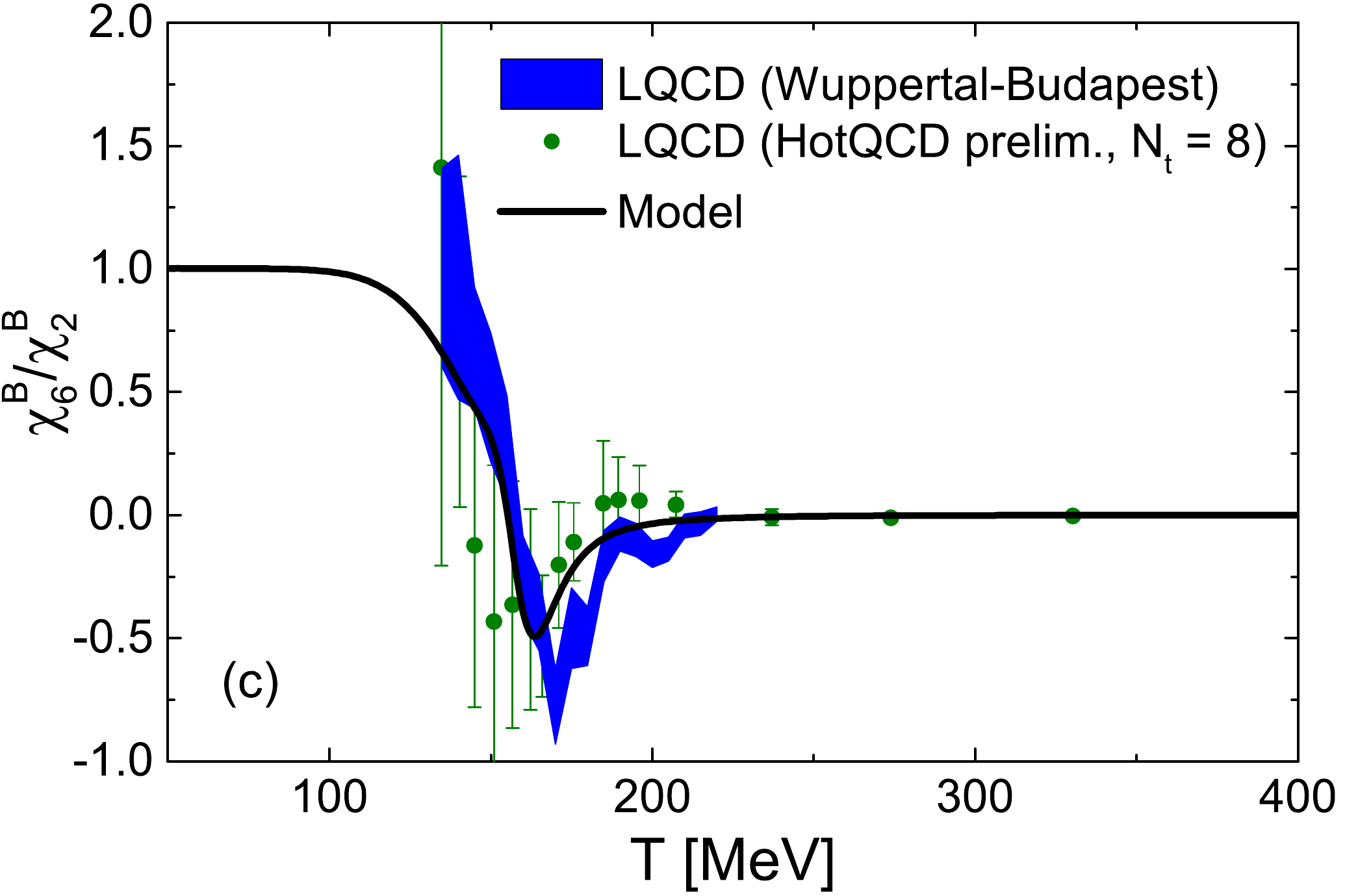}
  \includegraphics[width=.49\textwidth]{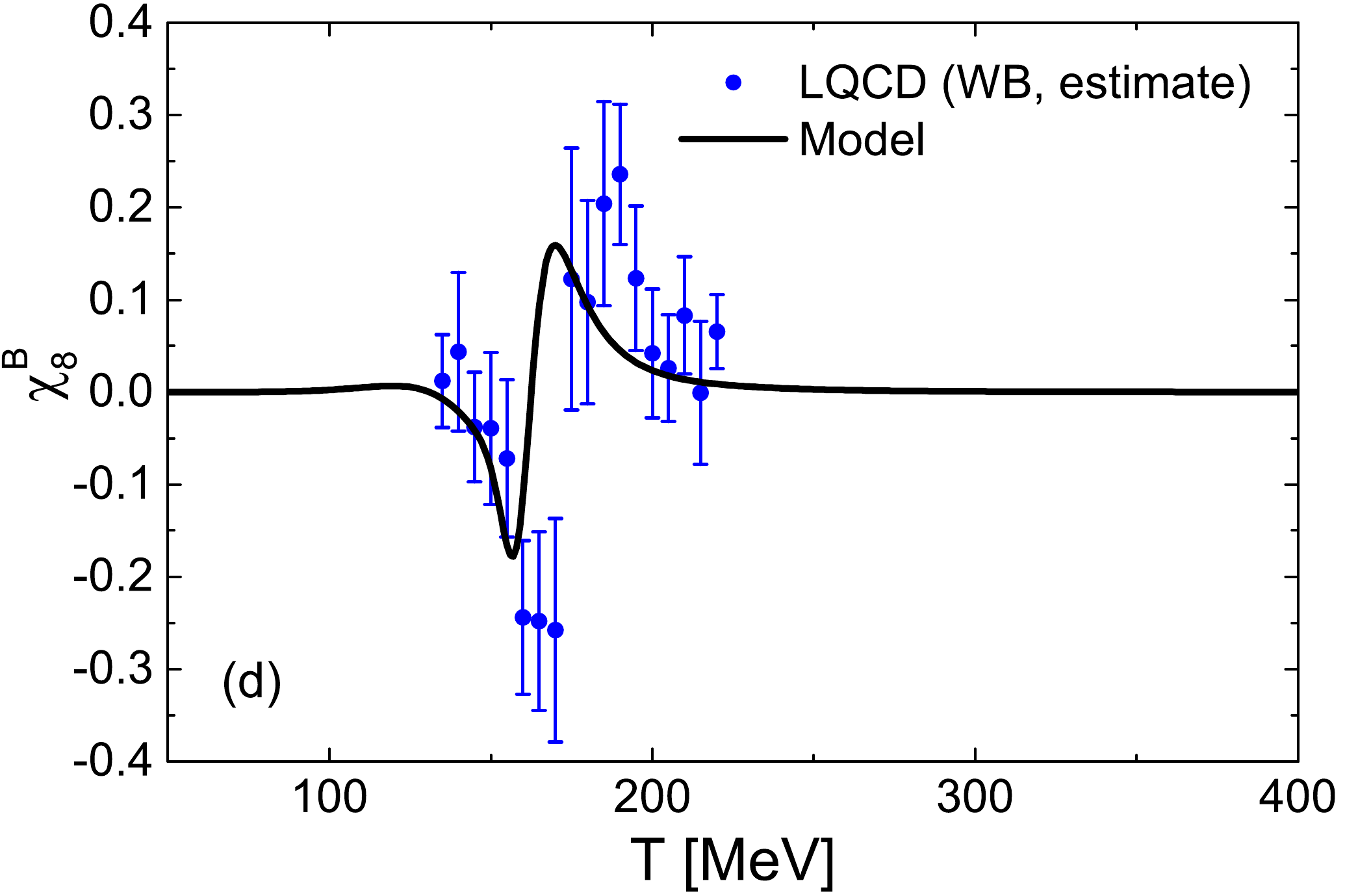}
  \caption{The temperature dependence of the conserved charges susceptibilities:
  (a) $\chi_4^B/\chi_2^B$,
  (b) $\chi_4^S/\chi_2^S$,
  (c) $\chi_6^B/\chi_2^B$,
  and
  (d) $\chi_8^B$,
  computed in the Hagedorn model with
  quark-gluon bags filled with massive quarks and gluons,
  and compared to the lattice QCD data
  of the Wuppertal-Budapest~\cite{Bellwied:2013cta,Borsanyi:2018grb} and
  HotQCD~\cite{Bazavov:2017tot,Bazavov:2017dus} collaborations.
  }
  \label{fig:MassiveHigherSusc}
\end{figure*}

Higher-order susceptibilities are expected to be particularly sensitive to crossing the crossover transition~\cite{Friman:2011pf}.
The higher-order susceptibilities, such as $\chi_4^B / \chi_2^B$, $\chi_4^S / \chi_2^S$, $\chi_6^B / \chi_2^B$, and $\chi_8^B$, were recently computed in lattice QCD~\cite{Bellwied:2013cta,Bazavov:2017dus,Bazavov:2017tot,Borsanyi:2018grb}.
Here we study the above observables in the Hagedorn bag-like model with massive quarks and gluons. 
The results are depicted in Fig.~\ref{fig:MassiveHigherSusc},
together with the lattice data.

The so-called kurtosis of the net baryon number fluctuations -- the $\chi_4^B / \chi_2^B$ ratio -- shows a rapid decrease from unity towards the Stefan-Boltzmann limiting value of $2/(3\pi^2)$ in the temperature range $T = 150-200$~MeV, see Fig.~\ref{fig:MassiveHigherSusc}(a).
In the conventional ideal HRG model this quantity is equal to unity, owing to the fact that no multi-baryon hadrons are known to exist.
This scenario is reasonable for low temperatures, where a dilute hadron gas is expected, and it is realized in the present model.
The deviation of this observable from unity in the vicinity of the pseudocritical temperature, seen in lattice QCD, is often interpreted as a signal for a rapid hadron melting and transition to a deconfined phase~\cite{Bazavov:2013dta}.
However, this onset of deviations from unity is well captured also in a HRG model with repulsive excluded volume interactions~\cite{Albright:2015uua,Vovchenko:2016rkn,Vovchenko:2017zpj,Huovinen:2017ogf}.
The present Hagedorn model extends the HRG model to include the exponentially increasing Hagedorn bag spectrum with massive quarks and gluons, as well as the excluded volume corrections.
The combined effect of these two extensions leads to the behavior shown in Fig.~\ref{fig:MassiveHigherSusc}(a) which is consistent with the lattice data on a \emph{quantitative} level.
At this point it is necessary to emphasize the importance of including both the exponential Hagedorn spectrum and the excluded-volume interactions.
Only when both of effects are included simultaneously is it possible to obtain a smooth transition to the quark-gluon plasma type behavior of $\chi_4^B / \chi_2^B$ at high temperatures.
Taking into account the Hagedorn spectrum, but not the excluded volume corrections, would lead to a monotonic increase of $\chi_4^B / \chi_2^B$ with temperature, with a subsequent divergence at the Hagedorn temperature~\cite{KaiCommun} -- a behavior inconsistent with lattice QCD.

The temperature dependence of the net strangeness kurtosis -- the $\chi_4^S / \chi_2^S$ ratio -- shows a peak at $T \simeq 160$~MeV, both in the model and in the lattice data.
It is also consistent with an approach to unity at low temperatures, and an approach to the Stefan-Boltzmann limit at high temperatures.
The presence of the peak in the temperature dependence of $\chi_4^S / \chi_2^S$ is an interplay of two effects: (i) the presence of the multi-strange hyperons in the list of known hadrons causes the initial increase of $\chi_4^S / \chi_2^S$ to above unity~\cite{Bellwied:2013cta}, whereas (ii) the presence of the excluded-volume corrections suppresses this ratio at higher temperatures~\cite{Vovchenko:2016rkn}.
The inclusion of the Hagedorn quark-gluon bag states enforces the quark-gluon plasma type behavior of this observable at high temperatures.
As shown by the dashed line in Fig.~\ref{fig:MassiveHigherSusc}(b), this observable is rather sensitive to the magnitude of the eigenvolumes taken for the PDG hadrons.
This sensitivity is less pronounced for $\chi_4^B / \chi_2^B$.

The behavior of the sixth- and eight-order net baryon susceptibilities $\chi_6^B / \chi_2^B$ and $\chi_8^B$, shown in Fig.~\ref{fig:MassiveHigherSusc} (c) and (d), shows a strong non-monotonic temperature dependence.
As far as the present level of accuracy in the lattice data is concerned, the Hagedorn model provides a reasonable quantitative description of these data.
As within its present formulation the model exhibits a crossover transition at both zero and non-zero baryon density, this agreement suggests that strong non-monotonic behavior seen in lattice data is not unambiguously related to possible critical phenomena, at least not directly.

\subsection{Fourier coefficients at imaginary $\mu_B$}

The model can also be applied to study observables at imaginary chemical potentials. This is achieved through the analytic continuation.
These observables can then be compared with lattice QCD data at imaginary chemical potentials.
Such a comparison with an independent set of lattice observables provides an important cross-check of the model validity.

Here we consider the behavior of the model at the imaginary baryochemical potential $\mu_B = i \theta_B \, T$, the electric and strangeness chemical potentials are set to zero.
The QCD partition function is an even function of $\mu_B$ because of the CP-symmetry, and it is periodic in the imaginary $\mu_B/T$ direction with the period of $2\pi$ -- the Roberge-Weiss symmetry~\cite{Roberge:1986mm}.
Therefore, it is sufficient to apply the model in the interval $0 < \theta_B < \pi$, where it exhibits analytic behavior\footnote{The Roberge-Weiss transition is expected at $\theta_B = \pi$ in the deconfined phase~\cite{Roberge:1986mm}.
Therefore, the functional form~\eqref{eq:MassiveQuarks} for the quantity $\sigma_Q$ 
should be considered at imaginary $\mu_B$ only up to this $\theta_B$ value.
}.

The QCD pressure at imaginary baryochemical potential can be written in terms of the Fourier series
\eq{
\left. \frac{p(T,\mu_B)}{T^4} \right|_{\mu_B = i \theta_B \, T} = p_0(T) + \sum_{k=1}^{\infty} \, p_k(T) \, \cos(k   \theta_B)~.
}
with the Fourier coefficients
\eq{\label{eq:pk}
p_k(T) = \frac{2}{ \pi \, (1+\delta_{k0}) } \, \int_{0}^{\pi} \, \frac{p(T,i \theta_B \, T)}{T^4} \, \cos(k \theta_B) \, d \theta_B~.
}

The net baryon density at imaginary $\mu_B$ reads
\eq{
\left. \frac{\rho_B(T,\mu_B)}{T^3} \right|_{\mu_B = i \theta_B \, T} \equiv
\left. \frac{\partial (p/T^4)}{ \partial (\mu_B/T)} \right|_{\mu_B = i \theta_B \, T} = i \, \sum_{k=1}^{\infty} \, b_k(T) \, \sin(k   \theta_B)~,
}
with $b_k \equiv k \, p_k$.

The leading four Fourier coefficients $b_k$ of the net baryon density at imaginary $\mu_B$ were recently calculated in lattice QCD and presented in Ref.~\cite{Vovchenko:2017xad}.
The coefficients were used to constrain the parameters of various phenomenological models, such as the excluded volume HRG model~\cite{Vovchenko:2017xad} or the cluster expansion model~\cite{Vovchenko:2017gkg}.
Here we do not use the Fourier coefficients to constrain the parameters of the Hagedorn model but rather test whether the behavior of $b_k$ in the model is generally compatible with the lattice data.
These Fourier coefficients are calculated in the Hagedorn model numerically, via Eq.~\eqref{eq:pk}.

\begin{figure*}[t]
  \centering
  \includegraphics[width=.69\textwidth]{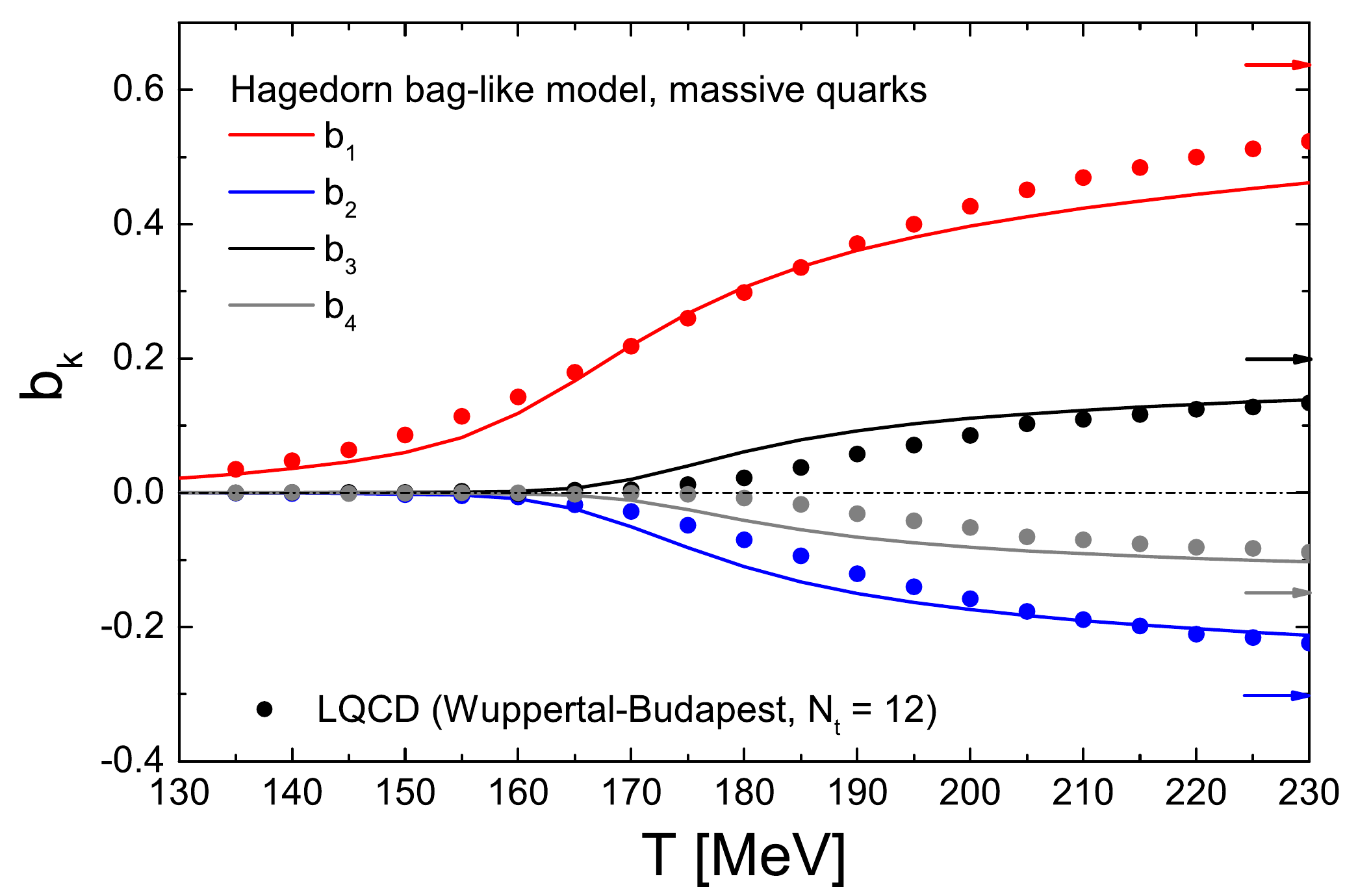}
  \caption{Temperature dependence of the leading four Fourier coefficients of the net baryon density at imaginary $\mu_B$ calculated within the Hagedorn model with quark-gluon bags filled with massive quarks and gluons~(solid lines) compared with the lattice QCD data~\cite{Vovchenko:2017xad}~(dots).
  The arrows depict the Stefan-Boltzmann limiting values~\cite{Vovchenko:2017xad}.
  }
  \label{fig:Fourier}
\end{figure*}

Comparison with the lattice data is presented in Fig.~\ref{fig:Fourier}.
Both the model and the lattice data predict $|b_k(T)/b_1(T)| \ll 1$ at $T \lesssim 160$~MeV. 
This is consistent with the picture of an uncorrelated gas of hadrons at low temperatures.
The higher-order coefficients start to visibly depart from zero as the temperature is increased to above the Hagedorn temperature $T_H$.
The coefficients show an alternating sign structure: the odd-order coefficients, $b_1$ and $b_3$, are positive, whereas the even-order ones, $b_2$ and $b_4$, are negative.
The emergence of this structure was explained in Ref.~\cite{Vovchenko:2017xad} in terms of a baryonic excluded-volume,
an alternating sign structure is also expected for an uncorrelated massless gas of quarks at high temperatures.

The present model agrees qualitatively with the available lattice data.
It does appear to underestimate $b_1$ and overestimate the higher-order coefficients at certain temperatures. 
The quantitative description can be improved by a variation of the model parameters.

\subsection{Some remarks on the chiral transition}

As presented, the description of a hadronic gas together with fluctuating Hagedorn bag-like states within the pressure ensemble including their (repulsive) eigenvolume interactions shows many agreements with the current state of the art lattice QCD equation of state. 
On the other hand, the chiral transition, taking place with increasing temperature in the crossover region, has not been discussed, as the present model is not suited
for a straightforward calculation of the chiral order parameter -- the chiral condensate $\langle \bar{q}q \rangle$.
The phenomenological picture of the chiral transition, however, can be given.
The bag-like states start to rapidly occupy nearly the whole system volume during the (crossover) transition within a small temperature interval [see the behavior of the filling fraction in Fig.~\ref{fig:MITBagAux}~(a)]. 
As the (MIT-)bags inside are chirally restored with $\langle \bar{q}q \rangle = 0$, the total overall order parameter should rapidly decrease towards 0, mimicking the chiral transition.

\begin{figure*}[t]
  \centering
  \includegraphics[width=.69\textwidth]{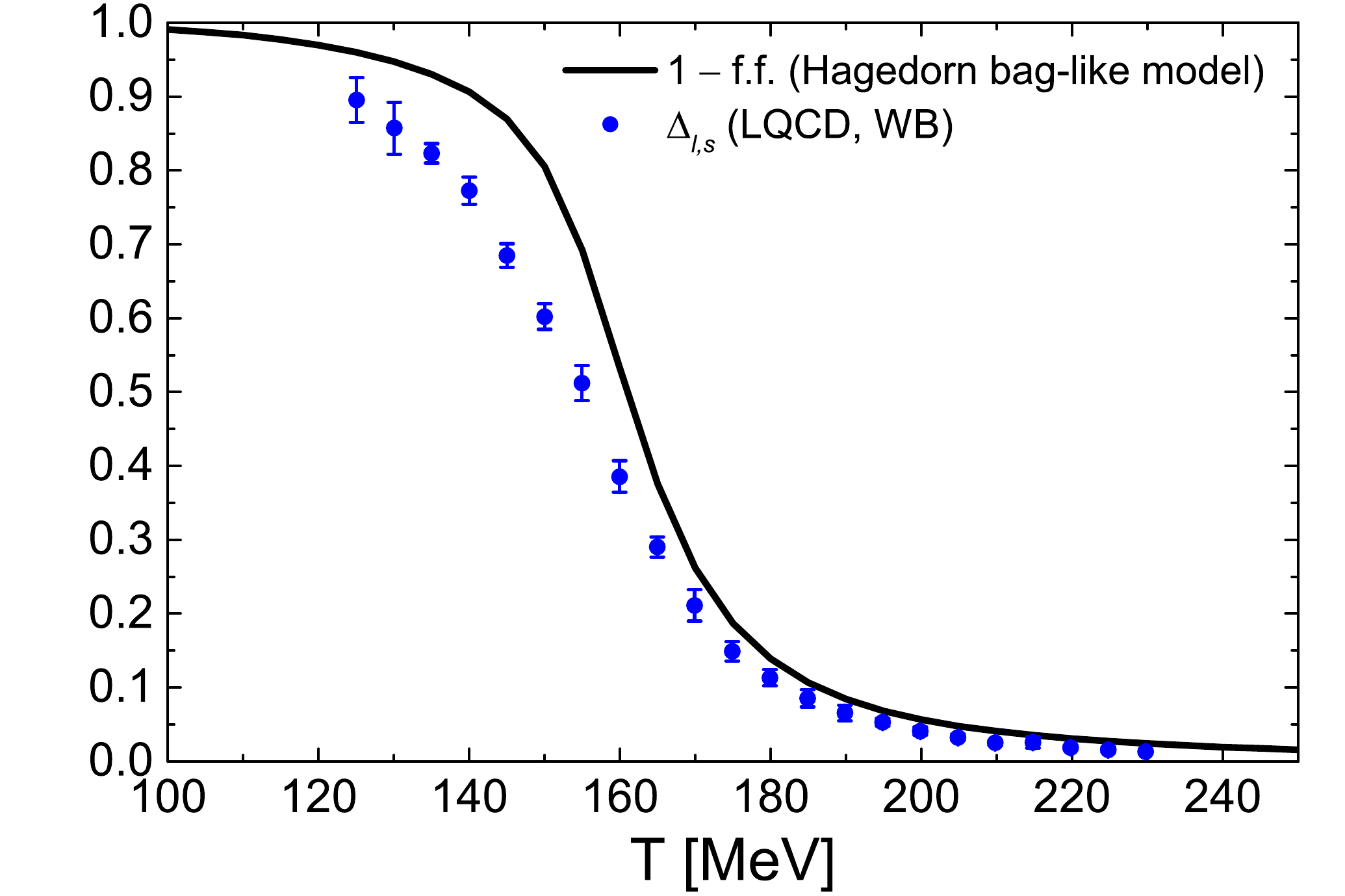}
  \caption{
  Temperature dependence of the available volume fraction, $1- f.f.$, calculated within the Hagedorn model with quark-gluon bags filled with massive quarks and gluons~(black solid line).
  The blue symbols depict the lattice QCD data of the Wuppertal-Budapest collaboration~\cite{Borsanyi:2010bp} for the subtracted chiral condensate $\Delta_{l,s}$.
  }
  \label{fig:ChiralCondensate}
\end{figure*}

The temperature dependence of the chiral transition in the presented picture can be characterized by the available volume fraction, $1 - f.f.$, which is depicted in Fig.~\ref{fig:ChiralCondensate} by the solid line.
In a chirally ''broken'' phase at very low temperatures the particle densities are small and this quantity is close to unity.
In a chirally ''restored'' phase at high temperatures, where the bags occupy almost the whole volume, the available volume fraction is close to zero.
The chiral transition takes place in a relatively narrow $T \sim 150-180$~MeV temperature range.
A related~(but not identical) quantity computed on the lattice is the subtracted  chiral condensate $\Delta_{l,s}$~\cite{Cheng:2007jq}, which is expected to show a similar qualitative behavior as one traverses the chiral transition.
The corresponding lattice QCD data of the Wuppertal-Budapest collaboration~\cite{Borsanyi:2010bp}, computed for physical quark masses, is depicted in Fig.~\ref{fig:ChiralCondensate} by the blue symbols.
This quantity exhibits a rather similar behavior to $1 - f.f.$ from the Hagedorn bag-like model.

In the further discussion of Chapter~\ref{chap:massive} we have assumed finite  quark masses, as motivated by various thermal field calculations. Strictly speaking the masses would be temperature dependent, as poles of the thermal quark propagators. 
For the time being we have approximated
the quark masses  inside the bags to be constant during the crossover.
This was done to pursue our new calculations, restricted at the moment to the equation of state properties. 
In principle, though, one can envisage a hard-thermal-loop-improved description for the intrinsic thermal pressure of the bags, Eq.~\eqref{eq:MassiveQuarks} [or Eq.~\eqref{eq:pBag}].
Hence, inside the bags, the quark condesate would
be vanishing if the perturbative masses, $m_q$, are taken to be zero.

If a true second order chiral phase transition occurs at vanishing
baryon chemical potential at the critical temperature,
the parameters $\gamma $ and $\delta$ appearing in Eq.~\eqref{eq:rhoQ} would have to be fine-tuned.
A theoretical explanation for the behaviour of those parameters
in the chiral limit, $m_q \rightarrow 0 $, is not obvious.
Also then, as $m_\pi \rightarrow 0$, the hadronic masses
in the present thermal and extended hadronic model have to be
shifted too. Mean field type studies, as e.g. in
elaborated chiral models~\cite{Skokov:2010sf},
cannot be straightforwardly overtaken.
We leave these ideas for future studies.

\section{Summary and outlook}
\label{sec:summary}

We have studied the behavior of thermodynamic functions, various conserved charges susceptibilities at zero chemical potentials, and the Fourier coefficients at imaginary $\mu_B$ in the Hagedorn quark-gluon bag model with a crossover transition.
To the best of our knowledge, the susceptibilities are considered within such an approach for the first time in the present paper.
The model behavior of susceptibilities is found to be qualitatively compatible with lattice QCD data already for the case of bags filled with massless quarks and gluons.
A simple phenomenological extension of the bag model to include constant but finite masses of quarks and gluons leads to a significantly improved agreement with the lattice data, remedying some of the known shortcomings using the standard MIT bag model approach, such as the peak in the temperature dependence of the scaled energy density.
This result lends support to the quasiparticle picture for the QCD equation of state at high temperatures.

\begin{figure*}[t]
  \centering
  \includegraphics[width=.69\textwidth]{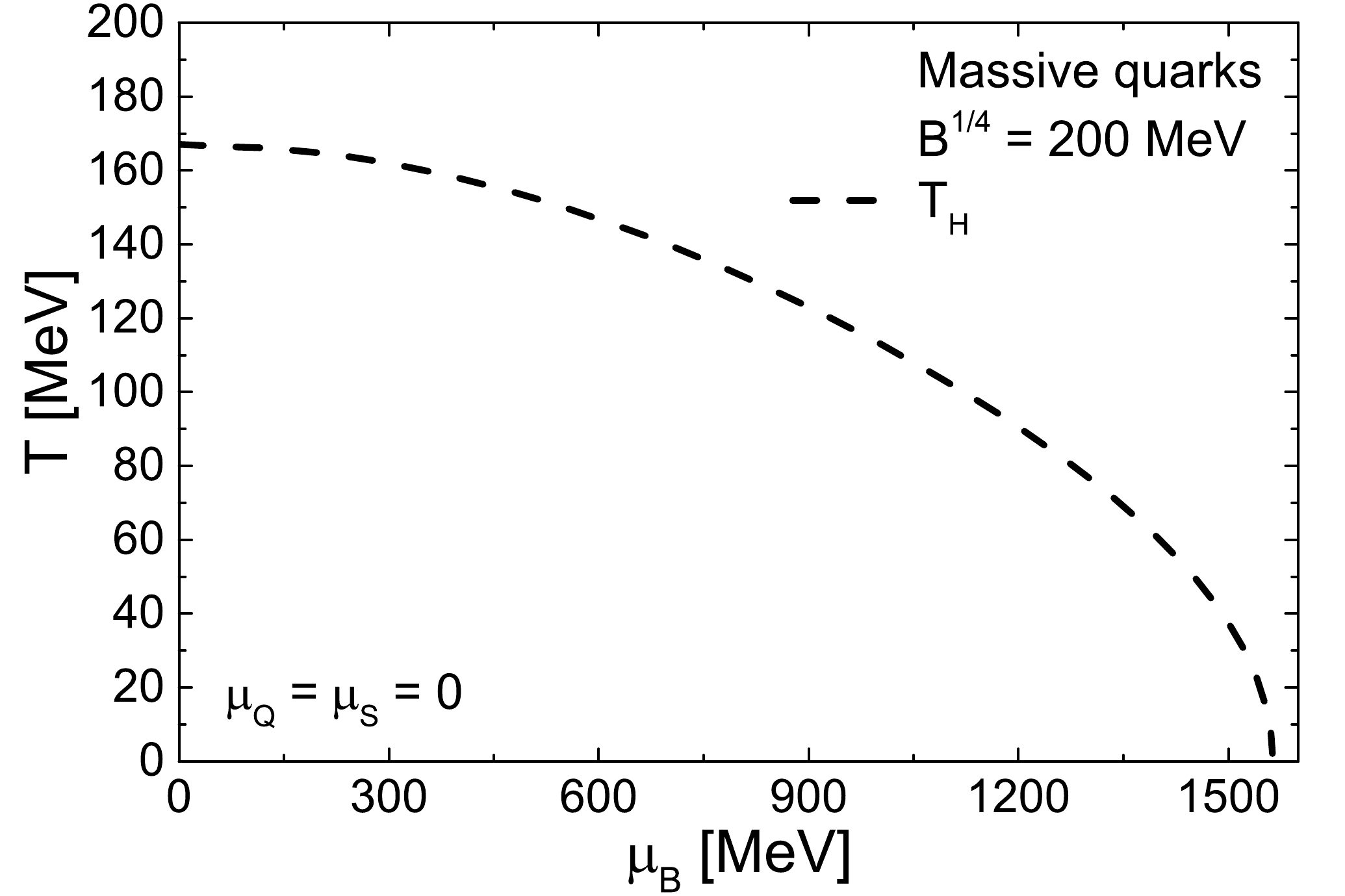}
  \caption{
  The $\mu_B$-dependence of the Hagedorn temperature, $T_H$~\eqref{eq:THMassive},
  computed for the massive quarks and gluons~\eqref{eq:QuarkMasses} and the bag constant $B^{1/4} = 200$~MeV.
  }
  \label{fig:MassiveTH}
\end{figure*}

The Hagedorn quark-gluon bag-like model,
introduced in Refs.~\cite{Gorenstein:1981fa,Gorenstein:1995vm}, 
is historically one of the first models for a hadron-parton transition in QCD.
The quantitative aspects of such a model have not been studied extensively before, the present model results and their comparison to the lattice data suggest that this approach is compatible with first-principle lattice QCD results.
One remarkable feature of such a model is that the whole transition, be it a real phase transition or a crossover, is described within a single partition function.
This is quite different from many conventional phenomenological models for the equation of state, where the hadronic and partonic phases are usually described by different partition functions that are then being matched, either via the Maxwell construction~\cite{Sollfrank:1996hd,Satarov:2009zx} or a smooth switching function~\cite{Albright:2014gva,Albright:2015uua}.

In the present work we have considered only the case where the crossover transition is realized, at all $\mu_B$.
The model can be generalized to incorporate first- and higher order phase transitions at finite baryon densities.
This can be achieved by consdering the $\mu_B$-dependent exponents $\gamma$ and $\delta$ in the pre-exponential factor of the quark-gluon bag mass-volume density~[Eq.~\eqref{eq:rhoQ}], as outlined in Ref.~\cite{Gorenstein:2005rc}.
The phase transition lines can then be expected to be located in the present approach in the vicinity of the Hagedorn temperature $T_H$~\eqref{eq:THMassive} at finite $\mu_B$, depicted for the finite quark and gluon masses and  bag constant used in the present work by the dashed line in Fig.~\ref{fig:MassiveTH}.
Such an extension would allow to look for signatures of the hypothetical hadron-parton phase transition at finite baryon density.


\begin{acknowledgments}

We thank J. Steinheimer for a suggestion to compare the available volume fraction with the chiral condensate from lattice QCD.
The work of M.I.G. was supported
by the Alexander von Humboldt Foundation and by the Program of Fundamental Research of the Department of
Physics and Astronomy of National Academy of Sciences of Ukraine.
This work was also supported by the Bundesministerium f\"ur Bildung und Forschung (BMBF), the Helmholtz International Center for FAIR within the framework of the LOEWE program launched by the State of Hesse, and by the Collaborative Research Center CRC-TR 211 “Strong-interaction matter under extreme conditions” funded by DFG.
H.St. acknowledges the support through the Judah M. Eisenberg Laureatus Chair at Goethe University, and the Walter Greiner Gesellschaft, Frankfurt.

\end{acknowledgments}

\section*{Appendix}

\subsection{The mass distribution of PDG hadrons and QGP bags}
\label{app:A}

In this Appendix we evaluate the mass distribution of PDG hadrons and of quark-gluon bags for the ``massive quarks'' parameter set used in the present work~[Eq.~\eqref{eq:MassiveQuarks}].
We also demonstrate the emergence of the exponential Hagedorn mass spectrum~[Eq.~\eqref{eq:rhoHagedorn}] at large masses.

To calculate the mass spectrum of PDG hadrons we smear their masses with relativistic Breit-Wigner distributions.
The Breit-Wigner widths are taken to be constant and correspond to the resonance widths listed in Particle Data Tables.
We also assume a width of 10~MeV for all stable hadrons, this is done for presentation purposes.
The mass spectrum of quark-gluon bags is calculated numerically from Eq.~\eqref{eq:rhoQm} using three different values of the $M_0$ parameter: 0.1, 0.5, and 1~GeV/$c^2$.
These calculations are compared with the Hagedorn mass spectrum~\eqref{eq:rhoHagedorn} with parameters given by Eq.~\eqref{eq:TH}.
The calculations are performed for the temperature $T = T_H \simeq 167.7$~MeV, this corresponds to $\sigma_Q \simeq 6.16$.

\begin{figure}
\centering
\includegraphics[width=.69\textwidth]{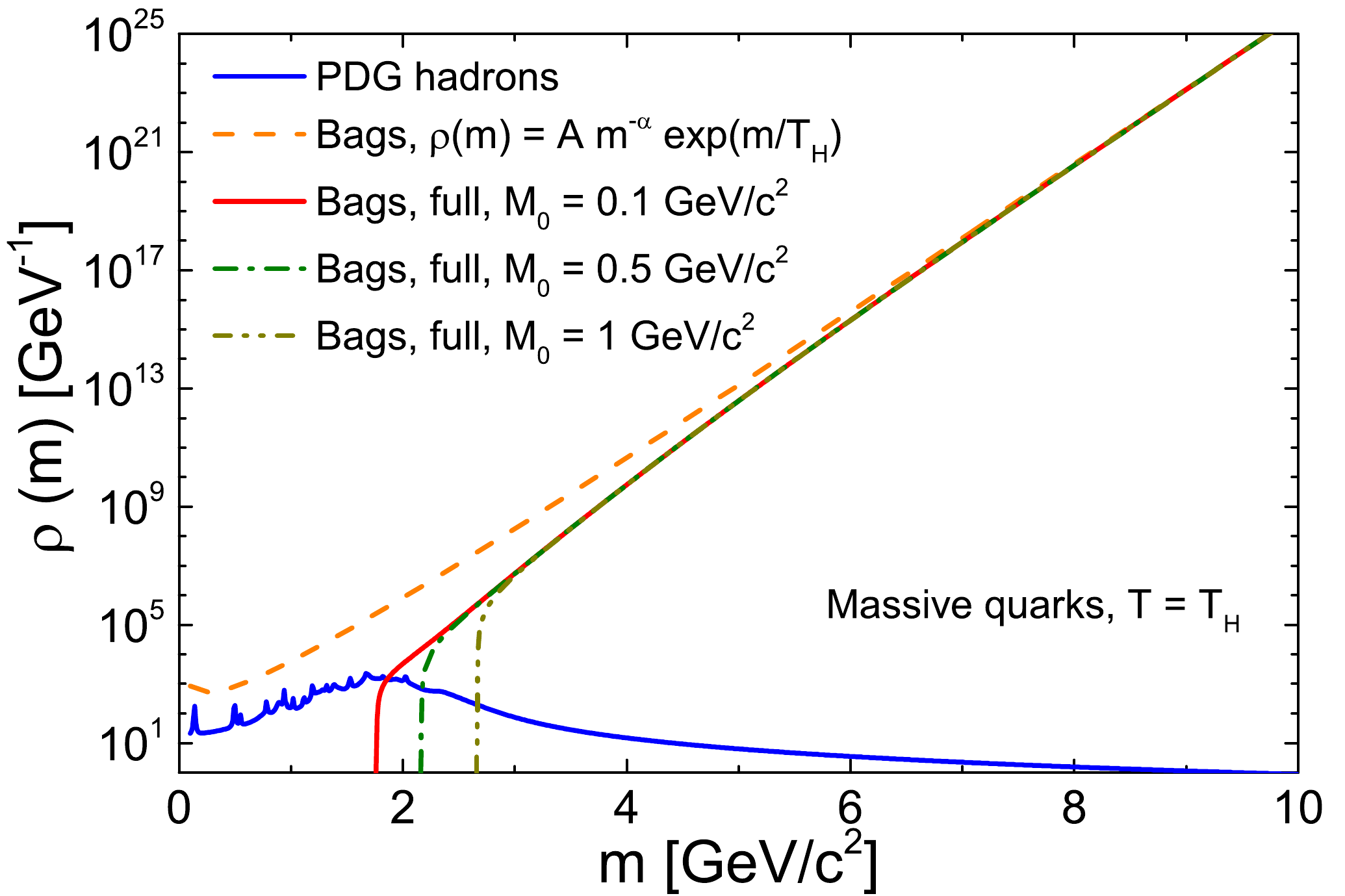}
\caption{\label{fig:MassSpectrumCompare}
The mass spectrum of the PDG hadrons (blue line) and of the QGP bags, using the Hagedorn mass spectrum form [Eq.~\eqref{eq:rhoHagedorn}]~(orange line), or evaluated numerically through Eq.~\eqref{eq:rhoQm} for three different values of $M_0$~(in GeV/$c^2$): 0.1~(red), 0.5~(green), and 1~(brown). Evaluations are done for the ``massive quarks'' parameter set~[Eq.~\eqref{eq:MassiveQuarks}] at the temperature $T = T_H \simeq 167.7$~MeV.
}
\end{figure}

Figure~\ref{fig:MassSpectrumCompare} presents the calculation results.
The numerical results for the quark-gluon bag spectrum approach the exponential Hagedorn form~\eqref{eq:rhoHagedorn} at large masses. This behavior is independent of the $M_0$ values considered.
Calculations show that the expression~\eqref{eq:rhoHagedorn} becomes accurate to within 10\% at $m \simeq 8$~GeV/$c^2$.
The behavior of the quark-gluon bag spectrum at smaller masses, $m \lesssim 3$~GeV/$c^2$, depends on the $M_0$ value.
As seen from the figure, for $M_0 = 0.1$~GeV/$c^2$ the mass spectra of PDG hadrons and quark-gluon bags appear to match with each other rather smoothly at $m \simeq 2$~GeV/$c^2$.
This is not the case for the other two $M_0$ values considered.

The picture for the ``massless quarks'' parameter sets~[Eq.~\eqref{eq:ParamSetScan}] turns out to be quite similar and not shown here. 

\subsection{Accuracy of the approximations}
\label{app:B}

In the present work the pressure has been determined as the solution of the transcendental equation~\eqref{eq:pfull}.
Two approximations were used in order to obtain the quark-gluon bag part of the r.h.s. of Eq.~\eqref{eq:pfull}:
\begin{itemize}
    \item The non-relativistic approximation~[Eq.~\eqref{eq:nonrel}]
    \item The Laplace's method to perform the integration over the mass~[Eqs.~\eqref{eq:fQ2}-\eqref{eq:fQ3}]
\end{itemize}

Both approximations are expected to be rather accurate for the heavy quark-gluon bags, as discussed in the main text.
Nevertheless, it can be useful to quantify the error introduced by these approximations.
In order to do that, we consider the exact transcendental equation for the model pressure without approximations:
\eq{\label{eq:pfullorig}
& p(T,\lambda_B,\lambda_Q,\lambda_S) =
T \sum_{i \in \rm HRG} d_i \, \phi(T,m) \, \lambda_B^{b_i} \, \lambda_Q^{q_i} \, \lambda_S^{s_i} \, \exp\left(- \frac{m_i p}{4 B T} \right) \nonumber \\
& + T \, C \, \int_{V_0} dv \, v^\gamma \, \exp\left(-\frac{vp}{T} \right) \int_{Bv + M_0} dm \, (m - Bv)^{\delta} \exp \left\{ \frac{4}{3} [\sigma_{Q}]^{1/4} \, v^{1/4} \, (m-Bv)^{3/4} \right\} \, \phi(T,m) \, .
}
Here is $\phi(T,m)$ is taken in the exact, relativistic form~\eqref{eq:phi}.
At each temperature value, we first obtain the approximate solution $p(\rm Approx)$ by solving Eq.~\eqref{eq:pfull}. This corresponds to the procedure employed in the main text.
Then, we use $p(\rm Approx)$ as the starting point to numerically solve ~\eqref{eq:pfullorig} and obtain the exact model pressure $p(\rm Full)$.
The combined accuracy of the non-relativistic and Laplace's method approximations can then be determined by comparing $p(\rm Approx)$ and $p(\rm Full)$.

\begin{figure}
\centering
\includegraphics[width=.69\textwidth]{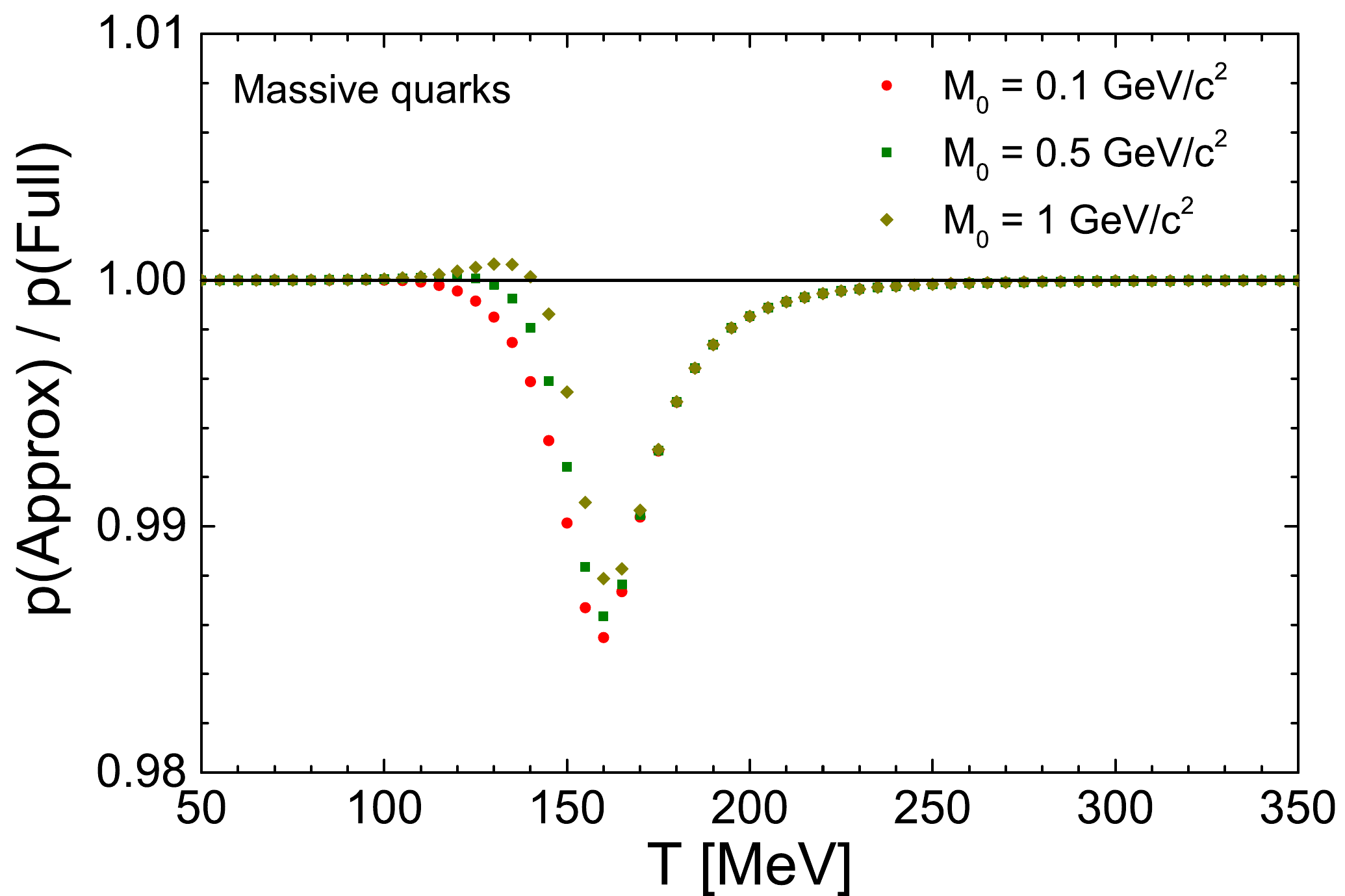}
\caption{\label{fig:pTestNonrel}
The temperature dependence of the ratio $p(\rm Approx)/p(\rm Full)$, evaluated for the "massive quarks" parameter set for three different values of $m_0$~(in GeV/$c^2$): 0.1~(red), 0.5~(green), and 1~(brown).
}
\end{figure}

Figure~\ref{fig:pTestNonrel} depicts the temperature dependence of the ratio $p(\rm Approx)/p(\rm Full)$ evaluated for the "massive quarks" parameter set~[Eq.~\eqref{eq:MassiveQuarks}] for $M_0 = 0.1$, 0.5, and 1~GeV/$c^2$.
Deviations of the ratio from unity are largest in the vicinity of $T_H$, but they do not exceed 2\%. The dependence on $M_0$ is very mild.
The results imply that the application of both, the non-relativistic approximation and Laplace's method, for calculating the pressure is well justified, at least for the parameter set considered.
This can be attributed to two reasons:
\begin{enumerate}
    \item At large temperatures the system is dominated by heavy bags. As already elaborated, the heavier the bags are, the more accurate are the approximations.
    \item At low temperatures the system is dominated by the PDG hadrons, which are evaluated without approximations. Therefore, possible inaccuracies in evaluating the contributions from the quark-gluon bags at these temperatures are irrelevant as these contributions are negligible anyhow.
\end{enumerate}

If only one of the two approximations discussed is preserved, i.e. if only the non-relativistic approximation is used but not Laplace's method, or vice versa, then the relative error in the calculated pressure is within 1\%.

We have performed similar checks for the ``massless quarks'' parameter sets~[Eq.~\eqref{eq:ParamSetScan}] and obtained a very similar result: the relative error in the calculated pressure does not exceed 2-3\%.

\bibliography{bag-model-crossover}

\begin{thebibliography}{74}%
\makeatletter
\providecommand \@ifxundefined [1]{%
 \@ifx{#1\undefined}
}%
\providecommand \@ifnum [1]{%
 \ifnum #1\expandafter \@firstoftwo
 \else \expandafter \@secondoftwo
 \fi
}%
\providecommand \@ifx [1]{%
 \ifx #1\expandafter \@firstoftwo
 \else \expandafter \@secondoftwo
 \fi
}%
\providecommand \natexlab [1]{#1}%
\providecommand \enquote  [1]{``#1''}%
\providecommand \bibnamefont  [1]{#1}%
\providecommand \bibfnamefont [1]{#1}%
\providecommand \citenamefont [1]{#1}%
\providecommand \href@noop [0]{\@secondoftwo}%
\providecommand \href [0]{\begingroup \@sanitize@url \@href}%
\providecommand \@href[1]{\@@startlink{#1}\@@href}%
\providecommand \@@href[1]{\endgroup#1\@@endlink}%
\providecommand \@sanitize@url [0]{\catcode `\\12\catcode `\$12\catcode
  `\&12\catcode `\#12\catcode `\^12\catcode `\_12\catcode `\%12\relax}%
\providecommand \@@startlink[1]{}%
\providecommand \@@endlink[0]{}%
\providecommand \url  [0]{\begingroup\@sanitize@url \@url }%
\providecommand \@url [1]{\endgroup\@href {#1}{\urlprefix }}%
\providecommand \urlprefix  [0]{URL }%
\providecommand \Eprint [0]{\href }%
\providecommand \doibase [0]{http://dx.doi.org/}%
\providecommand \selectlanguage [0]{\@gobble}%
\providecommand \bibinfo  [0]{\@secondoftwo}%
\providecommand \bibfield  [0]{\@secondoftwo}%
\providecommand \translation [1]{[#1]}%
\providecommand \BibitemOpen [0]{}%
\providecommand \bibitemStop [0]{}%
\providecommand \bibitemNoStop [0]{.\EOS\space}%
\providecommand \EOS [0]{\spacefactor3000\relax}%
\providecommand \BibitemShut  [1]{\csname bibitem#1\endcsname}%
\let\auto@bib@innerbib\@empty
\bibitem [{\citenamefont {Patrignani}\ \emph {et~al.}(2016)\citenamefont
  {Patrignani} \emph {et~al.}}]{Patrignani:2016xqp}%
  \BibitemOpen
  \bibfield  {author} {\bibinfo {author} {\bibfnamefont {C.}~\bibnamefont
  {Patrignani}} \emph {et~al.} (\bibinfo {collaboration} {Particle Data
  Group}),\ }\href {\doibase 10.1088/1674-1137/40/10/100001} {\bibfield
  {journal} {\bibinfo  {journal} {Chin. Phys.}\ }\textbf {\bibinfo {volume}
  {C40}},\ \bibinfo {pages} {100001} (\bibinfo {year} {2016})}\BibitemShut
  {NoStop}%
\bibitem [{\citenamefont {Hagedorn}(1965)}]{Hagedorn:1965st}%
  \BibitemOpen
  \bibfield  {author} {\bibinfo {author} {\bibfnamefont {R.}~\bibnamefont
  {Hagedorn}},\ }\href@noop {} {\bibfield  {journal} {\bibinfo  {journal}
  {Nuovo Cim. Suppl.}\ }\textbf {\bibinfo {volume} {3}},\ \bibinfo {pages}
  {147} (\bibinfo {year} {1965})}\BibitemShut {NoStop}%
\bibitem [{\citenamefont {Chodos}\ \emph {et~al.}(1974)\citenamefont {Chodos},
  \citenamefont {Jaffe}, \citenamefont {Johnson}, \citenamefont {Thorn},\ and\
  \citenamefont {Weisskopf}}]{Chodos:1974je}%
  \BibitemOpen
  \bibfield  {author} {\bibinfo {author} {\bibfnamefont {A.}~\bibnamefont
  {Chodos}}, \bibinfo {author} {\bibfnamefont {R.~L.}\ \bibnamefont {Jaffe}},
  \bibinfo {author} {\bibfnamefont {K.}~\bibnamefont {Johnson}}, \bibinfo
  {author} {\bibfnamefont {C.~B.}\ \bibnamefont {Thorn}}, \ and\ \bibinfo
  {author} {\bibfnamefont {V.~F.}\ \bibnamefont {Weisskopf}},\ }\href {\doibase
  10.1103/PhysRevD.9.3471} {\bibfield  {journal} {\bibinfo  {journal} {Phys.
  Rev.}\ }\textbf {\bibinfo {volume} {D9}},\ \bibinfo {pages} {3471} (\bibinfo
  {year} {1974})}\BibitemShut {NoStop}%
\bibitem [{\citenamefont {Kapusta}(1981)}]{Kapusta:1981ay}%
  \BibitemOpen
  \bibfield  {author} {\bibinfo {author} {\bibfnamefont {J.~I.}\ \bibnamefont
  {Kapusta}},\ }\href {\doibase 10.1103/PhysRevD.23.2444} {\bibfield  {journal}
  {\bibinfo  {journal} {Phys. Rev.}\ }\textbf {\bibinfo {volume} {D23}},\
  \bibinfo {pages} {2444} (\bibinfo {year} {1981})}\BibitemShut {NoStop}%
\bibitem [{\citenamefont {Cabibbo}\ and\ \citenamefont
  {Parisi}(1975)}]{Cabibbo:1975ig}%
  \BibitemOpen
  \bibfield  {author} {\bibinfo {author} {\bibfnamefont {N.}~\bibnamefont
  {Cabibbo}}\ and\ \bibinfo {author} {\bibfnamefont {G.}~\bibnamefont
  {Parisi}},\ }\href {\doibase 10.1016/0370-2693(75)90158-6} {\bibfield
  {journal} {\bibinfo  {journal} {Phys. Lett.}\ }\textbf {\bibinfo {volume}
  {59B}},\ \bibinfo {pages} {67} (\bibinfo {year} {1975})}\BibitemShut
  {NoStop}%
\bibitem [{\citenamefont {Aoki}\ \emph {et~al.}(2006)\citenamefont {Aoki},
  \citenamefont {Endrodi}, \citenamefont {Fodor}, \citenamefont {Katz},\ and\
  \citenamefont {Szabo}}]{Aoki:2006we}%
  \BibitemOpen
  \bibfield  {author} {\bibinfo {author} {\bibfnamefont {Y.}~\bibnamefont
  {Aoki}}, \bibinfo {author} {\bibfnamefont {G.}~\bibnamefont {Endrodi}},
  \bibinfo {author} {\bibfnamefont {Z.}~\bibnamefont {Fodor}}, \bibinfo
  {author} {\bibfnamefont {S.~D.}\ \bibnamefont {Katz}}, \ and\ \bibinfo
  {author} {\bibfnamefont {K.~K.}\ \bibnamefont {Szabo}},\ }\href {\doibase
  10.1038/nature05120} {\bibfield  {journal} {\bibinfo  {journal} {Nature}\
  }\textbf {\bibinfo {volume} {443}},\ \bibinfo {pages} {675} (\bibinfo {year}
  {2006})},\ \Eprint {http://arxiv.org/abs/hep-lat/0611014}
  {arXiv:hep-lat/0611014 [hep-lat]} \BibitemShut {NoStop}%
\bibitem [{\citenamefont {Borsanyi}\ \emph {et~al.}(2010)\citenamefont
  {Borsanyi}, \citenamefont {Fodor}, \citenamefont {Hoelbling}, \citenamefont
  {Katz}, \citenamefont {Krieg}, \citenamefont {Ratti},\ and\ \citenamefont
  {Szabo}}]{Borsanyi:2010bp}%
  \BibitemOpen
  \bibfield  {author} {\bibinfo {author} {\bibfnamefont {S.}~\bibnamefont
  {Borsanyi}}, \bibinfo {author} {\bibfnamefont {Z.}~\bibnamefont {Fodor}},
  \bibinfo {author} {\bibfnamefont {C.}~\bibnamefont {Hoelbling}}, \bibinfo
  {author} {\bibfnamefont {S.~D.}\ \bibnamefont {Katz}}, \bibinfo {author}
  {\bibfnamefont {S.}~\bibnamefont {Krieg}}, \bibinfo {author} {\bibfnamefont
  {C.}~\bibnamefont {Ratti}}, \ and\ \bibinfo {author} {\bibfnamefont {K.~K.}\
  \bibnamefont {Szabo}} (\bibinfo {collaboration} {Wuppertal-Budapest}),\
  }\href {\doibase 10.1007/JHEP09(2010)073} {\bibfield  {journal} {\bibinfo
  {journal} {JHEP}\ }\textbf {\bibinfo {volume} {09}},\ \bibinfo {pages} {073}
  (\bibinfo {year} {2010})},\ \Eprint {http://arxiv.org/abs/1005.3508}
  {arXiv:1005.3508 [hep-lat]} \BibitemShut {NoStop}%
\bibitem [{\citenamefont {Bazavov}\ \emph
  {et~al.}(2012{\natexlab{a}})\citenamefont {Bazavov} \emph
  {et~al.}}]{Bazavov:2011nk}%
  \BibitemOpen
  \bibfield  {author} {\bibinfo {author} {\bibfnamefont {A.}~\bibnamefont
  {Bazavov}} \emph {et~al.},\ }\href {\doibase 10.1103/PhysRevD.85.054503}
  {\bibfield  {journal} {\bibinfo  {journal} {Phys. Rev.}\ }\textbf {\bibinfo
  {volume} {D85}},\ \bibinfo {pages} {054503} (\bibinfo {year}
  {2012}{\natexlab{a}})},\ \Eprint {http://arxiv.org/abs/1111.1710}
  {arXiv:1111.1710 [hep-lat]} \BibitemShut {NoStop}%
\bibitem [{\citenamefont {Stoecker}\ \emph {et~al.}(1981)\citenamefont
  {Stoecker}, \citenamefont {Ogloblin},\ and\ \citenamefont
  {Greiner}}]{Stoecker:1981za}%
  \BibitemOpen
  \bibfield  {author} {\bibinfo {author} {\bibfnamefont {H.}~\bibnamefont
  {Stoecker}}, \bibinfo {author} {\bibfnamefont {A.~A.}\ \bibnamefont
  {Ogloblin}}, \ and\ \bibinfo {author} {\bibfnamefont {W.}~\bibnamefont
  {Greiner}},\ }\href {\doibase 10.1007/BF01421522} {\bibfield  {journal}
  {\bibinfo  {journal} {Z. Phys.}\ }\textbf {\bibinfo {volume} {A303}},\
  \bibinfo {pages} {259} (\bibinfo {year} {1981})}\BibitemShut {NoStop}%
\bibitem [{\citenamefont {Noronha-Hostler}\ \emph {et~al.}(2008)\citenamefont
  {Noronha-Hostler}, \citenamefont {Greiner},\ and\ \citenamefont
  {Shovkovy}}]{NoronhaHostler:2007jf}%
  \BibitemOpen
  \bibfield  {author} {\bibinfo {author} {\bibfnamefont {J.}~\bibnamefont
  {Noronha-Hostler}}, \bibinfo {author} {\bibfnamefont {C.}~\bibnamefont
  {Greiner}}, \ and\ \bibinfo {author} {\bibfnamefont {I.~A.}\ \bibnamefont
  {Shovkovy}},\ }\href {\doibase 10.1103/PhysRevLett.100.252301} {\bibfield
  {journal} {\bibinfo  {journal} {Phys. Rev. Lett.}\ }\textbf {\bibinfo
  {volume} {100}},\ \bibinfo {pages} {252301} (\bibinfo {year} {2008})},\
  \Eprint {http://arxiv.org/abs/0711.0930} {arXiv:0711.0930 [nucl-th]}
  \BibitemShut {NoStop}%
\bibitem [{\citenamefont {Noronha-Hostler}\ \emph {et~al.}(2010)\citenamefont
  {Noronha-Hostler}, \citenamefont {Beitel}, \citenamefont {Greiner},\ and\
  \citenamefont {Shovkovy}}]{NoronhaHostler:2009cf}%
  \BibitemOpen
  \bibfield  {author} {\bibinfo {author} {\bibfnamefont {J.}~\bibnamefont
  {Noronha-Hostler}}, \bibinfo {author} {\bibfnamefont {M.}~\bibnamefont
  {Beitel}}, \bibinfo {author} {\bibfnamefont {C.}~\bibnamefont {Greiner}}, \
  and\ \bibinfo {author} {\bibfnamefont {I.}~\bibnamefont {Shovkovy}},\ }\href
  {\doibase 10.1103/PhysRevC.81.054909} {\bibfield  {journal} {\bibinfo
  {journal} {Phys. Rev.}\ }\textbf {\bibinfo {volume} {C81}},\ \bibinfo {pages}
  {054909} (\bibinfo {year} {2010})},\ \Eprint {http://arxiv.org/abs/0909.2908}
  {arXiv:0909.2908 [nucl-th]} \BibitemShut {NoStop}%
\bibitem [{\citenamefont {Noronha-Hostler}\ \emph {et~al.}(2009)\citenamefont
  {Noronha-Hostler}, \citenamefont {Noronha},\ and\ \citenamefont
  {Greiner}}]{NoronhaHostler:2008ju}%
  \BibitemOpen
  \bibfield  {author} {\bibinfo {author} {\bibfnamefont {J.}~\bibnamefont
  {Noronha-Hostler}}, \bibinfo {author} {\bibfnamefont {J.}~\bibnamefont
  {Noronha}}, \ and\ \bibinfo {author} {\bibfnamefont {C.}~\bibnamefont
  {Greiner}},\ }\href {\doibase 10.1103/PhysRevLett.103.172302} {\bibfield
  {journal} {\bibinfo  {journal} {Phys. Rev. Lett.}\ }\textbf {\bibinfo
  {volume} {103}},\ \bibinfo {pages} {172302} (\bibinfo {year} {2009})},\
  \Eprint {http://arxiv.org/abs/0811.1571} {arXiv:0811.1571 [nucl-th]}
  \BibitemShut {NoStop}%
\bibitem [{\citenamefont {Noronha-Hostler}\ \emph {et~al.}(2012)\citenamefont
  {Noronha-Hostler}, \citenamefont {Noronha},\ and\ \citenamefont
  {Greiner}}]{NoronhaHostler:2012ug}%
  \BibitemOpen
  \bibfield  {author} {\bibinfo {author} {\bibfnamefont {J.}~\bibnamefont
  {Noronha-Hostler}}, \bibinfo {author} {\bibfnamefont {J.}~\bibnamefont
  {Noronha}}, \ and\ \bibinfo {author} {\bibfnamefont {C.}~\bibnamefont
  {Greiner}},\ }\href {\doibase 10.1103/PhysRevC.86.024913} {\bibfield
  {journal} {\bibinfo  {journal} {Phys. Rev.}\ }\textbf {\bibinfo {volume}
  {C86}},\ \bibinfo {pages} {024913} (\bibinfo {year} {2012})},\ \Eprint
  {http://arxiv.org/abs/1206.5138} {arXiv:1206.5138 [nucl-th]} \BibitemShut
  {NoStop}%
\bibitem [{\citenamefont {Beitel}\ \emph {et~al.}(2014)\citenamefont {Beitel},
  \citenamefont {Gallmeister},\ and\ \citenamefont {Greiner}}]{Beitel:2014kza}%
  \BibitemOpen
  \bibfield  {author} {\bibinfo {author} {\bibfnamefont {M.}~\bibnamefont
  {Beitel}}, \bibinfo {author} {\bibfnamefont {K.}~\bibnamefont {Gallmeister}},
  \ and\ \bibinfo {author} {\bibfnamefont {C.}~\bibnamefont {Greiner}},\ }\href
  {\doibase 10.1103/PhysRevC.90.045203} {\bibfield  {journal} {\bibinfo
  {journal} {Phys. Rev.}\ }\textbf {\bibinfo {volume} {C90}},\ \bibinfo {pages}
  {045203} (\bibinfo {year} {2014})},\ \Eprint {http://arxiv.org/abs/1402.1458}
  {arXiv:1402.1458 [hep-ph]} \BibitemShut {NoStop}%
\bibitem [{\citenamefont {Beitel}\ \emph {et~al.}(2016)\citenamefont {Beitel},
  \citenamefont {Greiner},\ and\ \citenamefont {Stoecker}}]{Beitel:2016ghw}%
  \BibitemOpen
  \bibfield  {author} {\bibinfo {author} {\bibfnamefont {M.}~\bibnamefont
  {Beitel}}, \bibinfo {author} {\bibfnamefont {C.}~\bibnamefont {Greiner}}, \
  and\ \bibinfo {author} {\bibfnamefont {H.}~\bibnamefont {Stoecker}},\ }\href
  {\doibase 10.1103/PhysRevC.94.021902} {\bibfield  {journal} {\bibinfo
  {journal} {Phys. Rev.}\ }\textbf {\bibinfo {volume} {C94}},\ \bibinfo {pages}
  {021902} (\bibinfo {year} {2016})},\ \Eprint
  {http://arxiv.org/abs/1601.02474} {arXiv:1601.02474 [hep-ph]} \BibitemShut
  {NoStop}%
\bibitem [{\citenamefont {Gallmeister}\ \emph {et~al.}(2018)\citenamefont
  {Gallmeister}, \citenamefont {Beitel},\ and\ \citenamefont
  {Greiner}}]{Gallmeister:2017ths}%
  \BibitemOpen
  \bibfield  {author} {\bibinfo {author} {\bibfnamefont {K.}~\bibnamefont
  {Gallmeister}}, \bibinfo {author} {\bibfnamefont {M.}~\bibnamefont {Beitel}},
  \ and\ \bibinfo {author} {\bibfnamefont {C.}~\bibnamefont {Greiner}},\ }\href
  {\doibase 10.1103/PhysRevC.98.024915} {\bibfield  {journal} {\bibinfo
  {journal} {Phys. Rev.}\ }\textbf {\bibinfo {volume} {C98}},\ \bibinfo {pages}
  {024915} (\bibinfo {year} {2018})},\ \Eprint
  {http://arxiv.org/abs/1712.04018} {arXiv:1712.04018 [hep-ph]} \BibitemShut
  {NoStop}%
\bibitem [{\citenamefont {Baacke}(1977)}]{Baacke:1976jv}%
  \BibitemOpen
  \bibfield  {author} {\bibinfo {author} {\bibfnamefont {J.}~\bibnamefont
  {Baacke}},\ }\href@noop {} {\bibfield  {journal} {\bibinfo  {journal} {Acta
  Phys. Polon.}\ }\textbf {\bibinfo {volume} {B8}},\ \bibinfo {pages} {625}
  (\bibinfo {year} {1977})}\BibitemShut {NoStop}%
\bibitem [{\citenamefont {Hagedorn}\ and\ \citenamefont
  {Rafelski}(1980)}]{Hagedorn:1980kb}%
  \BibitemOpen
  \bibfield  {author} {\bibinfo {author} {\bibfnamefont {R.}~\bibnamefont
  {Hagedorn}}\ and\ \bibinfo {author} {\bibfnamefont {J.}~\bibnamefont
  {Rafelski}},\ }\href {\doibase 10.1016/0370-2693(80)90566-3} {\bibfield
  {journal} {\bibinfo  {journal} {Phys. Lett.}\ }\textbf {\bibinfo {volume}
  {97B}},\ \bibinfo {pages} {136} (\bibinfo {year} {1980})}\BibitemShut
  {NoStop}%
\bibitem [{\citenamefont {Dixit}\ \emph {et~al.}(1981)\citenamefont {Dixit},
  \citenamefont {Karsch},\ and\ \citenamefont {Satz}}]{Dixit:1980zj}%
  \BibitemOpen
  \bibfield  {author} {\bibinfo {author} {\bibfnamefont {V.~V.}\ \bibnamefont
  {Dixit}}, \bibinfo {author} {\bibfnamefont {F.}~\bibnamefont {Karsch}}, \
  and\ \bibinfo {author} {\bibfnamefont {H.}~\bibnamefont {Satz}},\ }\href
  {\doibase 10.1016/0370-2693(81)90165-9} {\bibfield  {journal} {\bibinfo
  {journal} {Phys. Lett.}\ }\textbf {\bibinfo {volume} {101B}},\ \bibinfo
  {pages} {412} (\bibinfo {year} {1981})}\BibitemShut {NoStop}%
\bibitem [{\citenamefont {Gorenstein}\ \emph {et~al.}(1981)\citenamefont
  {Gorenstein}, \citenamefont {Petrov},\ and\ \citenamefont
  {Zinovev}}]{Gorenstein:1981fa}%
  \BibitemOpen
  \bibfield  {author} {\bibinfo {author} {\bibfnamefont {M.~I.}\ \bibnamefont
  {Gorenstein}}, \bibinfo {author} {\bibfnamefont {V.~K.}\ \bibnamefont
  {Petrov}}, \ and\ \bibinfo {author} {\bibfnamefont {G.~M.}\ \bibnamefont
  {Zinovev}},\ }\href {\doibase 10.1016/0370-2693(81)90546-3} {\bibfield
  {journal} {\bibinfo  {journal} {Phys. Lett.}\ }\textbf {\bibinfo {volume}
  {106B}},\ \bibinfo {pages} {327} (\bibinfo {year} {1981})}\BibitemShut
  {NoStop}%
\bibitem [{\citenamefont {Kapusta}\ and\ \citenamefont
  {Olive}(1983)}]{Kapusta:1982qd}%
  \BibitemOpen
  \bibfield  {author} {\bibinfo {author} {\bibfnamefont {J.~I.}\ \bibnamefont
  {Kapusta}}\ and\ \bibinfo {author} {\bibfnamefont {K.~A.}\ \bibnamefont
  {Olive}},\ }\href {\doibase 10.1016/0375-9474(83)90241-5} {\bibfield
  {journal} {\bibinfo  {journal} {Nucl. Phys.}\ }\textbf {\bibinfo {volume}
  {A408}},\ \bibinfo {pages} {478} (\bibinfo {year} {1983})}\BibitemShut
  {NoStop}%
\bibitem [{\citenamefont {Gorenstein}\ \emph {et~al.}(1998)\citenamefont
  {Gorenstein}, \citenamefont {Greiner},\ and\ \citenamefont
  {Yang}}]{Gorenstein:1998am}%
  \BibitemOpen
  \bibfield  {author} {\bibinfo {author} {\bibfnamefont {M.~I.}\ \bibnamefont
  {Gorenstein}}, \bibinfo {author} {\bibfnamefont {W.}~\bibnamefont {Greiner}},
  \ and\ \bibinfo {author} {\bibfnamefont {S.~N.}\ \bibnamefont {Yang}},\
  }\href {\doibase 10.1088/0954-3899/24/4/005} {\bibfield  {journal} {\bibinfo
  {journal} {J. Phys.}\ }\textbf {\bibinfo {volume} {G24}},\ \bibinfo {pages}
  {725} (\bibinfo {year} {1998})}\BibitemShut {NoStop}%
\bibitem [{\citenamefont {Gorenstein}\ \emph {et~al.}(2005)\citenamefont
  {Gorenstein}, \citenamefont {Gazdzicki},\ and\ \citenamefont
  {Greiner}}]{Gorenstein:2005rc}%
  \BibitemOpen
  \bibfield  {author} {\bibinfo {author} {\bibfnamefont {M.~I.}\ \bibnamefont
  {Gorenstein}}, \bibinfo {author} {\bibfnamefont {M.}~\bibnamefont
  {Gazdzicki}}, \ and\ \bibinfo {author} {\bibfnamefont {W.}~\bibnamefont
  {Greiner}},\ }\href {\doibase 10.1103/PhysRevC.72.024909} {\bibfield
  {journal} {\bibinfo  {journal} {Phys. Rev.}\ }\textbf {\bibinfo {volume}
  {C72}},\ \bibinfo {pages} {024909} (\bibinfo {year} {2005})},\ \Eprint
  {http://arxiv.org/abs/nucl-th/0505050} {arXiv:nucl-th/0505050 [nucl-th]}
  \BibitemShut {NoStop}%
\bibitem [{\citenamefont {Zakout}\ \emph {et~al.}(2007)\citenamefont {Zakout},
  \citenamefont {Greiner},\ and\ \citenamefont
  {Schaffner-Bielich}}]{Zakout:2006zj}%
  \BibitemOpen
  \bibfield  {author} {\bibinfo {author} {\bibfnamefont {I.}~\bibnamefont
  {Zakout}}, \bibinfo {author} {\bibfnamefont {C.}~\bibnamefont {Greiner}}, \
  and\ \bibinfo {author} {\bibfnamefont {J.}~\bibnamefont
  {Schaffner-Bielich}},\ }\href {\doibase 10.1016/j.nuclphysa.2006.10.064}
  {\bibfield  {journal} {\bibinfo  {journal} {Nucl. Phys.}\ }\textbf {\bibinfo
  {volume} {A781}},\ \bibinfo {pages} {150} (\bibinfo {year} {2007})},\ \Eprint
  {http://arxiv.org/abs/nucl-th/0605052} {arXiv:nucl-th/0605052 [nucl-th]}
  \BibitemShut {NoStop}%
\bibitem [{\citenamefont {Zakout}\ and\ \citenamefont
  {Greiner}(2008)}]{Zakout:2007nb}%
  \BibitemOpen
  \bibfield  {author} {\bibinfo {author} {\bibfnamefont {I.}~\bibnamefont
  {Zakout}}\ and\ \bibinfo {author} {\bibfnamefont {C.}~\bibnamefont
  {Greiner}},\ }\href {\doibase 10.1103/PhysRevC.78.034916} {\bibfield
  {journal} {\bibinfo  {journal} {Phys. Rev.}\ }\textbf {\bibinfo {volume}
  {C78}},\ \bibinfo {pages} {034916} (\bibinfo {year} {2008})},\ \Eprint
  {http://arxiv.org/abs/0709.0144} {arXiv:0709.0144 [nucl-th]} \BibitemShut
  {NoStop}%
\bibitem [{\citenamefont {Bugaev}(2007)}]{Bugaev:2007ww}%
  \BibitemOpen
  \bibfield  {author} {\bibinfo {author} {\bibfnamefont {K.~A.}\ \bibnamefont
  {Bugaev}},\ }\href {\doibase 10.1103/PhysRevC.76.014903} {\bibfield
  {journal} {\bibinfo  {journal} {Phys. Rev.}\ }\textbf {\bibinfo {volume}
  {C76}},\ \bibinfo {pages} {014903} (\bibinfo {year} {2007})},\ \Eprint
  {http://arxiv.org/abs/hep-ph/0703222} {arXiv:hep-ph/0703222 [hep-ph]}
  \BibitemShut {NoStop}%
\bibitem [{\citenamefont {Ferroni}\ and\ \citenamefont
  {Koch}(2009)}]{Ferroni:2008ej}%
  \BibitemOpen
  \bibfield  {author} {\bibinfo {author} {\bibfnamefont {L.}~\bibnamefont
  {Ferroni}}\ and\ \bibinfo {author} {\bibfnamefont {V.}~\bibnamefont {Koch}},\
  }\href {\doibase 10.1103/PhysRevC.79.034905} {\bibfield  {journal} {\bibinfo
  {journal} {Phys. Rev.}\ }\textbf {\bibinfo {volume} {C79}},\ \bibinfo {pages}
  {034905} (\bibinfo {year} {2009})},\ \Eprint {http://arxiv.org/abs/0812.1044}
  {arXiv:0812.1044 [nucl-th]} \BibitemShut {NoStop}%
\bibitem [{\citenamefont {Begun}\ \emph {et~al.}(2009)\citenamefont {Begun},
  \citenamefont {Gorenstein},\ and\ \citenamefont {Greiner}}]{Begun:2009an}%
  \BibitemOpen
  \bibfield  {author} {\bibinfo {author} {\bibfnamefont {V.~V.}\ \bibnamefont
  {Begun}}, \bibinfo {author} {\bibfnamefont {M.~I.}\ \bibnamefont
  {Gorenstein}}, \ and\ \bibinfo {author} {\bibfnamefont {W.}~\bibnamefont
  {Greiner}},\ }\href {\doibase 10.1088/0954-3899/36/9/095005} {\bibfield
  {journal} {\bibinfo  {journal} {J. Phys.}\ }\textbf {\bibinfo {volume}
  {G36}},\ \bibinfo {pages} {095005} (\bibinfo {year} {2009})},\ \Eprint
  {http://arxiv.org/abs/0906.3205} {arXiv:0906.3205 [nucl-th]} \BibitemShut
  {NoStop}%
\bibitem [{\citenamefont {Jeon}\ and\ \citenamefont
  {Koch}(2000)}]{Jeon:2000wg}%
  \BibitemOpen
  \bibfield  {author} {\bibinfo {author} {\bibfnamefont {S.}~\bibnamefont
  {Jeon}}\ and\ \bibinfo {author} {\bibfnamefont {V.}~\bibnamefont {Koch}},\
  }\href {\doibase 10.1103/PhysRevLett.85.2076} {\bibfield  {journal} {\bibinfo
   {journal} {Phys. Rev. Lett.}\ }\textbf {\bibinfo {volume} {85}},\ \bibinfo
  {pages} {2076} (\bibinfo {year} {2000})},\ \Eprint
  {http://arxiv.org/abs/hep-ph/0003168} {arXiv:hep-ph/0003168 [hep-ph]}
  \BibitemShut {NoStop}%
\bibitem [{\citenamefont {Asakawa}\ \emph {et~al.}(2000)\citenamefont
  {Asakawa}, \citenamefont {Heinz},\ and\ \citenamefont
  {Muller}}]{Asakawa:2000wh}%
  \BibitemOpen
  \bibfield  {author} {\bibinfo {author} {\bibfnamefont {M.}~\bibnamefont
  {Asakawa}}, \bibinfo {author} {\bibfnamefont {U.~W.}\ \bibnamefont {Heinz}},
  \ and\ \bibinfo {author} {\bibfnamefont {B.}~\bibnamefont {Muller}},\ }\href
  {\doibase 10.1103/PhysRevLett.85.2072} {\bibfield  {journal} {\bibinfo
  {journal} {Phys. Rev. Lett.}\ }\textbf {\bibinfo {volume} {85}},\ \bibinfo
  {pages} {2072} (\bibinfo {year} {2000})},\ \Eprint
  {http://arxiv.org/abs/hep-ph/0003169} {arXiv:hep-ph/0003169 [hep-ph]}
  \BibitemShut {NoStop}%
\bibitem [{\citenamefont {Albright}\ \emph {et~al.}(2014)\citenamefont
  {Albright}, \citenamefont {Kapusta},\ and\ \citenamefont
  {Young}}]{Albright:2014gva}%
  \BibitemOpen
  \bibfield  {author} {\bibinfo {author} {\bibfnamefont {M.}~\bibnamefont
  {Albright}}, \bibinfo {author} {\bibfnamefont {J.}~\bibnamefont {Kapusta}}, \
  and\ \bibinfo {author} {\bibfnamefont {C.}~\bibnamefont {Young}},\ }\href
  {\doibase 10.1103/PhysRevC.90.024915} {\bibfield  {journal} {\bibinfo
  {journal} {Phys. Rev.}\ }\textbf {\bibinfo {volume} {C90}},\ \bibinfo {pages}
  {024915} (\bibinfo {year} {2014})},\ \Eprint {http://arxiv.org/abs/1404.7540}
  {arXiv:1404.7540 [nucl-th]} \BibitemShut {NoStop}%
\bibitem [{\citenamefont {Albright}\ \emph {et~al.}(2015)\citenamefont
  {Albright}, \citenamefont {Kapusta},\ and\ \citenamefont
  {Young}}]{Albright:2015uua}%
  \BibitemOpen
  \bibfield  {author} {\bibinfo {author} {\bibfnamefont {M.}~\bibnamefont
  {Albright}}, \bibinfo {author} {\bibfnamefont {J.}~\bibnamefont {Kapusta}}, \
  and\ \bibinfo {author} {\bibfnamefont {C.}~\bibnamefont {Young}},\ }\href
  {\doibase 10.1103/PhysRevC.92.044904} {\bibfield  {journal} {\bibinfo
  {journal} {Phys. Rev.}\ }\textbf {\bibinfo {volume} {C92}},\ \bibinfo {pages}
  {044904} (\bibinfo {year} {2015})},\ \Eprint
  {http://arxiv.org/abs/1506.03408} {arXiv:1506.03408 [nucl-th]} \BibitemShut
  {NoStop}%
\bibitem [{\citenamefont {Vovchenko}\ \emph
  {et~al.}(2017{\natexlab{a}})\citenamefont {Vovchenko}, \citenamefont
  {Gorenstein},\ and\ \citenamefont {Stoecker}}]{Vovchenko:2016rkn}%
  \BibitemOpen
  \bibfield  {author} {\bibinfo {author} {\bibfnamefont {V.}~\bibnamefont
  {Vovchenko}}, \bibinfo {author} {\bibfnamefont {M.~I.}\ \bibnamefont
  {Gorenstein}}, \ and\ \bibinfo {author} {\bibfnamefont {H.}~\bibnamefont
  {Stoecker}},\ }\href {\doibase 10.1103/PhysRevLett.118.182301} {\bibfield
  {journal} {\bibinfo  {journal} {Phys. Rev. Lett.}\ }\textbf {\bibinfo
  {volume} {118}},\ \bibinfo {pages} {182301} (\bibinfo {year}
  {2017}{\natexlab{a}})},\ \Eprint {http://arxiv.org/abs/1609.03975}
  {arXiv:1609.03975 [hep-ph]} \BibitemShut {NoStop}%
\bibitem [{\citenamefont {Critelli}\ \emph {et~al.}(2017)\citenamefont
  {Critelli}, \citenamefont {Noronha}, \citenamefont {Noronha-Hostler},
  \citenamefont {Portillo}, \citenamefont {Ratti},\ and\ \citenamefont
  {Rougemont}}]{Critelli:2017oub}%
  \BibitemOpen
  \bibfield  {author} {\bibinfo {author} {\bibfnamefont {R.}~\bibnamefont
  {Critelli}}, \bibinfo {author} {\bibfnamefont {J.}~\bibnamefont {Noronha}},
  \bibinfo {author} {\bibfnamefont {J.}~\bibnamefont {Noronha-Hostler}},
  \bibinfo {author} {\bibfnamefont {I.}~\bibnamefont {Portillo}}, \bibinfo
  {author} {\bibfnamefont {C.}~\bibnamefont {Ratti}}, \ and\ \bibinfo {author}
  {\bibfnamefont {R.}~\bibnamefont {Rougemont}},\ }\href {\doibase
  10.1103/PhysRevD.96.096026} {\bibfield  {journal} {\bibinfo  {journal} {Phys.
  Rev.}\ }\textbf {\bibinfo {volume} {D96}},\ \bibinfo {pages} {096026}
  (\bibinfo {year} {2017})},\ \Eprint {http://arxiv.org/abs/1706.00455}
  {arXiv:1706.00455 [nucl-th]} \BibitemShut {NoStop}%
\bibitem [{\citenamefont {Huovinen}\ and\ \citenamefont
  {Petreczky}(2018)}]{Huovinen:2017ogf}%
  \BibitemOpen
  \bibfield  {author} {\bibinfo {author} {\bibfnamefont {P.}~\bibnamefont
  {Huovinen}}\ and\ \bibinfo {author} {\bibfnamefont {P.}~\bibnamefont
  {Petreczky}},\ }\href {\doibase 10.1016/j.physletb.2017.12.001} {\bibfield
  {journal} {\bibinfo  {journal} {Phys. Lett.}\ }\textbf {\bibinfo {volume}
  {B777}},\ \bibinfo {pages} {125} (\bibinfo {year} {2018})},\ \Eprint
  {http://arxiv.org/abs/1708.00879} {arXiv:1708.00879 [hep-ph]} \BibitemShut
  {NoStop}%
\bibitem [{\citenamefont {Vovchenko}\ \emph
  {et~al.}(2018{\natexlab{a}})\citenamefont {Vovchenko}, \citenamefont
  {Steinheimer}, \citenamefont {Philipsen},\ and\ \citenamefont
  {Stoecker}}]{Vovchenko:2017gkg}%
  \BibitemOpen
  \bibfield  {author} {\bibinfo {author} {\bibfnamefont {V.}~\bibnamefont
  {Vovchenko}}, \bibinfo {author} {\bibfnamefont {J.}~\bibnamefont
  {Steinheimer}}, \bibinfo {author} {\bibfnamefont {O.}~\bibnamefont
  {Philipsen}}, \ and\ \bibinfo {author} {\bibfnamefont {H.}~\bibnamefont
  {Stoecker}},\ }\href {\doibase 10.1103/PhysRevD.97.114030} {\bibfield
  {journal} {\bibinfo  {journal} {Phys. Rev.}\ }\textbf {\bibinfo {volume}
  {D97}},\ \bibinfo {pages} {114030} (\bibinfo {year} {2018}{\natexlab{a}})},\
  \Eprint {http://arxiv.org/abs/1711.01261} {arXiv:1711.01261 [hep-ph]}
  \BibitemShut {NoStop}%
\bibitem [{\citenamefont {Motornenko}\ \emph {et~al.}(2019)\citenamefont
  {Motornenko}, \citenamefont {Vovchenko}, \citenamefont {Steinheimer},
  \citenamefont {Schramm},\ and\ \citenamefont
  {Stoecker}}]{Motornenko:2018hjw}%
  \BibitemOpen
  \bibfield  {author} {\bibinfo {author} {\bibfnamefont {A.}~\bibnamefont
  {Motornenko}}, \bibinfo {author} {\bibfnamefont {V.}~\bibnamefont
  {Vovchenko}}, \bibinfo {author} {\bibfnamefont {J.}~\bibnamefont
  {Steinheimer}}, \bibinfo {author} {\bibfnamefont {S.}~\bibnamefont
  {Schramm}}, \ and\ \bibinfo {author} {\bibfnamefont {H.}~\bibnamefont
  {Stoecker}},\ }\bibfield  {booktitle} {\emph {\bibinfo {booktitle}
  {{Proceedings, 27th International Conference on Ultrarelativistic
  Nucleus-Nucleus Collisions (Quark Matter 2018): Venice, Italy, May 14-19,
  2018}}},\ }\href {\doibase 10.1016/j.nuclphysa.2018.11.028} {\bibfield
  {journal} {\bibinfo  {journal} {Nucl. Phys.}\ }\textbf {\bibinfo {volume}
  {A982}},\ \bibinfo {pages} {891} (\bibinfo {year} {2019})},\ \Eprint
  {http://arxiv.org/abs/1809.02000} {arXiv:1809.02000 [hep-ph]} \BibitemShut
  {NoStop}%
\bibitem [{\citenamefont {Rischke}\ \emph {et~al.}(1991)\citenamefont
  {Rischke}, \citenamefont {Gorenstein}, \citenamefont {Stoecker},\ and\
  \citenamefont {Greiner}}]{Rischke:1991ke}%
  \BibitemOpen
  \bibfield  {author} {\bibinfo {author} {\bibfnamefont {D.~H.}\ \bibnamefont
  {Rischke}}, \bibinfo {author} {\bibfnamefont {M.~I.}\ \bibnamefont
  {Gorenstein}}, \bibinfo {author} {\bibfnamefont {H.}~\bibnamefont
  {Stoecker}}, \ and\ \bibinfo {author} {\bibfnamefont {W.}~\bibnamefont
  {Greiner}},\ }\href {\doibase 10.1007/BF01548574} {\bibfield  {journal}
  {\bibinfo  {journal} {Z. Phys.}\ }\textbf {\bibinfo {volume} {C51}},\
  \bibinfo {pages} {485} (\bibinfo {year} {1991})}\BibitemShut {NoStop}%
\bibitem [{\citenamefont {Vovchenko}\ \emph
  {et~al.}(2018{\natexlab{b}})\citenamefont {Vovchenko}, \citenamefont
  {Gorenstein},\ and\ \citenamefont {Stoecker}}]{Vovchenko:2018fmh}%
  \BibitemOpen
  \bibfield  {author} {\bibinfo {author} {\bibfnamefont {V.}~\bibnamefont
  {Vovchenko}}, \bibinfo {author} {\bibfnamefont {M.~I.}\ \bibnamefont
  {Gorenstein}}, \ and\ \bibinfo {author} {\bibfnamefont {H.}~\bibnamefont
  {Stoecker}},\ }\href {\doibase 10.1103/PhysRevC.98.034906} {\bibfield
  {journal} {\bibinfo  {journal} {Phys. Rev.}\ }\textbf {\bibinfo {volume}
  {C98}},\ \bibinfo {pages} {034906} (\bibinfo {year} {2018}{\natexlab{b}})},\
  \Eprint {http://arxiv.org/abs/1807.02079} {arXiv:1807.02079 [nucl-th]}
  \BibitemShut {NoStop}%
\bibitem [{\citenamefont {Gorenstein}\ \emph {et~al.}(1982)\citenamefont
  {Gorenstein}, \citenamefont {Petrov}, \citenamefont {Shelest},\ and\
  \citenamefont {Zinovev}}]{Gorenstein:1982if}%
  \BibitemOpen
  \bibfield  {author} {\bibinfo {author} {\bibfnamefont {M.~I.}\ \bibnamefont
  {Gorenstein}}, \bibinfo {author} {\bibfnamefont {V.~K.}\ \bibnamefont
  {Petrov}}, \bibinfo {author} {\bibfnamefont {V.~P.}\ \bibnamefont {Shelest}},
  \ and\ \bibinfo {author} {\bibfnamefont {G.~M.}\ \bibnamefont {Zinovev}},\
  }\href {\doibase 10.1007/BF01038078} {\bibfield  {journal} {\bibinfo
  {journal} {Theor. Math. Phys.}\ }\textbf {\bibinfo {volume} {52}},\ \bibinfo
  {pages} {843} (\bibinfo {year} {1982})},\ \bibinfo {note} {[Teor. Mat.
  Fiz.52,346(1982)]}\BibitemShut {NoStop}%
\bibitem [{\citenamefont {Gorenstein}\ \emph {et~al.}(1984)\citenamefont
  {Gorenstein}, \citenamefont {Lipskikh},\ and\ \citenamefont
  {Zinovev}}]{Gorenstein:1983rm}%
  \BibitemOpen
  \bibfield  {author} {\bibinfo {author} {\bibfnamefont {M.~I.}\ \bibnamefont
  {Gorenstein}}, \bibinfo {author} {\bibfnamefont {S.~I.}\ \bibnamefont
  {Lipskikh}}, \ and\ \bibinfo {author} {\bibfnamefont {G.~M.}\ \bibnamefont
  {Zinovev}},\ }\href {\doibase 10.1007/BF01572171} {\bibfield  {journal}
  {\bibinfo  {journal} {Z. Phys.}\ }\textbf {\bibinfo {volume} {C22}},\
  \bibinfo {pages} {189} (\bibinfo {year} {1984})}\BibitemShut {NoStop}%
\bibitem [{\citenamefont {Gorenstein}\ \emph
  {et~al.}(1983{\natexlab{a}})\citenamefont {Gorenstein}, \citenamefont
  {Lipskikh}, \citenamefont {Petrov},\ and\ \citenamefont
  {Zinovev}}]{Gorenstein:1982ua}%
  \BibitemOpen
  \bibfield  {author} {\bibinfo {author} {\bibfnamefont {M.~I.}\ \bibnamefont
  {Gorenstein}}, \bibinfo {author} {\bibfnamefont {S.~I.}\ \bibnamefont
  {Lipskikh}}, \bibinfo {author} {\bibfnamefont {V.~K.}\ \bibnamefont
  {Petrov}}, \ and\ \bibinfo {author} {\bibfnamefont {G.~M.}\ \bibnamefont
  {Zinovev}},\ }\href {\doibase 10.1016/0370-2693(83)90988-7} {\bibfield
  {journal} {\bibinfo  {journal} {Phys. Lett.}\ }\textbf {\bibinfo {volume}
  {123B}},\ \bibinfo {pages} {437} (\bibinfo {year}
  {1983}{\natexlab{a}})}\BibitemShut {NoStop}%
\bibitem [{\citenamefont {Gorenstein}\ \emph
  {et~al.}(1983{\natexlab{b}})\citenamefont {Gorenstein}, \citenamefont
  {Mogilevsky}, \citenamefont {Petrov},\ and\ \citenamefont
  {Zinovev}}]{Gorenstein:1982ib}%
  \BibitemOpen
  \bibfield  {author} {\bibinfo {author} {\bibfnamefont {M.~I.}\ \bibnamefont
  {Gorenstein}}, \bibinfo {author} {\bibfnamefont {O.~A.}\ \bibnamefont
  {Mogilevsky}}, \bibinfo {author} {\bibfnamefont {V.~K.}\ \bibnamefont
  {Petrov}}, \ and\ \bibinfo {author} {\bibfnamefont {G.~M.}\ \bibnamefont
  {Zinovev}},\ }\href {\doibase 10.1007/BF01571699} {\bibfield  {journal}
  {\bibinfo  {journal} {Z. Phys.}\ }\textbf {\bibinfo {volume} {C18}},\
  \bibinfo {pages} {13} (\bibinfo {year} {1983}{\natexlab{b}})}\BibitemShut
  {NoStop}%
\bibitem [{\citenamefont {Yen}\ \emph {et~al.}(1997)\citenamefont {Yen},
  \citenamefont {Gorenstein}, \citenamefont {Greiner},\ and\ \citenamefont
  {Yang}}]{Yen:1997rv}%
  \BibitemOpen
  \bibfield  {author} {\bibinfo {author} {\bibfnamefont {G.~D.}\ \bibnamefont
  {Yen}}, \bibinfo {author} {\bibfnamefont {M.~I.}\ \bibnamefont {Gorenstein}},
  \bibinfo {author} {\bibfnamefont {W.}~\bibnamefont {Greiner}}, \ and\
  \bibinfo {author} {\bibfnamefont {S.-N.}\ \bibnamefont {Yang}},\ }\href
  {\doibase 10.1103/PhysRevC.56.2210} {\bibfield  {journal} {\bibinfo
  {journal} {Phys. Rev.}\ }\textbf {\bibinfo {volume} {C56}},\ \bibinfo {pages}
  {2210} (\bibinfo {year} {1997})},\ \Eprint
  {http://arxiv.org/abs/nucl-th/9711062} {arXiv:nucl-th/9711062 [nucl-th]}
  \BibitemShut {NoStop}%
\bibitem [{\citenamefont {Borsanyi}\ \emph {et~al.}(2014)\citenamefont
  {Borsanyi}, \citenamefont {Fodor}, \citenamefont {Hoelbling}, \citenamefont
  {Katz}, \citenamefont {Krieg},\ and\ \citenamefont
  {Szabo}}]{Borsanyi:2013bia}%
  \BibitemOpen
  \bibfield  {author} {\bibinfo {author} {\bibfnamefont {S.}~\bibnamefont
  {Borsanyi}}, \bibinfo {author} {\bibfnamefont {Z.}~\bibnamefont {Fodor}},
  \bibinfo {author} {\bibfnamefont {C.}~\bibnamefont {Hoelbling}}, \bibinfo
  {author} {\bibfnamefont {S.~D.}\ \bibnamefont {Katz}}, \bibinfo {author}
  {\bibfnamefont {S.}~\bibnamefont {Krieg}}, \ and\ \bibinfo {author}
  {\bibfnamefont {K.~K.}\ \bibnamefont {Szabo}},\ }\href {\doibase
  10.1016/j.physletb.2014.01.007} {\bibfield  {journal} {\bibinfo  {journal}
  {Phys. Lett.}\ }\textbf {\bibinfo {volume} {B730}},\ \bibinfo {pages} {99}
  (\bibinfo {year} {2014})},\ \Eprint {http://arxiv.org/abs/1309.5258}
  {arXiv:1309.5258 [hep-lat]} \BibitemShut {NoStop}%
\bibitem [{\citenamefont {Bazavov}\ \emph {et~al.}(2014)\citenamefont {Bazavov}
  \emph {et~al.}}]{Bazavov:2014pvz}%
  \BibitemOpen
  \bibfield  {author} {\bibinfo {author} {\bibfnamefont {A.}~\bibnamefont
  {Bazavov}} \emph {et~al.} (\bibinfo {collaboration} {HotQCD}),\ }\href
  {\doibase 10.1103/PhysRevD.90.094503} {\bibfield  {journal} {\bibinfo
  {journal} {Phys. Rev.}\ }\textbf {\bibinfo {volume} {D90}},\ \bibinfo {pages}
  {094503} (\bibinfo {year} {2014})},\ \Eprint {http://arxiv.org/abs/1407.6387}
  {arXiv:1407.6387 [hep-lat]} \BibitemShut {NoStop}%
\bibitem [{\citenamefont {Borsanyi}\ \emph {et~al.}(2012)\citenamefont
  {Borsanyi}, \citenamefont {Fodor}, \citenamefont {Katz}, \citenamefont
  {Krieg}, \citenamefont {Ratti},\ and\ \citenamefont
  {Szabo}}]{Borsanyi:2011sw}%
  \BibitemOpen
  \bibfield  {author} {\bibinfo {author} {\bibfnamefont {S.}~\bibnamefont
  {Borsanyi}}, \bibinfo {author} {\bibfnamefont {Z.}~\bibnamefont {Fodor}},
  \bibinfo {author} {\bibfnamefont {S.~D.}\ \bibnamefont {Katz}}, \bibinfo
  {author} {\bibfnamefont {S.}~\bibnamefont {Krieg}}, \bibinfo {author}
  {\bibfnamefont {C.}~\bibnamefont {Ratti}}, \ and\ \bibinfo {author}
  {\bibfnamefont {K.}~\bibnamefont {Szabo}},\ }\href {\doibase
  10.1007/JHEP01(2012)138} {\bibfield  {journal} {\bibinfo  {journal} {JHEP}\
  }\textbf {\bibinfo {volume} {01}},\ \bibinfo {pages} {138} (\bibinfo {year}
  {2012})},\ \Eprint {http://arxiv.org/abs/1112.4416} {arXiv:1112.4416
  [hep-lat]} \BibitemShut {NoStop}%
\bibitem [{\citenamefont {Bazavov}\ \emph
  {et~al.}(2012{\natexlab{b}})\citenamefont {Bazavov} \emph
  {et~al.}}]{Bazavov:2012jq}%
  \BibitemOpen
  \bibfield  {author} {\bibinfo {author} {\bibfnamefont {A.}~\bibnamefont
  {Bazavov}} \emph {et~al.} (\bibinfo {collaboration} {HotQCD}),\ }\href
  {\doibase 10.1103/PhysRevD.86.034509} {\bibfield  {journal} {\bibinfo
  {journal} {Phys. Rev.}\ }\textbf {\bibinfo {volume} {D86}},\ \bibinfo {pages}
  {034509} (\bibinfo {year} {2012}{\natexlab{b}})},\ \Eprint
  {http://arxiv.org/abs/1203.0784} {arXiv:1203.0784 [hep-lat]} \BibitemShut
  {NoStop}%
\bibitem [{\citenamefont {Peshier}\ \emph {et~al.}(1994)\citenamefont
  {Peshier}, \citenamefont {Kampfer}, \citenamefont {Pavlenko},\ and\
  \citenamefont {Soff}}]{Peshier:1994zf}%
  \BibitemOpen
  \bibfield  {author} {\bibinfo {author} {\bibfnamefont {A.}~\bibnamefont
  {Peshier}}, \bibinfo {author} {\bibfnamefont {B.}~\bibnamefont {Kampfer}},
  \bibinfo {author} {\bibfnamefont {O.~P.}\ \bibnamefont {Pavlenko}}, \ and\
  \bibinfo {author} {\bibfnamefont {G.}~\bibnamefont {Soff}},\ }\href {\doibase
  10.1016/0370-2693(94)90969-5} {\bibfield  {journal} {\bibinfo  {journal}
  {Phys. Lett.}\ }\textbf {\bibinfo {volume} {B337}},\ \bibinfo {pages} {235}
  (\bibinfo {year} {1994})}\BibitemShut {NoStop}%
\bibitem [{\citenamefont {Gorenstein}\ and\ \citenamefont
  {Yang}(1995)}]{Gorenstein:1995vm}%
  \BibitemOpen
  \bibfield  {author} {\bibinfo {author} {\bibfnamefont {M.~I.}\ \bibnamefont
  {Gorenstein}}\ and\ \bibinfo {author} {\bibfnamefont {S.-N.}\ \bibnamefont
  {Yang}},\ }\href {\doibase 10.1103/PhysRevD.52.5206} {\bibfield  {journal}
  {\bibinfo  {journal} {Phys. Rev.}\ }\textbf {\bibinfo {volume} {D52}},\
  \bibinfo {pages} {5206} (\bibinfo {year} {1995})}\BibitemShut {NoStop}%
\bibitem [{\citenamefont {Peshier}\ \emph {et~al.}(1996)\citenamefont
  {Peshier}, \citenamefont {Kampfer}, \citenamefont {Pavlenko},\ and\
  \citenamefont {Soff}}]{Peshier:1995ty}%
  \BibitemOpen
  \bibfield  {author} {\bibinfo {author} {\bibfnamefont {A.}~\bibnamefont
  {Peshier}}, \bibinfo {author} {\bibfnamefont {B.}~\bibnamefont {Kampfer}},
  \bibinfo {author} {\bibfnamefont {O.~P.}\ \bibnamefont {Pavlenko}}, \ and\
  \bibinfo {author} {\bibfnamefont {G.}~\bibnamefont {Soff}},\ }\href {\doibase
  10.1103/PhysRevD.54.2399} {\bibfield  {journal} {\bibinfo  {journal} {Phys.
  Rev.}\ }\textbf {\bibinfo {volume} {D54}},\ \bibinfo {pages} {2399} (\bibinfo
  {year} {1996})}\BibitemShut {NoStop}%
\bibitem [{\citenamefont {Levai}\ and\ \citenamefont
  {Heinz}(1998)}]{Levai:1997yx}%
  \BibitemOpen
  \bibfield  {author} {\bibinfo {author} {\bibfnamefont {P.}~\bibnamefont
  {Levai}}\ and\ \bibinfo {author} {\bibfnamefont {U.~W.}\ \bibnamefont
  {Heinz}},\ }\href {\doibase 10.1103/PhysRevC.57.1879} {\bibfield  {journal}
  {\bibinfo  {journal} {Phys. Rev.}\ }\textbf {\bibinfo {volume} {C57}},\
  \bibinfo {pages} {1879} (\bibinfo {year} {1998})},\ \Eprint
  {http://arxiv.org/abs/hep-ph/9710463} {arXiv:hep-ph/9710463 [hep-ph]}
  \BibitemShut {NoStop}%
\bibitem [{\citenamefont {Peshier}\ \emph {et~al.}(2000)\citenamefont
  {Peshier}, \citenamefont {Kampfer},\ and\ \citenamefont
  {Soff}}]{Peshier:1999ww}%
  \BibitemOpen
  \bibfield  {author} {\bibinfo {author} {\bibfnamefont {A.}~\bibnamefont
  {Peshier}}, \bibinfo {author} {\bibfnamefont {B.}~\bibnamefont {Kampfer}}, \
  and\ \bibinfo {author} {\bibfnamefont {G.}~\bibnamefont {Soff}},\ }\href
  {\doibase 10.1103/PhysRevC.61.045203} {\bibfield  {journal} {\bibinfo
  {journal} {Phys. Rev.}\ }\textbf {\bibinfo {volume} {C61}},\ \bibinfo {pages}
  {045203} (\bibinfo {year} {2000})},\ \Eprint
  {http://arxiv.org/abs/hep-ph/9911474} {arXiv:hep-ph/9911474 [hep-ph]}
  \BibitemShut {NoStop}%
\bibitem [{\citenamefont {Bluhm}\ and\ \citenamefont
  {Kampfer}(2008)}]{Bluhm:2007cp}%
  \BibitemOpen
  \bibfield  {author} {\bibinfo {author} {\bibfnamefont {M.}~\bibnamefont
  {Bluhm}}\ and\ \bibinfo {author} {\bibfnamefont {B.}~\bibnamefont
  {Kampfer}},\ }\href {\doibase 10.1103/PhysRevD.77.034004} {\bibfield
  {journal} {\bibinfo  {journal} {Phys. Rev.}\ }\textbf {\bibinfo {volume}
  {D77}},\ \bibinfo {pages} {034004} (\bibinfo {year} {2008})},\ \Eprint
  {http://arxiv.org/abs/0711.0590} {arXiv:0711.0590 [hep-ph]} \BibitemShut
  {NoStop}%
\bibitem [{\citenamefont {Plumari}\ \emph {et~al.}(2011)\citenamefont
  {Plumari}, \citenamefont {Alberico}, \citenamefont {Greco},\ and\
  \citenamefont {Ratti}}]{Plumari:2011mk}%
  \BibitemOpen
  \bibfield  {author} {\bibinfo {author} {\bibfnamefont {S.}~\bibnamefont
  {Plumari}}, \bibinfo {author} {\bibfnamefont {W.~M.}\ \bibnamefont
  {Alberico}}, \bibinfo {author} {\bibfnamefont {V.}~\bibnamefont {Greco}}, \
  and\ \bibinfo {author} {\bibfnamefont {C.}~\bibnamefont {Ratti}},\ }\href
  {\doibase 10.1103/PhysRevD.84.094004} {\bibfield  {journal} {\bibinfo
  {journal} {Phys. Rev.}\ }\textbf {\bibinfo {volume} {D84}},\ \bibinfo {pages}
  {094004} (\bibinfo {year} {2011})},\ \Eprint {http://arxiv.org/abs/1103.5611}
  {arXiv:1103.5611 [hep-ph]} \BibitemShut {NoStop}%
\bibitem [{\citenamefont {Ivanov}\ \emph {et~al.}(2005)\citenamefont {Ivanov},
  \citenamefont {Khvorostukhin}, \citenamefont {Kolomeitsev}, \citenamefont
  {Skokov}, \citenamefont {Toneev},\ and\ \citenamefont
  {Voskresensky}}]{Ivanov:2005be}%
  \BibitemOpen
  \bibfield  {author} {\bibinfo {author} {\bibfnamefont {{\relax Yu}.~B.}\
  \bibnamefont {Ivanov}}, \bibinfo {author} {\bibfnamefont {A.~S.}\
  \bibnamefont {Khvorostukhin}}, \bibinfo {author} {\bibfnamefont {E.~E.}\
  \bibnamefont {Kolomeitsev}}, \bibinfo {author} {\bibfnamefont {V.~V.}\
  \bibnamefont {Skokov}}, \bibinfo {author} {\bibfnamefont {V.~D.}\
  \bibnamefont {Toneev}}, \ and\ \bibinfo {author} {\bibfnamefont {D.~N.}\
  \bibnamefont {Voskresensky}},\ }\href {\doibase 10.1103/PhysRevC.72.025804}
  {\bibfield  {journal} {\bibinfo  {journal} {Phys. Rev.}\ }\textbf {\bibinfo
  {volume} {C72}},\ \bibinfo {pages} {025804} (\bibinfo {year} {2005})},\
  \Eprint {http://arxiv.org/abs/astro-ph/0501254} {arXiv:astro-ph/0501254
  [astro-ph]} \BibitemShut {NoStop}%
\bibitem [{\citenamefont {Braaten}\ and\ \citenamefont
  {Pisarski}(1992)}]{Braaten:1991gm}%
  \BibitemOpen
  \bibfield  {author} {\bibinfo {author} {\bibfnamefont {E.}~\bibnamefont
  {Braaten}}\ and\ \bibinfo {author} {\bibfnamefont {R.~D.}\ \bibnamefont
  {Pisarski}},\ }\href {\doibase 10.1103/PhysRevD.45.R1827} {\bibfield
  {journal} {\bibinfo  {journal} {Phys. Rev.}\ }\textbf {\bibinfo {volume}
  {D45}},\ \bibinfo {pages} {R1827} (\bibinfo {year} {1992})}\BibitemShut
  {NoStop}%
\bibitem [{\citenamefont {Andersen}\ \emph {et~al.}(2000)\citenamefont
  {Andersen}, \citenamefont {Braaten},\ and\ \citenamefont
  {Strickland}}]{Andersen:1999sf}%
  \BibitemOpen
  \bibfield  {author} {\bibinfo {author} {\bibfnamefont {J.~O.}\ \bibnamefont
  {Andersen}}, \bibinfo {author} {\bibfnamefont {E.}~\bibnamefont {Braaten}}, \
  and\ \bibinfo {author} {\bibfnamefont {M.}~\bibnamefont {Strickland}},\
  }\href {\doibase 10.1103/PhysRevD.61.014017} {\bibfield  {journal} {\bibinfo
  {journal} {Phys. Rev.}\ }\textbf {\bibinfo {volume} {D61}},\ \bibinfo {pages}
  {014017} (\bibinfo {year} {2000})},\ \Eprint
  {http://arxiv.org/abs/hep-ph/9905337} {arXiv:hep-ph/9905337 [hep-ph]}
  \BibitemShut {NoStop}%
\bibitem [{\citenamefont {Karsch}\ \emph {et~al.}(2016)\citenamefont {Karsch},
  \citenamefont {Morita},\ and\ \citenamefont {Redlich}}]{Karsch:2015zna}%
  \BibitemOpen
  \bibfield  {author} {\bibinfo {author} {\bibfnamefont {F.}~\bibnamefont
  {Karsch}}, \bibinfo {author} {\bibfnamefont {K.}~\bibnamefont {Morita}}, \
  and\ \bibinfo {author} {\bibfnamefont {K.}~\bibnamefont {Redlich}},\ }\href
  {\doibase 10.1103/PhysRevC.93.034907} {\bibfield  {journal} {\bibinfo
  {journal} {Phys. Rev.}\ }\textbf {\bibinfo {volume} {C93}},\ \bibinfo {pages}
  {034907} (\bibinfo {year} {2016})},\ \Eprint
  {http://arxiv.org/abs/1508.02614} {arXiv:1508.02614 [hep-ph]} \BibitemShut
  {NoStop}%
\bibitem [{\citenamefont {Koch}\ \emph {et~al.}(2005)\citenamefont {Koch},
  \citenamefont {Majumder},\ and\ \citenamefont {Randrup}}]{Koch:2005vg}%
  \BibitemOpen
  \bibfield  {author} {\bibinfo {author} {\bibfnamefont {V.}~\bibnamefont
  {Koch}}, \bibinfo {author} {\bibfnamefont {A.}~\bibnamefont {Majumder}}, \
  and\ \bibinfo {author} {\bibfnamefont {J.}~\bibnamefont {Randrup}},\ }\href
  {\doibase 10.1103/PhysRevLett.95.182301} {\bibfield  {journal} {\bibinfo
  {journal} {Phys. Rev. Lett.}\ }\textbf {\bibinfo {volume} {95}},\ \bibinfo
  {pages} {182301} (\bibinfo {year} {2005})},\ \Eprint
  {http://arxiv.org/abs/nucl-th/0505052} {arXiv:nucl-th/0505052 [nucl-th]}
  \BibitemShut {NoStop}%
\bibitem [{\citenamefont {Friman}\ \emph {et~al.}(2011)\citenamefont {Friman},
  \citenamefont {Karsch}, \citenamefont {Redlich},\ and\ \citenamefont
  {Skokov}}]{Friman:2011pf}%
  \BibitemOpen
  \bibfield  {author} {\bibinfo {author} {\bibfnamefont {B.}~\bibnamefont
  {Friman}}, \bibinfo {author} {\bibfnamefont {F.}~\bibnamefont {Karsch}},
  \bibinfo {author} {\bibfnamefont {K.}~\bibnamefont {Redlich}}, \ and\
  \bibinfo {author} {\bibfnamefont {V.}~\bibnamefont {Skokov}},\ }\href
  {\doibase 10.1140/epjc/s10052-011-1694-2} {\bibfield  {journal} {\bibinfo
  {journal} {Eur. Phys. J.}\ }\textbf {\bibinfo {volume} {C71}},\ \bibinfo
  {pages} {1694} (\bibinfo {year} {2011})},\ \Eprint
  {http://arxiv.org/abs/1103.3511} {arXiv:1103.3511 [hep-ph]} \BibitemShut
  {NoStop}%
\bibitem [{\citenamefont {Bellwied}\ \emph {et~al.}(2013)\citenamefont
  {Bellwied}, \citenamefont {Borsanyi}, \citenamefont {Fodor}, \citenamefont
  {Katz},\ and\ \citenamefont {Ratti}}]{Bellwied:2013cta}%
  \BibitemOpen
  \bibfield  {author} {\bibinfo {author} {\bibfnamefont {R.}~\bibnamefont
  {Bellwied}}, \bibinfo {author} {\bibfnamefont {S.}~\bibnamefont {Borsanyi}},
  \bibinfo {author} {\bibfnamefont {Z.}~\bibnamefont {Fodor}}, \bibinfo
  {author} {\bibfnamefont {S.~D.}\ \bibnamefont {Katz}}, \ and\ \bibinfo
  {author} {\bibfnamefont {C.}~\bibnamefont {Ratti}},\ }\href {\doibase
  10.1103/PhysRevLett.111.202302} {\bibfield  {journal} {\bibinfo  {journal}
  {Phys. Rev. Lett.}\ }\textbf {\bibinfo {volume} {111}},\ \bibinfo {pages}
  {202302} (\bibinfo {year} {2013})},\ \Eprint {http://arxiv.org/abs/1305.6297}
  {arXiv:1305.6297 [hep-lat]} \BibitemShut {NoStop}%
\bibitem [{\citenamefont {Bazavov}\ \emph
  {et~al.}(2017{\natexlab{a}})\citenamefont {Bazavov} \emph
  {et~al.}}]{Bazavov:2017dus}%
  \BibitemOpen
  \bibfield  {author} {\bibinfo {author} {\bibfnamefont {A.}~\bibnamefont
  {Bazavov}} \emph {et~al.},\ }\href {\doibase 10.1103/PhysRevD.95.054504}
  {\bibfield  {journal} {\bibinfo  {journal} {Phys. Rev.}\ }\textbf {\bibinfo
  {volume} {D95}},\ \bibinfo {pages} {054504} (\bibinfo {year}
  {2017}{\natexlab{a}})},\ \Eprint {http://arxiv.org/abs/1701.04325}
  {arXiv:1701.04325 [hep-lat]} \BibitemShut {NoStop}%
\bibitem [{\citenamefont {Bazavov}\ \emph
  {et~al.}(2017{\natexlab{b}})\citenamefont {Bazavov} \emph
  {et~al.}}]{Bazavov:2017tot}%
  \BibitemOpen
  \bibfield  {author} {\bibinfo {author} {\bibfnamefont {A.}~\bibnamefont
  {Bazavov}} \emph {et~al.} (\bibinfo {collaboration} {HotQCD}),\ }\href
  {\doibase 10.1103/PhysRevD.96.074510} {\bibfield  {journal} {\bibinfo
  {journal} {Phys. Rev.}\ }\textbf {\bibinfo {volume} {D96}},\ \bibinfo {pages}
  {074510} (\bibinfo {year} {2017}{\natexlab{b}})},\ \Eprint
  {http://arxiv.org/abs/1708.04897} {arXiv:1708.04897 [hep-lat]} \BibitemShut
  {NoStop}%
\bibitem [{\citenamefont {Borsanyi}\ \emph {et~al.}(2018)\citenamefont
  {Borsanyi}, \citenamefont {Fodor}, \citenamefont {Guenther}, \citenamefont
  {Katz}, \citenamefont {Szabó}, \citenamefont {Pasztor}, \citenamefont
  {Portillo},\ and\ \citenamefont {Ratti}}]{Borsanyi:2018grb}%
  \BibitemOpen
  \bibfield  {author} {\bibinfo {author} {\bibfnamefont {S.}~\bibnamefont
  {Borsanyi}}, \bibinfo {author} {\bibfnamefont {Z.}~\bibnamefont {Fodor}},
  \bibinfo {author} {\bibfnamefont {J.~N.}\ \bibnamefont {Guenther}}, \bibinfo
  {author} {\bibfnamefont {S.~K.}\ \bibnamefont {Katz}}, \bibinfo {author}
  {\bibfnamefont {K.~K.}\ \bibnamefont {Szabó}}, \bibinfo {author}
  {\bibfnamefont {A.}~\bibnamefont {Pasztor}}, \bibinfo {author} {\bibfnamefont
  {I.}~\bibnamefont {Portillo}}, \ and\ \bibinfo {author} {\bibfnamefont
  {C.}~\bibnamefont {Ratti}},\ }\href@noop {} {\  (\bibinfo {year} {2018})},\
  \Eprint {http://arxiv.org/abs/1805.04445} {arXiv:1805.04445 [hep-lat]}
  \BibitemShut {NoStop}%
\bibitem [{\citenamefont {Bazavov}\ \emph {et~al.}(2013)\citenamefont {Bazavov}
  \emph {et~al.}}]{Bazavov:2013dta}%
  \BibitemOpen
  \bibfield  {author} {\bibinfo {author} {\bibfnamefont {A.}~\bibnamefont
  {Bazavov}} \emph {et~al.},\ }\href {\doibase 10.1103/PhysRevLett.111.082301}
  {\bibfield  {journal} {\bibinfo  {journal} {Phys. Rev. Lett.}\ }\textbf
  {\bibinfo {volume} {111}},\ \bibinfo {pages} {082301} (\bibinfo {year}
  {2013})},\ \Eprint {http://arxiv.org/abs/1304.7220} {arXiv:1304.7220
  [hep-lat]} \BibitemShut {NoStop}%
\bibitem [{\citenamefont {Vovchenko}\ \emph
  {et~al.}(2017{\natexlab{b}})\citenamefont {Vovchenko}, \citenamefont
  {Motornenko}, \citenamefont {Alba}, \citenamefont {Gorenstein}, \citenamefont
  {Satarov},\ and\ \citenamefont {Stoecker}}]{Vovchenko:2017zpj}%
  \BibitemOpen
  \bibfield  {author} {\bibinfo {author} {\bibfnamefont {V.}~\bibnamefont
  {Vovchenko}}, \bibinfo {author} {\bibfnamefont {A.}~\bibnamefont
  {Motornenko}}, \bibinfo {author} {\bibfnamefont {P.}~\bibnamefont {Alba}},
  \bibinfo {author} {\bibfnamefont {M.~I.}\ \bibnamefont {Gorenstein}},
  \bibinfo {author} {\bibfnamefont {L.~M.}\ \bibnamefont {Satarov}}, \ and\
  \bibinfo {author} {\bibfnamefont {H.}~\bibnamefont {Stoecker}},\ }\href
  {\doibase 10.1103/PhysRevC.96.045202} {\bibfield  {journal} {\bibinfo
  {journal} {Phys. Rev.}\ }\textbf {\bibinfo {volume} {C96}},\ \bibinfo {pages}
  {045202} (\bibinfo {year} {2017}{\natexlab{b}})},\ \Eprint
  {http://arxiv.org/abs/1707.09215} {arXiv:1707.09215 [nucl-th]} \BibitemShut
  {NoStop}%
\bibitem [{Kai()}]{KaiCommun}%
  \BibitemOpen
  \href@noop {} {}\bibinfo {note} {K. Gallmeister, private
  communication.}\BibitemShut {Stop}%
\bibitem [{\citenamefont {Roberge}\ and\ \citenamefont
  {Weiss}(1986)}]{Roberge:1986mm}%
  \BibitemOpen
  \bibfield  {author} {\bibinfo {author} {\bibfnamefont {A.}~\bibnamefont
  {Roberge}}\ and\ \bibinfo {author} {\bibfnamefont {N.}~\bibnamefont
  {Weiss}},\ }\href {\doibase 10.1016/0550-3213(86)90582-1} {\bibfield
  {journal} {\bibinfo  {journal} {Nucl. Phys.}\ }\textbf {\bibinfo {volume}
  {B275}},\ \bibinfo {pages} {734} (\bibinfo {year} {1986})}\BibitemShut
  {NoStop}%
\bibitem [{\citenamefont {Vovchenko}\ \emph
  {et~al.}(2017{\natexlab{c}})\citenamefont {Vovchenko}, \citenamefont
  {Pasztor}, \citenamefont {Fodor}, \citenamefont {Katz},\ and\ \citenamefont
  {Stoecker}}]{Vovchenko:2017xad}%
  \BibitemOpen
  \bibfield  {author} {\bibinfo {author} {\bibfnamefont {V.}~\bibnamefont
  {Vovchenko}}, \bibinfo {author} {\bibfnamefont {A.}~\bibnamefont {Pasztor}},
  \bibinfo {author} {\bibfnamefont {Z.}~\bibnamefont {Fodor}}, \bibinfo
  {author} {\bibfnamefont {S.~D.}\ \bibnamefont {Katz}}, \ and\ \bibinfo
  {author} {\bibfnamefont {H.}~\bibnamefont {Stoecker}},\ }\href {\doibase
  10.1016/j.physletb.2017.10.042} {\bibfield  {journal} {\bibinfo  {journal}
  {Phys. Lett.}\ }\textbf {\bibinfo {volume} {B775}},\ \bibinfo {pages} {71}
  (\bibinfo {year} {2017}{\natexlab{c}})},\ \Eprint
  {http://arxiv.org/abs/1708.02852} {arXiv:1708.02852 [hep-ph]} \BibitemShut
  {NoStop}%
\bibitem [{\citenamefont {Cheng}\ \emph {et~al.}(2008)\citenamefont {Cheng}
  \emph {et~al.}}]{Cheng:2007jq}%
  \BibitemOpen
  \bibfield  {author} {\bibinfo {author} {\bibfnamefont {M.}~\bibnamefont
  {Cheng}} \emph {et~al.},\ }\href {\doibase 10.1103/PhysRevD.77.014511}
  {\bibfield  {journal} {\bibinfo  {journal} {Phys. Rev.}\ }\textbf {\bibinfo
  {volume} {D77}},\ \bibinfo {pages} {014511} (\bibinfo {year} {2008})},\
  \Eprint {http://arxiv.org/abs/0710.0354} {arXiv:0710.0354 [hep-lat]}
  \BibitemShut {NoStop}%
\bibitem [{\citenamefont {Skokov}\ \emph {et~al.}(2010)\citenamefont {Skokov},
  \citenamefont {Friman}, \citenamefont {Nakano}, \citenamefont {Redlich},\
  and\ \citenamefont {Schaefer}}]{Skokov:2010sf}%
  \BibitemOpen
  \bibfield  {author} {\bibinfo {author} {\bibfnamefont {V.}~\bibnamefont
  {Skokov}}, \bibinfo {author} {\bibfnamefont {B.}~\bibnamefont {Friman}},
  \bibinfo {author} {\bibfnamefont {E.}~\bibnamefont {Nakano}}, \bibinfo
  {author} {\bibfnamefont {K.}~\bibnamefont {Redlich}}, \ and\ \bibinfo
  {author} {\bibfnamefont {B.~J.}\ \bibnamefont {Schaefer}},\ }\href {\doibase
  10.1103/PhysRevD.82.034029} {\bibfield  {journal} {\bibinfo  {journal} {Phys.
  Rev.}\ }\textbf {\bibinfo {volume} {D82}},\ \bibinfo {pages} {034029}
  (\bibinfo {year} {2010})},\ \Eprint {http://arxiv.org/abs/1005.3166}
  {arXiv:1005.3166 [hep-ph]} \BibitemShut {NoStop}%
\bibitem [{\citenamefont {Sollfrank}\ \emph {et~al.}(1997)\citenamefont
  {Sollfrank}, \citenamefont {Huovinen}, \citenamefont {Kataja}, \citenamefont
  {Ruuskanen}, \citenamefont {Prakash},\ and\ \citenamefont
  {Venugopalan}}]{Sollfrank:1996hd}%
  \BibitemOpen
  \bibfield  {author} {\bibinfo {author} {\bibfnamefont {J.}~\bibnamefont
  {Sollfrank}}, \bibinfo {author} {\bibfnamefont {P.}~\bibnamefont {Huovinen}},
  \bibinfo {author} {\bibfnamefont {M.}~\bibnamefont {Kataja}}, \bibinfo
  {author} {\bibfnamefont {P.~V.}\ \bibnamefont {Ruuskanen}}, \bibinfo {author}
  {\bibfnamefont {M.}~\bibnamefont {Prakash}}, \ and\ \bibinfo {author}
  {\bibfnamefont {R.}~\bibnamefont {Venugopalan}},\ }\href {\doibase
  10.1103/PhysRevC.55.392} {\bibfield  {journal} {\bibinfo  {journal} {Phys.
  Rev.}\ }\textbf {\bibinfo {volume} {C55}},\ \bibinfo {pages} {392} (\bibinfo
  {year} {1997})},\ \Eprint {http://arxiv.org/abs/nucl-th/9607029}
  {arXiv:nucl-th/9607029 [nucl-th]} \BibitemShut {NoStop}%
\bibitem [{\citenamefont {Satarov}\ \emph {et~al.}(2009)\citenamefont
  {Satarov}, \citenamefont {Dmitriev},\ and\ \citenamefont
  {Mishustin}}]{Satarov:2009zx}%
  \BibitemOpen
  \bibfield  {author} {\bibinfo {author} {\bibfnamefont {L.~M.}\ \bibnamefont
  {Satarov}}, \bibinfo {author} {\bibfnamefont {M.~N.}\ \bibnamefont
  {Dmitriev}}, \ and\ \bibinfo {author} {\bibfnamefont {I.~N.}\ \bibnamefont
  {Mishustin}},\ }\href {\doibase 10.1134/S1063778809080146} {\bibfield
  {journal} {\bibinfo  {journal} {Phys. Atom. Nucl.}\ }\textbf {\bibinfo
  {volume} {72}},\ \bibinfo {pages} {1390} (\bibinfo {year} {2009})},\ \Eprint
  {http://arxiv.org/abs/0901.1430} {arXiv:0901.1430 [hep-ph]} \BibitemShut
  {NoStop}%
\end{thebibliography}%


\end{document}